\documentclass[5p,twocolumn,sort&compress,lefttitle]{elsarticle}
\usepackage{graphicx,graphics}
\usepackage{amsmath,amsfonts,amssymb}
\usepackage{color}
\definecolor{erase}{gray}{0.75}
\newcommand{\newc}{black}
\newcommand{\erase}[1]{\!\!}

\usepackage{subfigure}
\renewcommand{\thesubfigure}{\alph{subfigure}}
\makeatletter
  \renewcommand{\@thesubfigure}{(\thesubfigure)\hskip\subfiglabelskip}
\makeatother

\usepackage{subfloat}
\usepackage{float}
\usepackage[format=plain, justification=centerlast]{caption}

\usepackage{mylatexcommands}
\usepackage{submath}

\newcommand{\figdir}{./figs/}
\newcommand{\phient}{\phiy_{ent}}
\newcommand{\phienep}{\splus{\phiy_{ene}}}
\newcommand{\phienem}{\sminus{\phiy_{ene}}}

\newcommand{\rA}{\tilde{r}_{\ssub{0.5}{A}}}

\newcommand{\rE}{\tilde{r}_{\ssub{0.5}{E}}}     

\newcommand{\rK}{\tilde{r}_{\ssub{0.5}{K}}}    

\begin{document}
\title{Fluid statics of a self-gravitating perfect-gas isothermal sphere}
\date{\today}
\author[1]{Domenico Giordano\corref{ca}}\ead{dg.esa.retired@gmail.com}       
\author[2]{Pierluigi Amodio}         
\author[2]{Felice Iavernaro}          
\author[2]{Arcangelo Labianca}     
\author[2]{Monica Lazzo}            
\author[3]{Francesca Mazzia}        
\author[2]{Lorenzo Pisani}

\cortext[ca]{Corresponding author}

\address[1]{European Space Agency - ESTEC (retired), The Netherlands}
\address[2]{Dipartimento di Matematica, Universit\`a di Bari, Italy}
\address[3]{Dipartimento di Informatica, Universit\`a di Bari, Italy}

\begin{abstract}
We open the paper with introductory considerations describing the motivations of our long-term research plan targeting gravitomagnetism, illustrating the fluid-dynamics numerical test case selected for that purpose, that is, a perfect-gas sphere contained in a solid shell located in empty space sufficiently away from other masses, and defining the main objective of this study: the determination of the gravitofluid-static field required as initial field ($t=0$) in forthcoming fluid-dynamics calculations.
The determination of the gravitofluid-static field requires the solution of the isothermal-sphere Lane-Emden equation. 
We do not follow the habitual approach of the literature based on the prescription of the central density as boundary condition; we impose the gravitational field at the solid-shell internal wall. 
As the discourse develops, we point out differences and similarities between the literature's and our approach.
We show that the nondimensional formulation of the problem hinges on a unique physical characteristic number that we call gravitational number because it gauges the self-gravity effects on the gas' fluid statics.
We illustrate and discuss numerical results; some peculiarities, such as gravitational-number upper bound and multiple solutions, lead us to investigate the thermodynamics of the physical system, particularly entropy and energy, and preliminarily explore whether or not thermodynamic-stability reasons could provide justification for either selection or exclusion of multiple solutions.
We close the paper with a summary of the present study in which we draw conclusions and describe future work.
\end{abstract}

\maketitle

\section{Introduction}\label{intro}
\flushbottom
The motivation and the inspiration for the research activities that we begin to describe in this paper originate in a somewhat peculiar manner from aerothermodynamics, a physical/engineering discipline rather distant from gravitation.
In 2001, the aerothermodynamics section of the European Space Agency initiated a series of studies driven by interest in investigating the influence of electromagnetic fields on the heat transfer to a body in hypersonic flow.
The studies addressed theoretical, numerical and experimental aspects of the complex physics encompassed in electromagnetic fluid dynamics; the most challenging part was undoubtedly the one aiming at the development of numerical algorithms for the solution of the fully coupled Maxwell equations and Navier-Stokes equations, the latter with due inclusion of the physics characterizing flows in strong thermochemical nonequilibrium.
The coupling is conceptually smooth and straightforward from a theoretical point of view; 
on the other hand, coupling the two sets of equations from a numerical point of view is a different story in consequence of the huge disparity in orders of magnitude between the speeds of light, on the electromagnetism side, and sound, on the fluid-dynamics side. 
As a matter of fact, the numerical challenge turned out to be formidable and it is still unbeaten nowadays, many assaults by practitioners of computational fluid dynamics notwithstanding. 
The core difficulty lies mainly in the diffusion and relaxation processes taking place in thermochemical nonequilibrium that add numerical heaviness to the field equations with terms of such different orders of magnitude as to make their interplay practically uncontrollable, at least until now, by all attempted numerical-algorithm machineries.

How about gravitational fields? 
Terrestrial fluid dynamicists are prevalently occupied with applications on the surface of the planet; for them, the gravitational field is externally imposed, uniform and, above all, known because the contribution to the field from the fluid mass under consideration and from the other masses in the universe is absolutely negligible with respect to the planet's contribution. 
Astrophysical fluid dynamicists dealing with non-relativistic applications resort to Newton's theory of gravity to study the dynamics of (self-)gravitating fluid masses. 
Yet, Newton's theory of gravity is an action-at-distance theory; its incompatibility with the relativity theory has been overwhelmingly detected, addressed and resolved by well-known eminent physicists. 
More humbly, a fluid dynamicist could add that Newton's theory of gravity smoothly matches situations belonging to fluid statics and steady-state fluid dynamics but presents conceptual rugosities when confronted with time-dependent fluid-dynamics circumstances. 
Straightforward analysis suggests that a theory of gravity based on a single-vector field, like Newton's, is not sufficient when the non-relativistic dynamics of a (self-)gravitating fluid mass is unsteady; two vector fields are necessary and they are governed by differential equations whose mathematical structure coincide with the Maxwell equations in which densities of matter's mass and linear momentum play the role of field sources \cite{jh2007ajp,jh2009ejop,gn2015jmp,gl2018jamp,ds2018mfm}. 
Interestingly enough, the bottom-up conclusions achieved in a fluid-dynamics context are confirmed by a top-down approach that starts from Einstein's theory of gravity and, in the weak but time-dependent field approximation, leads into the realm of the fiercely debated subject known as gravitomagnetism \cite{jm1865ptrsl,gh1870zmp,ft1872cras,ft1890cras,oh1893te1,oh1893te2,oh1971v1,hl1900vknaw,hl1900pknaw,hp1905cras,hp1906rcmp,ae1912vms,gn1914pz,gn1914ofvsf,ma1915jre,ma2007,ht1918phz,jl1918pz,ht1918pz,ht2012grg,rf1961pire,vb1977prd,wr1977,db1985ajp,pk1987ajp,hk1988ajp,eh1991ajp,gc1991pe,rj1992aop,kmd1996pn,kmd1997ajp,sc2000cqg,jps2001,li2007,bm2008cqg,vb2011gc,db2015grg,hb2015,ds2015jdde}, 
the battleground in which scientists struggle to chase experimentally the proof of existence of the Lorentz-type force for gravity.

Gravitational fluid dynamics offers, therefore, the potential opportunity to skirmish with the numerical difficulties of the Maxwell-Navier-Stokes system of differential equations but without the numerical heaviness deriving from the complex physics of flows in thermochemical nonequilibrium; a perfect-gas model in standard conditions of pressure and temperature will do! 
\begin{figure}[h]
	\includegraphics[keepaspectratio=true, trim= 5ex 8ex 4ex 10ex , clip , width=\columnwidth]{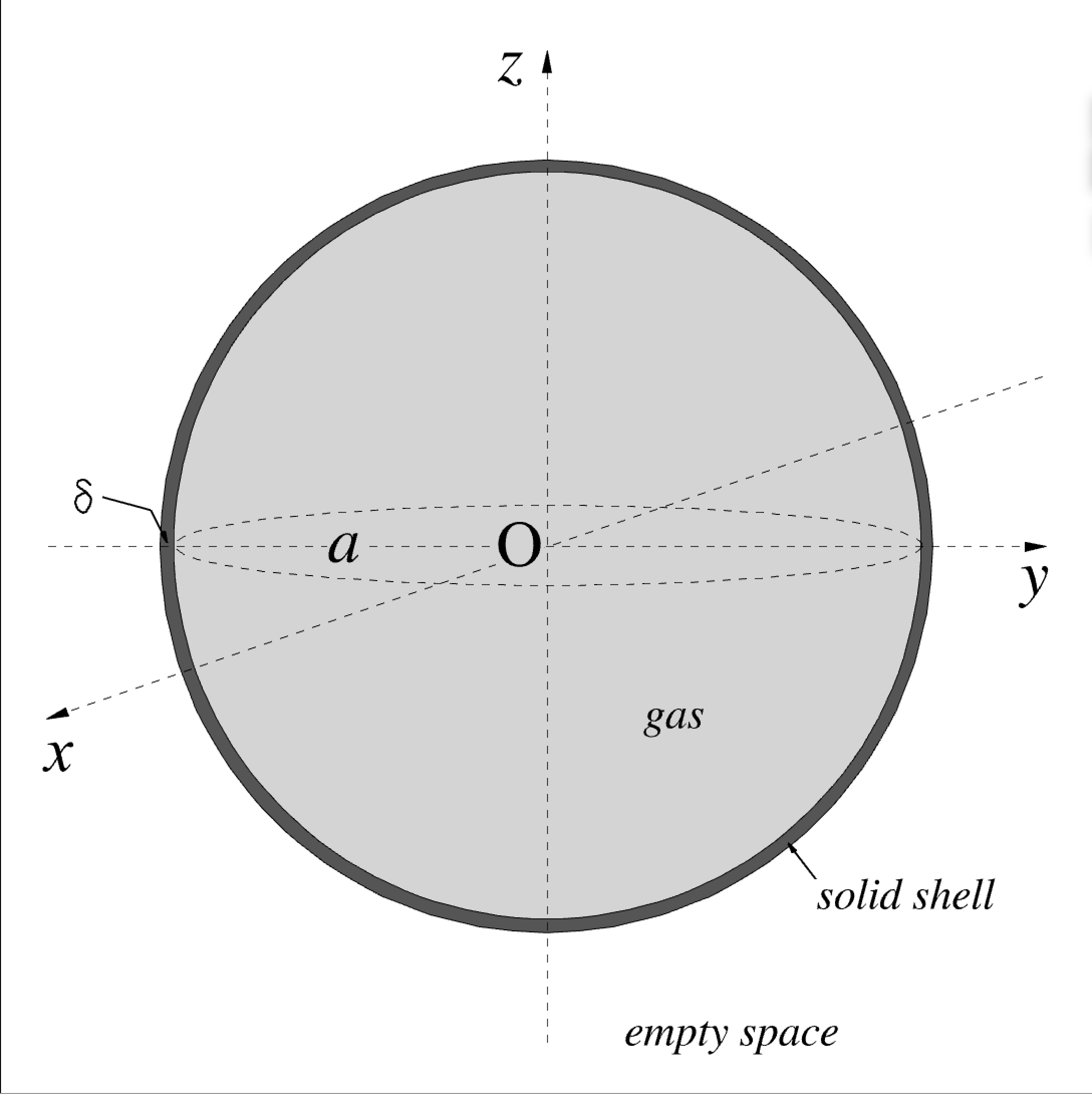}
	\caption{Study test case: fluid dynamics of a self-gravitating perfect gas inside a spherical solid shell.\hfill\ }\label{sphere}
\end{figure}
This is the challenge we settled to deal with and the test case we selected for this purpose is illustrated in Fig.~\ref{sphere}. 
A perfect gas is contained in a spherical solid shell of radius $a$ and thickness $\deltay$ placed in empty space far away from other masses in the universe. 
The gas is self-gravitating and we imagine to induce non-relativistic motion in it by adequately changing the internal-wall temperature distribution, in an axisymmetric manner to conveniently simplify the flow pattern. 
We intend to calculate the ensuing flow field with both theories of gravity: Newton's and gravitomagnetic Maxwell's.

The main objective of this paper is the determination of the initial fluid and gravitational fields ($t=0$) required by the time-dependent calculation of the dynamics ($t > 0$).
We assume the gas initially quiescent [$\V(\R,0)=0$]; then, the pursuit of the initial fields leads straightforwardly to the kind of Lane-Emden equations, a well-known family of differential equations recurrent in astrophysics 
\cite{
hl1870ajs,eb1880inc, ar1882adp, wt1887pma, gh1888aom, gd1889ptrs,jj1902ptrs, re1907,ae1930, rf1930mnras, rf1931qjom,re1955zfa, wb1956mnras, sc1957,
dl1968mnras,ws1987,tp1990pr,gh2004,cc2007,jb2008,rk2012} 
and in terrestrial applications such as fuel ignition \cite{ab2000sjsc} and thermal explosions \cite{kr2013amm,rvg2011na}, that describes the so-called isothermal-gas sphere.
The main difference between our manner to deal with such a differential equation and the approach followed, more or less systematically, in the astrophysical community resides in the boundary conditions.
In our problem, gas confinement rules and the gravitational side of the physics provides the required boundary condition at the shell internal wall ($r=a$); there is no need to, and in fact we do not, impose the gas density either at the center or at any other radial position in the gas sphere.
In the Lane-Emden problem, there is no gas confinement and a general consensus exists supporting the idea that the imposition of the density at the center of the gas sphere constitutes a legitimate boundary condition.
We have grappled with such an idea for a long time but have not been able to reconcile its presumed legitimacy with the arbitrariness descending from the undeniable lack of any independent physical information fixing the density at the center of the gas sphere.

The paper is logically organized in two parts.
In the first part (\Rse{fsgsf}), we describe the solution procedure to obtain the initial fields for our test case; we also briefly review the salient aspects of the Lane-Emden solution for the isothermal gas sphere and emphasize differences/equivalences with our approach (\Rse{ap-le}).
In the second part (\Rse{thd}), we concentrate on the thermodynamics of the physical system and discuss entropy, energy and aspects of thermodynamic stability.


\section{Gravitofluid-static fields} \label{fsgsf}

\subsection{Governing equations and boundary conditions} \label{ge+bc}
Let us consider a \textit{prescribed} mass $m_{g}$ of perfect gas in mechanical equilibrium [$\V(\R,t)=0$] contained in the spherical solid shell illustrated in \Rfi{sphere}. 
The shell has internal radius $a$, contains a volume $V = 4/3\, \piy a^{3}$ and has thickness $\deltay$.
The gravitational field occupies all space and is governed by
\begin{subequations} \label{gfeqs}
	\begin{equation}
	  \Lap\psiy = 
	              \begin{cases} \; 4\piy G \rhoy      \quad & \text{gas}  \\[2.5ex] 
	                              \; 4\piy G \rhoy_{s} & \text{shell} \\[2.5ex] 
	                              \; 0                    & \text{empty space}   
	              \end{cases} \label{gpot}
	\end{equation} \vspace*{.5\baselineskip}
	\begin{equation}
	  \g = - \Grad\psiy    \label{gvec}
	\end{equation}
\end{subequations}
according to Newton's theory of gravity.
The fluid-static field is governed by the static versions of the traditional balance equations of mass, momentum and energy
\begin{subequations} \label{ffeqs}
  \begin{align}
    \pder{\rhoy}{t}                                                         & = 0 \label{mass} \\[.25\baselineskip]
    \Grad p - \rhoy \g                                                     & = 0 \label{mom} \\[.25\baselineskip]
    \rhoy c_{v}\pder{T}{t} - \Dive\left( \lambdas \Grad T \right)   & = 0 \label{ene}    
  \end{align}
\end{subequations}
together with the thermodynamic state equation 
\begin{equation}\label{pgse}
  p = \rhoy RT
\end{equation}
In \REqs{gfeqs}{pgse},
\begin{tabbing}
	i \= xx \= xxxxxxxxxx \kill
	\> $\psiy$        \> gravitational potential \\
	\> $G$            \> gravitational constant, $6.67428\cdot10^{-11}$m$^{3}\cdot$kg$^{-1}\cdot$s$^{-2}$ \\
	\> $\rhoy$       \> mass density (gas) \\
	\> $\rhoy_{s}$   \> mass density (shell)\\
	\> $\g$            \> gravitational field \\
	\> $p$            \> pressure \\
	\> $c_{v}$        \> constant-volume specific heat \\
	\> $T$            \> temperature \\
	\> $\lambdas$    \> thermal conductivity \\
	\> $R$            \> gas constant 
\end{tabbing}
The continuity equation [\REq{mass}] enforces the time independence of the gas density.
The mechanical equilibrium exists also in the shell and, therefore, its density must also be time independent $(\pdet{\rhoy_{s}}{t}=0)$;
moreover, we consider negligible the deformation field and assume the shell density uniform and prescribed.
Consequently, time independence is sequentially passed on to gravitational potential and field by \REqd{gpot}{gvec}, to pressure by the momentum equation [\REq{mom}], and to temperature by the state equation [\REq{pgse}].
The temperature time derivative in the energy equation [\REq{ene}] vanishes identically (we left it in there only for formal consistency and clarity) and the heat flux, assumed according to Fourier law 
\begin{equation}\label{fourier}
   \hf = - \lambdas \Grad T
\end{equation}
becomes necessarily solenoidal
\begin{equation}\label{ene.ss}
  \Dive\left( \lambdas \Grad T \right) = 0
\end{equation}
The mechanical equilibrium of the gas implies temperature uniformity in the containing shell;
otherwise, temperature gradients would settle in at the gas/solid wall and would induce motion in the gas through the pressure gradient in the momentum equation.
An important consequence ensuing from the shell-temperature uniformity is the spherical symmetry of the physical system; hence, we conveniently select standard polar coordinates $r,\thetay,\phiy$ with origin coincident with the center of the spherical geometry (\Rfi{sphere}).
Before further processing the governing equations, it is convenient to pay a bit of attention to the spherically-symmetric Laplace operator
\begin{equation}\label{lap.rad}
   \Lap  = \frac{1}{r^{2}}\pder{}{r}\left( r^{2} \pder{}{r} \right) = \psde{}{r} +\frac{2}{r} \pder{}{r}
\end{equation}
because it may become unbounded when $r\rightarrow 0$. 
Clearly, we must assume 
\begin{equation} \label{fder.0}
   {\left. \pder{\ }{r} \right|}_{r=0} = 0
\end{equation}
in order to avoid the physically unacceptable discontinuity; then, the second term on the rightmost-hand side of \REq{lap.rad} becomes an indeterminate form that can be cured by de l'H\^{o}pital's theorem 
\begin{equation}\label{st.lap.rad.0}
   \lim_{r \rightarrow 0} \; \frac{1}{r} \pder{}{r} = {\left. \psde{\ }{r} \right|}_{r=0}  
\end{equation}
and the spherically-symmetric Laplace operator [\REq{lap.rad}] becomes discontinuity-free
\begin{equation}\label{lap.rad.0}
   \lim_{r \rightarrow 0} \; \Lap  =  3{\left. \psde{\ }{r} \right|}_{r=0} 
\end{equation}

The differential equations [\REqs{gfeqs}{ffeqs}] governing the statics of the physical system, composed by gas, shell and gravitational field, must be supplemented with the required boundary conditions.
Let us begin with the gravitational field which, of course, has only the radial component. 
The expansion of \REq{gpot} (gas)
\begin{equation} \label{gpot.gas.exp}
   \psde{\psiy}{r} +\frac{2}{r} \pder{\psiy}{r} = 4\piy G \rhoy
\end{equation}
provides, according to \REq{fder.0}, the boundary condition
\begin{subequations}\label{gvec.bc}
   \begin{equation} \label{fder.0.gpot}
      {\left. \pder{\psiy}{r} \right|}_{r=0} = - g(0) = 0
   \end{equation}
   enforcing the vanishing of the gravitational field at the sphere center.
   The other boundary conditions descend from the gravitational-field continuity at the shell internal wall
   \begin{equation}\label{gvec.c.g-s}
      g(\sminus{a}) = g(\splus{a})
   \end{equation}
   at the shell external wall
   \begin{equation}\label{gvec.c.s-es}
      g[\sminus{(a+\deltay)}] = g[\splus{(a+\deltay)}]
   \end{equation}
   and from the gravitational-field asymptotic vanishing
   \begin{equation}\label{gvec.c.inf}
      g(\infty) = 0
   \end{equation}
\end{subequations}
The spherical symmetry of the physical system allows to shortcut the integration of \REqq{gfeqs} by taking advantage of Gauss' theorem, whose application to concentric spherical surfaces inside the gas ($r<a$), inside ($a<r<a+\deltay$) and outside ($a+\deltay<r$) the shell yields respectively
\begin{equation} \label{gf.gauss}
	 g(r) =  
	 \begin{cases} \displaystyle -\frac{G}{r^{2}}         \left( 4\piy \int_{0}^{r}\rhoy(x)x^{2} dx \right) & \text{gas} \\[4ex]
                     \displaystyle -\frac{G m_{g}}{r^{2}} \left( 1 + \frac{m_{s}}{m_{g}} \frac{r^{3} - a^{3}}{(a + \deltay)^{3}- a^{3}}\right) & \text{shell} \\[4ex]
                     \displaystyle -\frac{G m_{g}}{r^{2}} \left( 1 + \frac{m_{s}}{m_{g}} \right) & \parbox{2em}{\centering empty \\ space}\end{cases}  \\[0ex]
\end{equation}
The term in parentheses in \REq{gf.gauss} (gas) is the mass of gas contained in the sphere of radius $r \; (< a)$; the variable $x$ represents a dummy integration variable.
The masses of gas and shell in \REq{gf.gauss} (shell \& empty space) are expressible in terms of the corresponding densities 
\begin{subequations} \label{masses}
  \begin{align}
    m_{g} & =  4\piy \int_{0}^{a}\rhoy(r)r^{2} dr \label{m.gas} \\[.5\baselineskip]
    m_{s} & =  \frac{4}{3} \piy \, \rhoy_{s} \left( (a+\deltay)^{3} - a^{3} \right) \label{m.shell} 
  \end{align}    
\end{subequations}
\REqb{m.gas} should be read from right to left: it is a constraint to which the density distribution must comply because the mass on its left-hand side is \textit{prescribed}.
It is also convenient at this point to introduce average density
\begin{subequations}\label{agdp}
	\begin{equation}\label{agd}
	   \brhoy = \dfrac{m_{g}}{V} = \dfrac{m_{g}}{\tfrac{4}{3} \piy a^{3}}
	\end{equation}
	and average pressure
	\begin{equation}\label{agp}
	   \bpres = \brhoy R T
	\end{equation}
\end{subequations}
of the gas in view of the forthcoming nondimensional analysis of \Rse{ge+bc.nd}. 
Compliance of the gravitational-field solution [\REq{gf.gauss}] with the gravitational boundary conditions [\REq{gvec.bc}] is easily verified.
The verification of \REq{fder.0.gpot} calls for a bit of attention because the limit of \REq{gf.gauss} (gas)
\begin{equation}\label{lim.gf.gauss.gas}
   g(0) = - \lim_{r \rightarrow 0} \; \frac{G}{r^{2}} \cdot \left[ 4\piy \int_{0}^{r}\rhoy(x)x^{2} dx \right] = 0
\end{equation}
requires repeated application of de l'H\^{o}pital's theorem but verification of field-continuity and asymptotic-vanishing boundary conditions [\REqs{gvec.c.g-s}{gvec.c.inf}] is straightforward.
Of course, \REq{gf.gauss} (gas) becomes operational only when the density distribution is known; however, it nails down the gravitational field at the shell internal wall to the immutable value 
\begin{equation} \label{gf.gauss.r=a}
	 g(a) = -\frac{G m_{g}}{a^{2}}  
\end{equation}
regardless of the specific density distribution that settles in the gas.
\REqb{gf.gauss.r=a} plays an important role for the determination of the fluid-static field to which we move on now.

%
The energy equation [\REq{ene.ss}] reduces to the spherically-symmetric form
\begin{equation} \label{ene.ss.rs}
	\frac{1}{r^{2}}\pder{}{r}\left( \lambdas r^{2} \pder{T}{r} \right) = 0
\end{equation}
that can be easily integrated starting from $r=0$
\begin{equation} \label{ene.ss.rs.i}
	\lambdas r^{2} \pder{T}{r} = \left( \lambdas r^{2} \pder{T}{r} \right)_{r=0} =0
\end{equation}
to yield
\begin{equation} \label{ene.ss.rs.i.i}
	\pder{T}{r} = 0
\end{equation}
According to \REq{ene.ss.rs.i.i}, the perfect gas cannot sustain thermal gradients in a spherically-symmetric steady state and must necessarily be isothermal. 
The thermal boundary condition must be prescribed at the shell internal wall because, in compliance with shell-temperature uniformity, the gas temperature there must coincide with the shell's
\begin{equation} \label{bc.Twall}
	T(a) = \Ts
\end{equation}
regardless of the angular position.
From the imposition of \REq{bc.Twall} we obtain
\begin{equation} \label{Tprofile}
	T(r) = \Ts
\end{equation}
and the state equation [\REq{pgse}] simplifies to the isothermal form
\begin{equation}\label{pgse.isoT}
	p = \rhoy R \Ts
\end{equation}
Hereinafter, we will drop the subscript `$s$' from the temperature symbol to simplify the notation.

The next step consists in introducing the isothermal state equation [\REq{pgse.isoT}] into the momentum equation [\REq{mom}] and solving the latter for the gravitational field
\begin{equation}
    \g =  R T \Grad\ln\rhoy \label{mom.isoT}
\end{equation}
Given the spherical symmetry, \REq{mom.isoT} is equivalent to the scalar form
\begin{equation}
    g = R T \pder{\ln\rhoy}{r} \label{mom.isoT.r}
\end{equation}
By comparing \REq{gf.gauss} (gas) and \REq{mom.isoT.r}, {the attentive reader may wonder whether or not those differently looking expressions of the gravitational field are equivalent; we provide the answer at the end of this section.}
\\ \noindent
The comparison of \REq{mom.isoT} with \REq{gvec} leads to
\begin{equation}
     \Grad \left( \psiy + R T \ln\rhoy \right) =  0   \label{gvec.r}
\end{equation}
from which we deduce the gravitational potential
\begin{equation}
     \psiy   =  A - R T \ln\rhoy   \label{gpot.rho}
\end{equation}
in terms of the gas density, save for an arbitrary and inessential constant $A$.
Then, the substitution of \REq{gpot.rho} into \REq{gpot} (gas) leads to the isothermal Lane-Emden equation
\begin{subequations} \label{LE-isoT}
	\begin{equation}\label{LE-isoT.3d}
	    \Lap\ln\rhoy + \frac{4\piy G}{RT}\rhoy = 0
	\end{equation}
that, according to \REq{lap.rad}, expands into the spherically-symmetric form
\begin{equation}\label{LE-isoT.r}
    \frac{1}{r^{2}}\pder{}{r}\left( r^{2} \pder{\ln\rhoy}{r} \right) + \frac{4\piy G}{RT}\rhoy = 0
\end{equation}
\end{subequations}
This differential equation must be integrated with the \textit{gravitational} boundary conditions [\REqd{fder.0.gpot}{gf.gauss.r=a}] reformulated, by account of \REq{mom.isoT.r}, in terms of the gas-density radial gradient
\begin{subequations} \label{bc.rho}
  \begin{align}
    {\left. \pder{\ln\rhoy}{r} \right|}_{r=0} & =  0 \label{bc.rho.r=0} \\[.5\baselineskip]
    {\left. \pder{\ln\rhoy}{r} \right|}_{r=a} & =  -\frac{G m_{g}}{a^{2}RT} \label{bc.rho.r=a} 
  \end{align}    
\end{subequations}
The mathematical problem [\REqs{LE-isoT.r}{bc.rho}] is thus well posed: 
we have a second-order differential equation and its two required boundary conditions whose physical meaning is clear and unambiguous. 
It is not necessary to think the density as assigned either at the sphere center 
or at any other radial position; as a matter of fact, the value $\rhoy(0)$ can, actually must, be obtained \textit{as a result} after that the density distribution has been determined from the integration of \REq{LE-isoT.r}.  
With that distribution in hand, the pressure distribution follows straightforwardly from the isothermal state equation [\REq{pgse.isoT}].

Before proceeding to cast the mathematical problem in nondimensional form, we believe a few brief considerations regarding physical consistency are in order.
Multiplication of \REq{LE-isoT.r} by $r^{2}$, subsequent integration from the sphere center up to the generic radial position $r$, and slight rearrangement of constant factors give
\begin{equation}\label{LE-isoT.r.int}
      RT \pder{\ln\rhoy}{r}  + \frac{G}{r^{2}}  \left[ 4\piy \int_{0}^{r}\rhoy(x) x^{2} dx \right] = 0
\end{equation}
\REqb{LE-isoT.r.int} shows the equivalence between the integral expression [\REq{gf.gauss} (gas)] originated from Gauss' theorem and the differential expression [\REq{mom.isoT.r}] of the gravitational field that follows from the fluid-statics momentum equation [\REq{mom}] of the isothermal gas.
This is not surprising: the equivalence was established when we substituted \REq{gpot.rho} into \REq{gpot} (gas).
\REqb{LE-isoT.r.int} also clearly indicates that $\pdet{\ln\rhoy}{r} < 0$ as a consequence of the attractive nature of the gravitational field; therefore, density and pressure [\REq{pgse.isoT}] are monotonically decreasing from the sphere center ($r=0$) to the shell internal wall ($r=a$) 
\begin{subequations} \label{rho.p.dec}
  \begin{eqnarray}
     \rhoy(a) < & \rhoy(r) & < \rhoy(0) \label{rho.dec} \\[.5\baselineskip]
     p(a)      < & p(r)          & < p(0)  \label{p.dec} 
  \end{eqnarray}    
\end{subequations}
If calculated at the shell internal wall, \REq{LE-isoT.r.int}  gives
\begin{equation}\label{LE-isoT.r=a.int}
      RT {\left. \pder{\ln\rhoy}{r} \right|}_{r=a}  + \frac{G}{a^{2}}  \left[ 4\piy \int_{0}^{a}\rhoy(x) x^{2} dx \right] = 0
\end{equation}
Substitution of \REq{bc.rho.r=a} into \REq{LE-isoT.r=a.int} and simplification of constant factors yield
\begin{equation}\label{LE-isoT.r=a.int.sim}
      -m_{g}  +  4\piy \int_{0}^{a}\rhoy(x) x^{2} dx  = 0
\end{equation}
\REqb{LE-isoT.r=a.int.sim} indicates that the gas-mass constraint [\REq{m.gas}] is identically satisfied; in other words, the imposition of the second boundary condition [\REq{bc.rho.r=a}] ensures that whatever density distribution be extracted from \REq{LE-isoT.r}, it complies automatically with the prescribed mass of the gas, a somewhat comforting result.

\subsection{Nondimensional formulation} \label{ge+bc.nd}
According to standard practice, we introduce nondimensional radial coordinate and density together with corresponding dimensional scale factors marked with a tilde
\begin{subequations} \label{nd.vars}
  \begin{align}
    r          & = \tilde{r}\,\etay \label{nd.d} \\[.5\baselineskip]
    \rhoy(r) & =  \trhoy\,\xiy(\etay) \label{nd.r} 
  \end{align}        
\end{subequations}
in order to cast the mathematical problem [\REqs{LE-isoT.r}{bc.rho}] in nondimensional form
\begin{subequations} \label{nd.prob.sf}
  \begin{equation}\label{LE-isoT.r.nd.sf}
      \frac{1}{\etay^{2}}\pder{}{\etay}\left( \etay^{2} \pder{\ln\xiy}{\etay} \right) + \frac{4\piy G {\tilde{r}}^{2} \trhoy}{RT}\,\xiy = 0
  \end{equation}
  \begin{align}
    {\left. \pder{\ln\xiy}{\etay} \right|}_{\setay=0} & =  0 \label{bc.rho.r=0.nd.sf} \\[.5\baselineskip]
    {\left. \pder{\ln\xiy}{\etay} \right|}_{\setay=a/\tilde{r}} & =  - \dfrac{G m_{g}\tilde{r}}{a^{2} R T} \label{bc.rho.r=a.nd.sf} 
  \end{align}    
\end{subequations}
The set of \REqq{nd.prob.sf} features three characteristic numbers
\begin{equation} \label{cn.sf}
    \begin{bmatrix} \; 
        \Piy_{1} = \dfrac{4\piy G {\tilde{r}}^{2} \trhoy}{RT} \quad & 
        \Piy_{2} = \dfrac{G m_{g}\tilde{r}}{a^{2} R T}         \quad & 
        \Piy_{4} = \dfrac{a}{\tilde{r}}                               \;  
    \end{bmatrix}
\end{equation}
In general, the scale factors are chosen according to, hopefully judicious, criteria of convenience and the $\Piy$-type characteristic numbers are deduced therefrom.
A more preferable approach is the other way around: it is more convenient to resolve the scale factors 
\begin{equation} \label{sf.cn}
    \begin{bmatrix} \; 
        \trhoy = \dfrac{\Piy_{1} \Piy_{4}^{2}}{3N} \brhoy   \quad  & 
        \Piy_{2} \Piy_{4} =\dfrac{G m_{g}}{aRT}               \quad  & 
        \tilde{r} = \dfrac{1}{\Piy_{4}}  a                         \;  
    \end{bmatrix}
\end{equation}
in terms of the $\Piy$-type characteristic numbers and to set the latter to convenient values.
The second approach has two indisputable advantages. 
The first is that the dependent $\Piy$-type characteristic numbers are disclosed with clear evidence;
indeed, \REq{sf.cn} reveals that $\Piy_{1}$ is unrestricted but $\Piy_{2}$ and $\Piy_{4}$ cannot be assigned independently.
True, this could have been detected easily also from attentive scrutiny of \REq{cn.sf} but just because we are dealing here only with two scale factors and three $\Piy$-type characteristic numbers.
If the set of differential equations and boundary conditions becomes overcrowded with scale factors and characteristic numbers then processing the analog of \REq{cn.sf} for the purpose of setting the scale factors to convenient values turns into an unmanageable task prone to unavoidable mistakes.
The second advantage of the second approach is that it brings to light the physical characteristic numbers that control the problem at hand.
In this case, \REq{sf.cn} displays only (the one we call) the gravitational number 
\begin{subequations}\label{gcn}
\begin{equation}\label{gcn1}
   N = \frac{G m_{g}}{aRT}
\end{equation}
Its definition is unique because it involves only supposedly known and definitely \textit{controllable} variables belonging to the physical system.
Its meaning becomes evident if we rewrite \REq{gcn1} as
\begin{equation}\label{gcn2}
   N = \frac{G m_{g}^{2}/a}{m_{g}RT}
\end{equation}
Hence, the gravitational number expresses the ratio of orders of magnitude of energies: the gravitational one goes in the numerator and the thermodynamic one goes in the denominator. 
Therefore, the role of the gravitational effects on the fluid-static field in the gas sphere should be expected to be negligible if $N \ll 1$ and to acquire importance for increasing $N$. 
Other interpretations have been attached to this characteristic number in the literature \cite{gd1889ptrs,dl1968mnras,tp1990pr,phc2002aa,jb2008}. 
\end{subequations}

With the scale factors enforced by \REq{sf.cn}, the nondimensional mathematical problem [\REqq{nd.prob.sf}] can be rephrased accordingly
\begin{subequations} \label{nd.prob.cn}
\begin{equation}\label{LE-isoT.r.nd.cn}
    \frac{1}{\etay^{2}}\pder{}{\etay}\left( \etay^{2} \pder{\ln\xiy}{\etay} \right) + \Piy_{1}\xiy = 0
\end{equation}
  \begin{align}
    {\left. \pder{\ln\xiy}{\etay} \right|}_{\setay=0}          & =  0 \label{bc.rho.r=0.nd.cn} \\[.5\baselineskip]
    {\left. \pder{\ln\xiy}{\etay} \right|}_{\setay=\tPis_{4}} & =  -\Piy_{2} \label{bc.rho.r=a.nd.cn} 
  \end{align}    
\end{subequations}
and the mass constraint [\REq{m.gas}] turns into the normalization condition
\begin{equation} \label{m.gas.nd}
    \frac{1}{N}       \frac{\Piy_{1}}{\Piy_{4}} \int_{0}^{\tPis_{4}} \etay^{2}\,\xiy(\etay) \,d\etay = 1
\end{equation}
Remembering the considerations at the end of \Rse{ge+bc} [\REqd{LE-isoT.r=a.int}{LE-isoT.r=a.int.sim}], we should expect \REq{m.gas.nd} to be identically satisfied due to the imposition of \REq{bc.rho.r=a.nd.cn}.

The next step consists in choosing the two independent $\Piy$-type characteristic numbers. 
In view of the necessarily numerical solution of \REq{LE-isoT.r.nd.cn}, we set $\Piy_{4}=1$ to keep the computational domain for $\etay$ fixed to the interval [0,1], a desirable feature when doing numerical calculations; we then set $\Piy_{1} = 3 N$ to remove the dependence of the density scale factor on the gravitational number.
With these choices, \REq{sf.cn} simplifies to
\begin{equation} \label{sf.cn.m}
    \begin{bmatrix} \; 
        \trhoy    = \brhoy \quad & 
        \Piy_{2} = N       \quad & 
        \tilde{r}  = a       \;  
    \end{bmatrix}
\end{equation}
and \REqq{nd.vars} yield the nondimensional variables
\begin{subequations} \label{nd.vars.f}
  \begin{align}
    r          & =  a\,\etay \label{nd.d.f} \\[.5\baselineskip]
    \rhoy(r) & =  \brhoy\,\xiy(\etay) \label{nd.r.f} 
  \end{align}        
\end{subequations}
that were also selected by Darwin \cite{gd1889ptrs}; the mathematical problem [\REqq{nd.prob.cn}] attains its final form 
\begin{subequations} \label{nd.prob.cn.m}
\begin{equation}\label{LE-isoT.r.nd.cn.m}
    \frac{1}{\etay^{2}}\pder{}{\etay}\left( \etay^{2} \pder{\ln\xiy}{\etay} \right) + 3N\xiy = 0
\end{equation}
  \begin{align}
    {\left. \pder{\ln\xiy}{\etay} \right|}_{\setay=0} & =  0    \label{bc.rho.r=0.nd.cn.m} \\[.5\baselineskip]
    {\left. \pder{\ln\xiy}{\etay} \right|}_{\setay=1} & =  - N \label{bc.rho.r=a.nd.cn.m} 
  \end{align}    
\end{subequations}
conveniently predisposed for numerical processing and the accompanying normalization condition [\REq{m.gas.nd}] becomes
\begin{equation} \label{m.gas.nd.m}
    3 \int_{0}^{1} \etay^{2}\,\xiy(\etay) \,d\etay = 1
\end{equation}
Obviously, every solution of \REqq{nd.prob.cn.m} depends parametrically on the gravitational number
\begin{equation}\label{xi.sol}
   \xiy = \xiy(\etay,N)
\end{equation}
but we will explicitly indicate that dependence only if required by the context.
We wish to point out what we believe to be fine features of the nondimensional problem [\REqq{nd.vars.f} and \REqq{nd.prob.cn.m}] based on \REq{sf.cn.m}: it includes boundary conditions whose physical legitimization is unambiguous; it is governed by one single, clearly identified physical characteristic number, the gravitational number $N$;
it involves nondimensional variables scaled with factors constructed with variables characterizing the physical system, known a priori and as precisely controllable as the gravitational number is. 

To complete the nondimensional formulation, we scale the gravitational field [\REqd{gf.gauss}{mom.isoT.r}] with the value [\REq{gf.gauss.r=a}] attained at the shell internal wall
\begin{equation} \label{gf.gauss.nd}
	 \frac{a^{2} g}{G m_{g}} = 
	 \begin{cases} \displaystyle -\frac{1}{\etay^{2}} \left( 3 \int_{0}^{\setay}\xiy(u)u^{2} du \right) & \text{gas} \\[4ex] 
                     \displaystyle -\frac{1}{\etay^{2}} \left( 1 + \frac{m_{s}}{m_{g}} \frac{\etay^{3} - 1}{(1 + \deltay/a)^{3}- 1}\right) & \text{shell} \\[4ex] 
                     \displaystyle -\frac{1}{\etay^{2}} \left( 1 + \frac{m_{s}}{m_{g}} \right) & \parbox{2em}{empty \\ space}\end{cases}  \\[2ex]
\end{equation}
\begin{equation}\label{gf.nd}
    \subeqn{\frac{a^{2} g}{G m_{g}} = \frac{1}{N} \pder{\ln\xiy}{\etay}}{(\;<0\;)}{}
\end{equation}
precisely as Ritter \cite{ar1882adp} and Darwin \cite{gd1889ptrs} also did,
and the pressure with its average value [\REq{pgse.isoT}]
\begin{equation}\label{p.nd}
   \frac{p}{\bpres} = \xiy(\etay)
\end{equation}

\subsection{Some analytical results}\label{sar}
It is somewhat interesting to analyze the qualitative behavior of the solutions of \REqq{nd.prob.cn.m} before proceeding to discuss numerical operations and results.

Given the selection $\Piy_{1}=3N, \Piy_{4}=1$, the inequalities shown in \REqq{rho.p.dec} merge into
\begin{equation}\label{rho.p.dec.nd}
     \xiy(1) <  \xiy(\etay) < \xiy(0) 
\end{equation}
Multiplication of \REq{rho.p.dec.nd} by $3\,\etay^{2}$ and integration on the interval [0,1], with due account of the normalization condition [\REq{m.gas.nd.m}],  leads to the interesting result
\begin{equation}\label{rho.p.dec.nd.strong}
     \xiy(1) <  1 < \xiy(0) 
\end{equation}
Thus, the density is always greater or lower than its average value at, respectively, the sphere center or the shell internal wall; therefore, a radial position $\obetay$ must exists at which the density attains its average value [\REq{agd}] 
\begin{equation}\label{etabar}
    \xiy(\obetay) = 1
\end{equation}
and the conceptual boundary between a denser (than average) core (\mbox{$ 0 \leq \etay < \obetay $}) and a less dense peripheral layer (\mbox{$ \obetay < \etay \leq 1 $}) can be set.

When $N \rightarrow \epsilon \ll 1$, \REqq{nd.prob.cn.m} provide infinite solutions
\begin{equation}\label{s.N=0}
   \xiy(\etay) \simeq C
\end{equation}
The normalization condition [\REq{m.gas.nd.m}] fixes the value of the constant to $C=1$ and singles out the only physical solution
\begin{equation}\label{s.N=0.p}
   \xiy(\etay) = \frac{\rhoy(r)}{\brhoy} = \frac{p(r)}{\bpres} \simeq 1
\end{equation}
Therefore, density and pressure are uniform throughout the gas sphere and the self-gravitating condition becomes negligible, as expected. 
The gravitational field in the gas sphere is easily obtained from \REq{gf.gauss.nd} (gas) with $\xiy(u)\simeq 1$ and turns out to decrease linearly from the sphere center to the shell internal wall
\begin{equation}\label{gf.nd.N=0}
    \subeqn{\lim_{N \rightarrow \epsilon} \frac{a^{2} g}{G m_{g}} = - \etay}{(\text{gas})}{}
\end{equation}
consistently with the existence of a uniform density.
Taking into account \REq{gf.nd.N=0}, we also deduce from the alternative equivalent expression [\REq{gf.nd}] of the gravitational field that
\begin{subequations}\label{gf.nd.N=0.ae}
   \begin{equation}\label{gf.nd.N=0.ae.1}
       \lim_{N \rightarrow \epsilon} \frac{1}{N} \pd{}{\ln\xiy}{\etay} = - \etay  
   \end{equation}
as well as slope and curvature
  \begin{align}
    \lim_{N \rightarrow \epsilon} \dfrac{1}{N} \pd{2}{\ln\xiy}{\etay} & =  - 1  \label{gf.nd.N=0.ae.2}\\[.5\baselineskip]
    \lim_{N \rightarrow \epsilon} \dfrac{1}{N} \pd{3}{\ln\xiy}{\etay} & =    0  \label{gf.nd.N=0.ae.3} 
  \end{align}    
\end{subequations}
in view of forthcoming considerations. 

\begin{table}[h] 
  \caption{physical meaning of partial derivatives\label{derivs.pm}}
  \begin{tabular*}{\columnwidth}{@{\extracolsep{\fill}}ccc} \hline\hline \\[-2ex]
      & {\parbox[c]{.15\textwidth}{gas density \\[.2\baselineskip] (logarithmic)}} & $N\times \left({\parbox[c]{.11\textwidth}{gravitational \\ field}} \right) $ \\[1.5ex] \hline \\[-2ex]
      $\pd{}{\ln\xiy}{\etay}$               & slope       &       \\[5ex]
      $\pd{2}{\ln\xiy}{\etay}$             & curvature  & slope \\[5ex] 
      $\pd{3}{\ln\xiy}{\etay}$             &              & curvature \\[2.5ex]
      \hline \hline
  \end{tabular*}
\end{table}
Considering $N \neq \epsilon \ll 1 $ now, more qualitative information can be gathered from the evaluation of the logarithmic-density derivatives, up to third order, evaluated at the boundary points of the interval [0,1].
The physical meaning of these derivatives is summarized for convenience in \Rta{derivs.pm}; the third-order derivative 
\begin{equation}\label{deriv3}
  \pd{3}{\ln\xiy}{\etay} = - \frac{2}{\etay^{2}} \left( \etay \pd{2}{\ln\xiy}{\etay} - \pd{}{\ln\xiy}{\etay} \right) - 3 N \xiy \pd{}{\ln\xiy}{\etay}
\end{equation}
is easily obtained from further derivation of \REq{LE-isoT.r.nd.cn.m}.
The situation at the sphere center
\begin{subequations} \label{derivs.eta=0}
  \begin{align}
    \pdat{}{\ln\xiy}{\etay}{\setay=0} & =  0 \label{derivs.eta=0.1} \\[.5\baselineskip]
    \pdat{2}{\ln\xiy}{\etay}{\setay=0} & =  - N \xiy(0) < 0 \label{derivs.eta=0.2}\\[.5\baselineskip]
    \pdat{3}{\ln\xiy}{\etay}{\setay=0} & =  0 \label{derivs.eta=0.3} 
  \end{align}    
is rather uneventful. 
\end{subequations}
First and second derivatives [\REqd{derivs.eta=0.1}{derivs.eta=0.2}] inform that the density attains a maximum there in consequence of the vanishing of the gravitational field. 
The latter's profile has negative slope and infinite curvature [\REq{derivs.eta=0.3}; evaluation requires application of de l'H\^{o}pital's theorem to first term on right-hand side of \REq{deriv3}]. 
The situation at the shell internal wall 
\begin{subequations} \label{derivs.eta=1}
\begin{align}
    \pdat{}{\ln\xiy}{\etay}{\setay=1}   & =  - N                                                      \label{derivs.eta=1.1} \\[.5\baselineskip]
    \pdat{2}{\ln\xiy}{\etay}{\setay=1} & =  - 3N \left[\xiy(1,N)-\frac{2}{3}\right]           \label{derivs.eta=1.2} \\[.5\baselineskip]
    \pdat{3}{\ln\xiy}{\etay}{\setay=1} & =  3N(N+2) \left[\xiy(1,N)-\frac{2}{N+2}\right]  \label{derivs.eta=1.3}
\end{align}
\end{subequations}
is more informative. 
According to \REqd{derivs.eta=0.2}{derivs.eta=1.2}, the function $\ln\xiy$ preserves its negative curvature on the whole interval [0,1] up to a value $N'$ of the gravitational number at which
\begin{equation}\label{ip}
  \xiy(1,N') = \frac{2}{3}
\end{equation}
If \mbox{$N=N'$} then the second derivative [\REq{derivs.eta=1.2}] vanishes; hence, an inflection point and a minimum appear for, respectively, logarithmic density and gravitational-field profiles.
The third derivative [\REq{derivs.eta=1.3}] takes on the value
\begin{equation}\label{3rdd}
   \pdat{3}{\ln\xiy}{\etay}{\setay=1} = 2 N' (N'-1)
\end{equation}
and its necessary positivity (gravitational-field minimum) leads to $N'>1$.
If $N'< N$ then the positions of both inflection point and minimum shift leftward inside the interval [0,1].
Another noticeable value $N''$ of the gravitational number is the one at which
\begin{equation}\label{ip.gf}
  \xiy(1,N'') = \frac{2}{N''+2}
\end{equation}
If $N=N''$ then the third derivative [\REq{derivs.eta=1.3}] vanishes and an inflection point appears for the gravitational-field profile.

\subsection{Numerical solution and results} \label{ge+bc.nd.res}
As repeatedly mentioned and well emphasized in the literature, the differential equation [\REq{LE-isoT.r.nd.cn.m}] governing the mathematical problem is not amenable to analytical integration and has to be dealt with numerically. 
In this regard, we have used three different numerical algorithms (labels F, FP, D in the forthcoming figures) independently coded in three different programming languages (R, matlab and fortran). 
The first algorithm (F) is based on the standard approach of recasting the second-order differential equation [\REq{LE-isoT.r.nd.cn.m}]  as a first-order system and then applying a general-purpose code for boundary-value problems \cite{ACMBVPTWP,TOM,bvpSolve,CONDMESH1,CONDMESH2}. 
The second algorithm (FP) \cite{ASjnaiam,ASaip} is based on high-order (up to eight) finite-difference schemes to solve general second-order ordinary differential equations subjected to Neumann, Dirichlet or mixed boundary conditions; the ideas underlying this approach have also been adapted to solve Sturm-Liouville problems with one and two parameters \cite{ALSWjamc,ALSWcpc}. 
The third algorithm (D) is based on a finite-difference scheme whose nonlinear algebraic system is treated with a multidimensional Newton-Raphson iterative method.
All the algorithms need initial guesses of the solution from which the iterative procedure starts and proceeds to the converged solution.
\begin{figure}[h]
  \includegraphics[keepaspectratio=true, trim= 5ex 8ex 2.9ex 20ex , clip , width=\columnwidth]{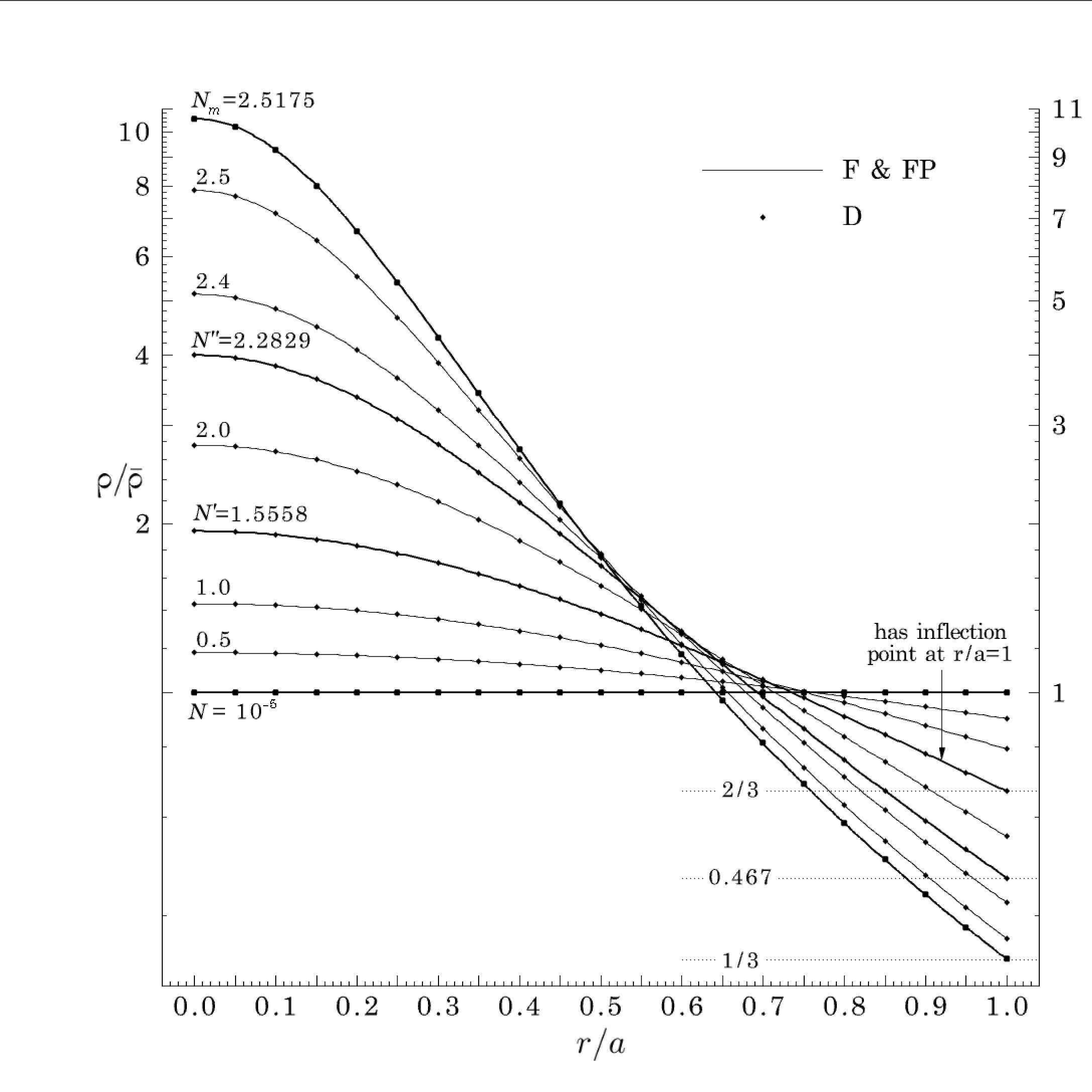}\\
  \caption{Gas-density profiles in logarithmic scale for selected values of the gravitational number\hfill\ }
  \label{p.xi}
\end{figure}
\begin{figure}[H]
  {\includegraphics[keepaspectratio=true, trim= 1ex 7ex 5ex 25ex , clip , width=\columnwidth]{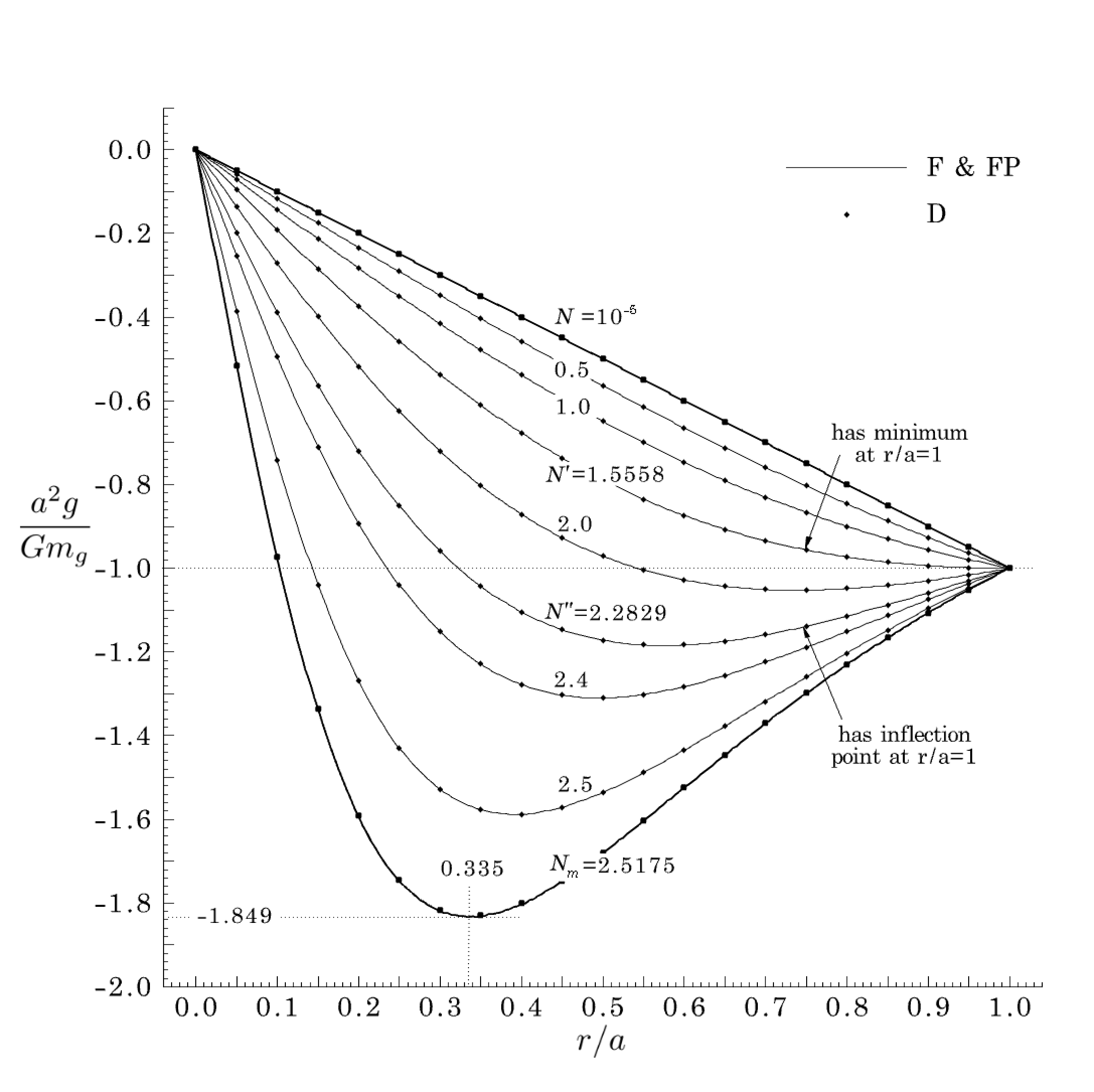}} \\
  \caption{Gravitational-field profiles inside the gas sphere for selected values of the gravitational number\hfill\ }
  \label{p.gf.g}
\end{figure}
The agreement among the results produced by the different algorithms is very satisfactory, as evidenced in \Rfid{p.xi}{p.gf.g};
unless otherwise specified, the data-representation convention displayed in the legends (solid line for algorithms F \& FP; solid circles for algorithm D) applies systematically in all figures.

\Rfib{p.xi} shows the gas-density profiles for selected values of the gravitational number, in logarithmic scale to illustrate more clearly the situation in the vicinity of the shell internal wall. 
In view of the discussion about the Lane-Emden solution (\Rse{ap-le}), we call the reader's attention to an important detail: the density values attained at sphere center and shell internal wall are unequivocally determined by the integration of \REqq{nd.prob.cn.m}; no knowledge of those values is required \textit{before} integration.
\Rfib{p.gf.g} shows the corresponding profiles of the gravitational field inside the gas sphere.
All the analytical characteristics described in \Rse{sar} are exhibited by the numerical results illustrated in \Rfid{p.xi}{p.gf.g}.
The density profiles feature the expected monotonic behavior [\REq{rho.p.dec.nd}] from a denser central core to a less dense peripheral layer in the neighborhood of the shell internal wall; the average-density boundary location  $\obetay$ [\REq{etabar}] is clearly detectable and, expectedly, it decreases with increasing $N$.
The calculation with \mbox{$N=10^{-5}\ll 1$} served to crosscheck whether or not the numerical algorithms would reproduce the analytical solution [\REqd{s.N=0.p}{gf.nd.N=0}]; they clearly do.
The density inflection point and gravitational-field minimum appear at \mbox{$N'\simeq 1.5558$}; the gravitational-field inflection point appears at \mbox{$N''\simeq 2.2829$}. 
Their locations shift leftward and the minimum becomes deeper as the gravitational number increases.
\begin{figure}[h]
  {\includegraphics[keepaspectratio=true, trim= 1ex 7ex 5ex 25ex , clip , width=\columnwidth]{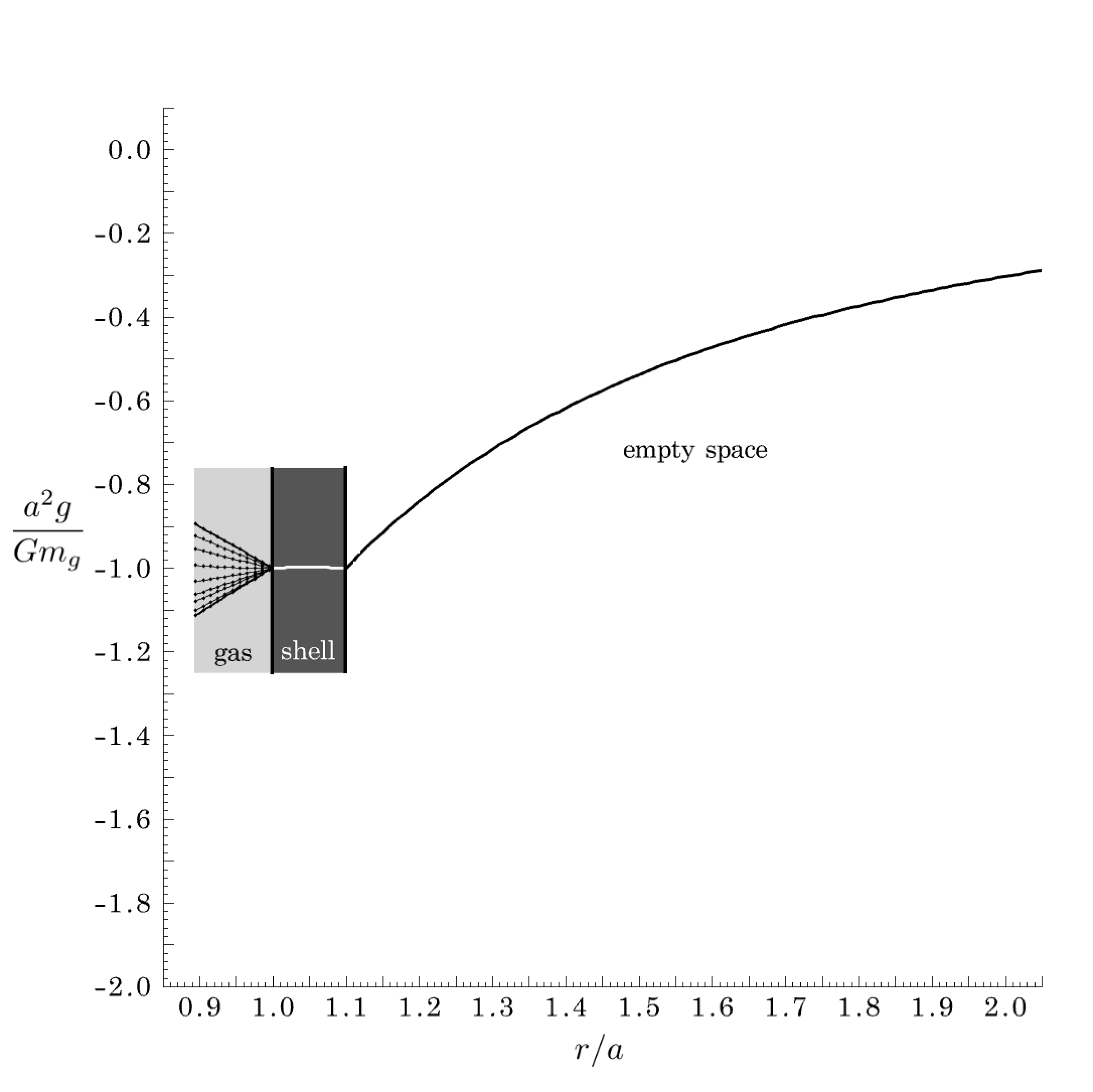}}
  \caption{Gravitational-field profile inside the shell and in empty space\hfill\ }
  \label{p.gf.ses}
\end{figure}
The rightward continuation of \Rfi{p.gf.g} is presented in \Rfi{p.gf.ses} which shows the analytical profile of the gravitational field inside the shell and in empty space up to two sphere radii produced by \REq{gf.gauss.nd} (shell, empty space) with $\deltay/a=0.1$ and $m_{s}/m_{g}=0.21$; the latter values were selected for the convenience of keeping the gravitational field inside the shell practically constant. 
The profile in \Rfi{p.gf.ses} does not depend on the gravitational number and is unique on account of the spherical symmetry of the physical system, regardless of how matter is radially distributed inside the sphere.

Similarly to what already found out in the literature, particularly Darwin \cite{gd1889ptrs}, Lynden-Bell and Wood \cite{dl1968mnras} and Padmanabhan \cite{tp1990pr}, with regard to the Lane-Emden solution, although detected and described within different perspectives, we also discovered two peculiar aspects of the physical problem: the existence of gravitational-number upper bound and of multiple solutions.
They are clearly illustrated in the diagram of the density at the shell internal wall, peripheral density hereinafter, as function of the gravitational number shown in \Rfi{p.xi1}.
\begin{figure}[h]
  \includegraphics[keepaspectratio=true, trim= 5ex 8ex 4ex 20ex , clip , width=\columnwidth]{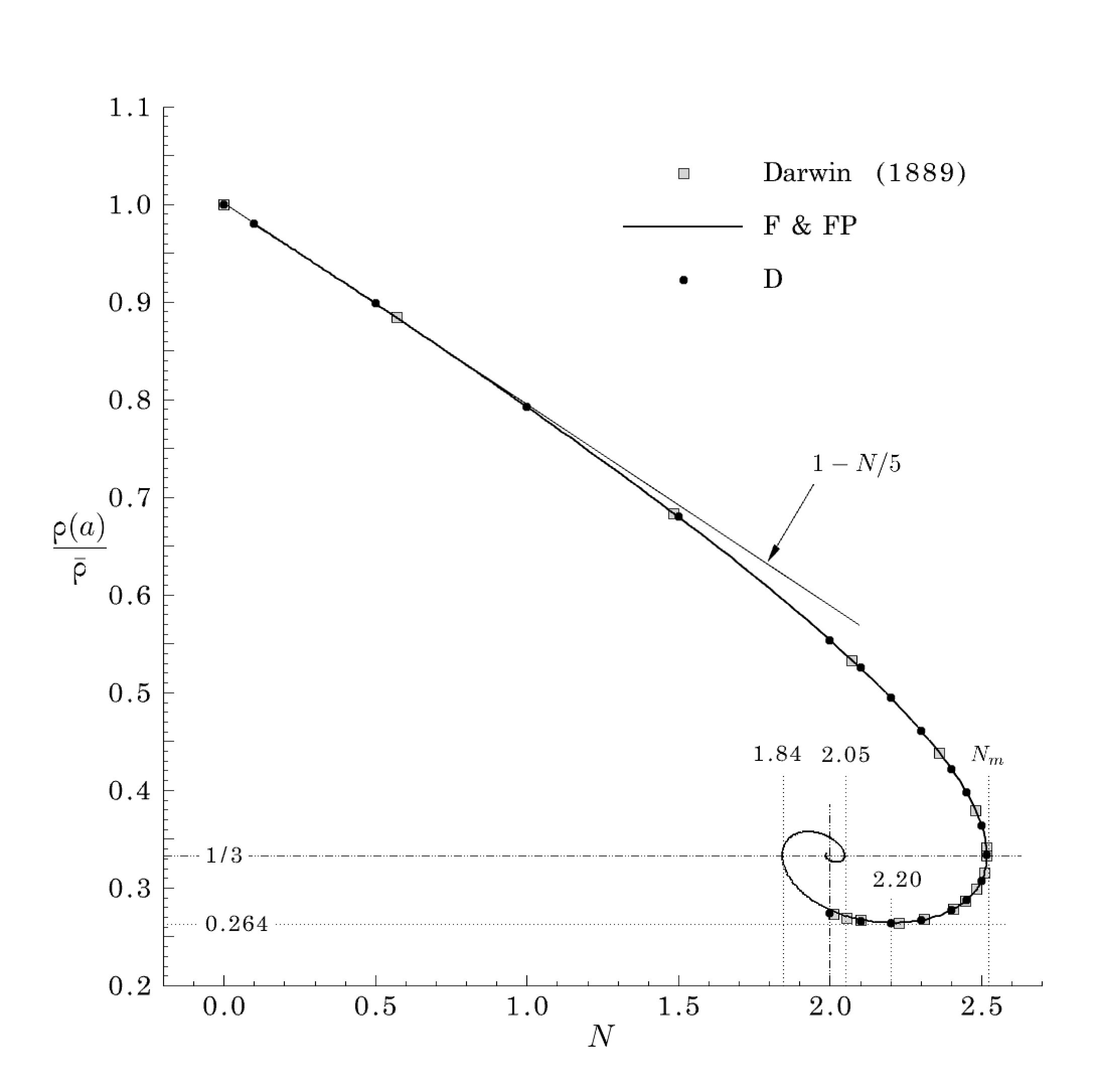}
  \caption{Peripheral density as function of the gravitational number according to Darwin \cite{gd1889ptrs} and to our results\hfill\ }
  \label{p.xi1}
\end{figure}
The profile practically follows a linear trend
\begin{equation}\label{linear}
    \frac{\rhoy(a)}{\brhoy} = \xiy(1,N) \simeq 1 - N/5 
\end{equation}
for $N\leq 1$, then it bends downward until it reaches vertical slope and marks the presence of an upper bound $N_{m}$ in correspondence to which the peripheral density reduces to 1/3 \cite{gd1889ptrs}; after that the profile spirals around the point \mbox{$\,N=2, \xiy(1,2) = 1/3\,$} as a consequence of the existence of multiple solutions when \mbox{$N>1.84$}.
At \mbox{$N\simeq 2.20$}, the profile goes through the absolute minimum \mbox{$\xiy(1,2.20) \simeq 0.264$} which represents a 74\% drop below the average density; there is no solution of \REqq{nd.prob.cn.m} less dense  than this one at the shell internal wall.
Another interesting diagram, shown in \Rfi{p.xi0}, is the one that illustrates the density at the sphere center, central density hereinafter, as function of the gravitational number.
The profile reveals no upper bound and goes through an infinite series of oscillations with decreasing amplitude about the vertical line \mbox{$N=2$} as a consequence of the existence of multiple solutions.
The value \mbox{$N=2$} is rather peculiar: it presupposes the existence of infinite solutions, the most extreme of which has infinite central density.
\begin{figure}[h]  
  \includegraphics[keepaspectratio=true, trim= 5ex 8ex 4ex 20ex , clip , width=\columnwidth]{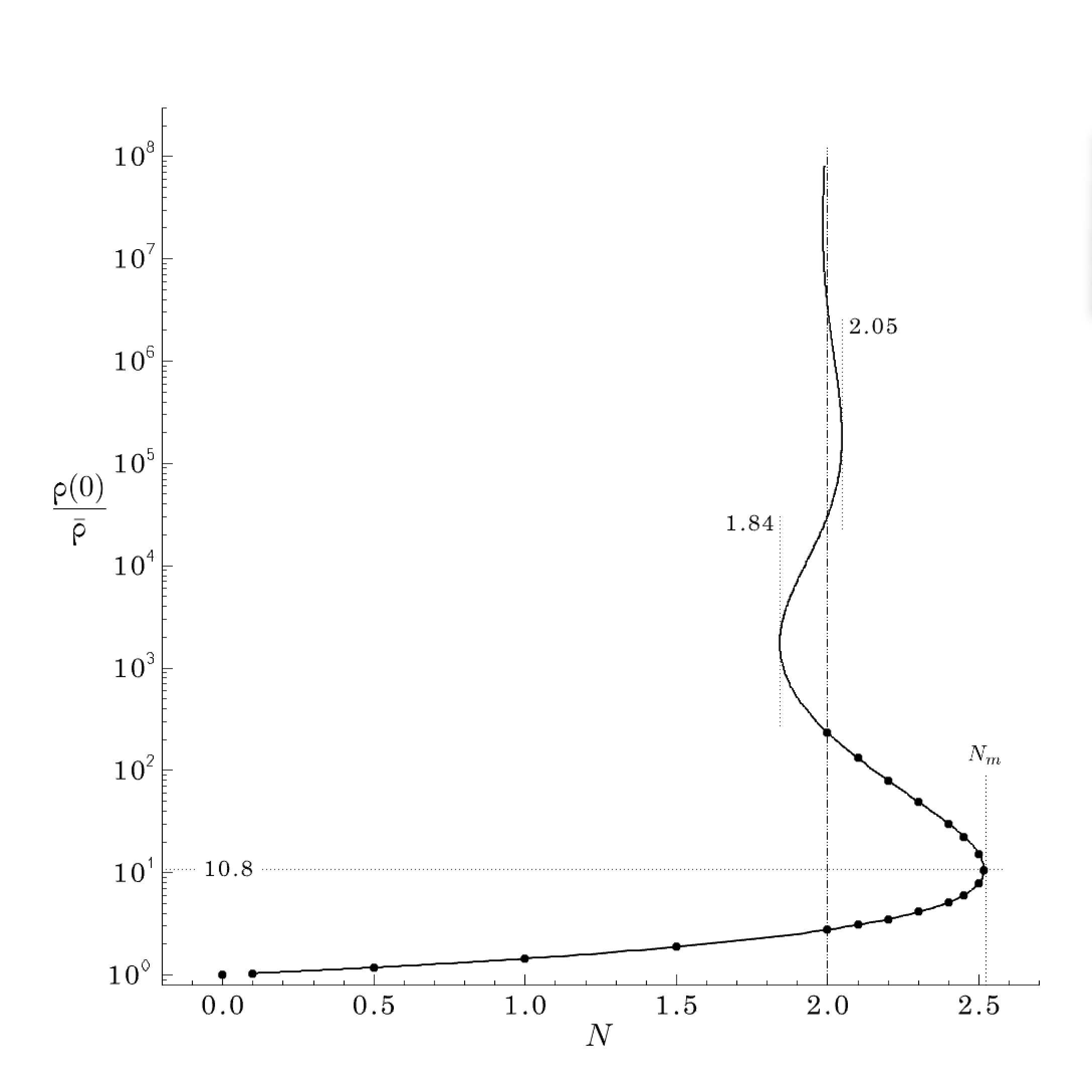}%
  \caption{Central density as function of the gravitational number\hfill\ }%
  \label{p.xi0}
\end{figure}%
However, we do not share the commonly accepted opinion that this asymptotic solution coincides with the singular isothermal sphere \cite{jb2008}
\begin{equation}\label{sitm}
    \xiy^{\ast}(\etay,N) = \frac{2}{3N}\frac{1}{\etay^{2}}
\end{equation}
This particular solution of \REq{LE-isoT.r.nd.cn.m} matches the boundary condition at the shell internal wall [\REq{bc.rho.r=a.nd.cn.m}] only if \mbox{$N=2$}, in that case even verifies smoothly the normalization condition [\REq{m.gas.nd.m}] and attains the correct value $\xiy^{\ast}(1,2) = 1/3$ of the peripheral density, but does not match at all the central boundary condition [\REq{bc.rho.r=0.nd.cn.m}]
\begin{equation}\label{bc.rho.r=0.nd.cn.m.sit}
    {\left. \pder{\ln\xiy^{\ast}}{\etay} \right|}_{\setay=0} = - 2 {\left. \frac{1}{\etay} \right|}_{\setay=0} = -\infty \neq 0   
\end{equation}
a detail correctly emphasized also by Saslaw \cite{ws1987}.
So, the singular isothermal sphere described by \REq{sitm} and the asymptotic isothermal sphere to which the profile of \Rfi{p.xi0} tends to for \mbox{$N=2$} have central gravitational fields that are, respectively, either infinitely great (and unphysical) or vanishing and there will always be a neighborhood of $\etay=0$, however small, in which they will differ.
The density contrast, widely preferred and utilized in the literature, can be obtained as
\begin{equation}\label{mdc}
        \frac{\rhoy(0)}{\rhoy(a)} = \frac{\rhoy(0)}{\brhoy} \frac{\brhoy}{\rhoy(a)} = \frac{\xiy(0)}{\xiy(1)}  
\end{equation}
and presents a profile similar to the central density's (\Rfi{p.xi0}).

No solution of \REqq{nd.prob.cn.m} exists beyond the upper bound $ N_{m} $ and, therefore, no fluid-static field can exist in the gas if \mbox{$N > N_{m}$}; with numerical accuracy preset to eight significant digits, we obtained $N_{m}=2.51755148$ (rounded off to 2.5175 in the figures) as the greatest value for which it was possible to obtain a solution of \REqq{nd.prob.cn.m}.
Characteristic values corresponding to $N_{m}$ can be read off from \Rfis{p.xi}{p.gf.g} and \Rfis{p.xi1}{p.xi0}.
The central density rises to $\,\xiy(0,N_{m}) \simeq {\color{black}10.8}\,$ and the peripheral one lowers to $\,\xiy(1,N_{m}) = 1/3\,$, so that the density contrast turns out to be $\,\xiy(0,N_{m})/\xiy(1,N_{m})  \simeq {\color{black} 32.4}$.
The average density [\REq{etabar}] is attained at \mbox{$ \obetay \simeq 0.640$}.
The gravitational-field minimum is located at \mbox{$ \etay \simeq 0.335$} and its intensity is about 85\% stronger than the value attained at the shell internal wall.
From the gravitational-number definition [\REq{gcn1}], we can extract critical values of radius
\begin{equation}\label{c.rad}
   a_{c} = \frac{1}{N_{m}}\frac{G m_{g}}{RT}
\end{equation}
temperature
\begin{equation}\label{c.temp}
   T_{c} = \frac{1}{N_{m}}\frac{G m_{g}}{aR}
\end{equation}
and mass 
\begin{equation}\label{c.mass}
   m_{g,c}  = N_{m}\frac{aRT}{G }
\end{equation}
and assert that no fluid-static field can exist in the gas sphere if \mbox{$ a < a_{c} $} or \mbox{$T < T_{c}$} or \mbox{$ m_{g} > m_{g,c} $}.
To the best of our understanding, the existence of the maximum value $ N_{m} $ is somewhat surprising because there does not seem to appear any sign either in \REqq{nd.prob.cn.m} or in the profiles of \Rfid{p.xi}{p.gf.g} 
that explicitly hints at the occurrence of limitations constraining the gravitational number. 
\begin{figure}[H]
  \includegraphics[keepaspectratio=true, trim= 5ex 8ex 4ex 20ex , clip , width=\columnwidth]{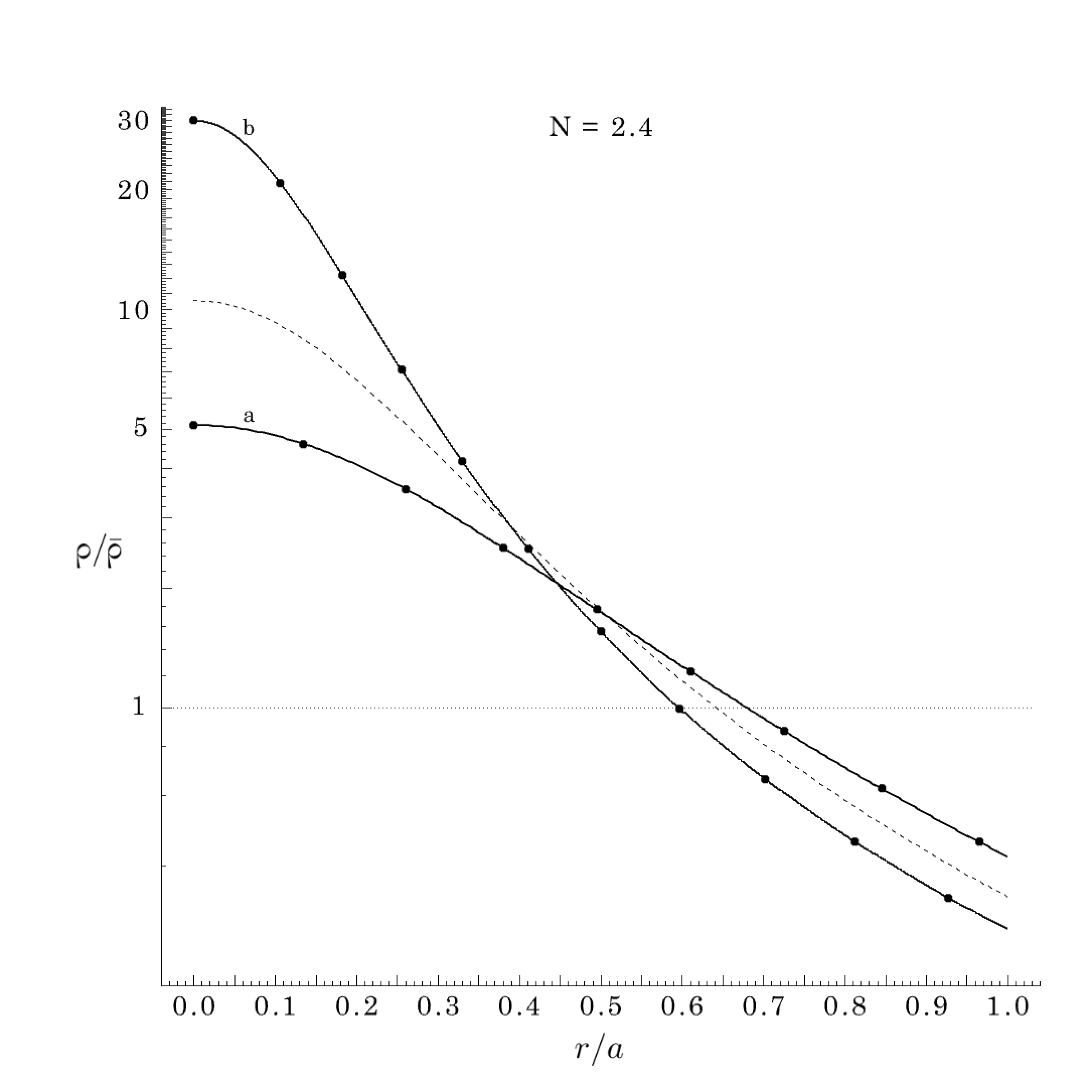}
  \caption{The two possible gas-density profiles at $ N=2.4 $; the dashed line corresponds to $N=N_{m}$ and is included as reference.\hfill\ }
  \label{p.xi.2.4}
\end{figure}
\noindent The, at least mathematical, existence of multiple solutions for a given gravitational number is also perplexing.
As representative example, we show the two solutions relative to \mbox{$N=2.4$} in \Rfid{p.xi.2.4}{p.gf.g.2.4}; the curves corresponding to \mbox{$N=N_{m}$} (dashed lines) are included as reference.
The b-solution is more extreme than the a-solution. 
Central cores and peripheral layers have more or less commensurate extension (\mbox{$ \obetay_{a} \simeq 0.68$ and $ \obetay_{b} \simeq 0.59$})  but the density of the b-solution is almost six times greater at the sphere center and about 0.66 times smaller at the shell internal wall with respect to that of the a-solution.
The position of the gravitational-field minimum shifts from about 48\% to 20\% of the radius and the intensity becomes almost 2.5 times stronger.
\begin{figure}[h]
  {\includegraphics[keepaspectratio=true, trim= 1ex 7ex 5ex 25ex , clip , width=\columnwidth]{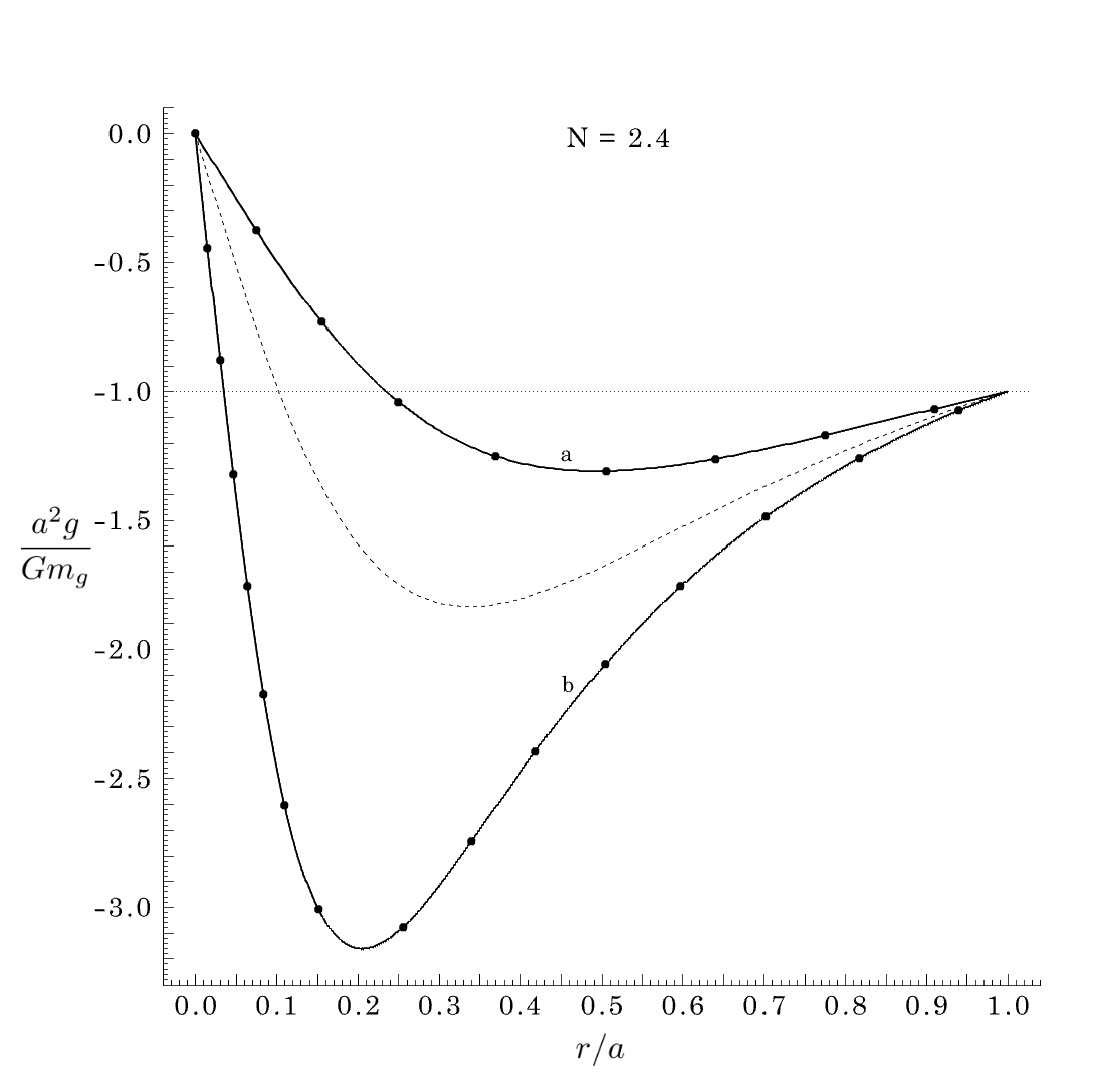}}
  \caption{The two possible gravitational-field profiles inside the gas sphere at $ N=2.4 $; the dashed line corresponds to $N=N_{m}$ and is included as reference.\hfill\ }
  \label{p.gf.g.2.4}
\end{figure}
Inevitable questions arise when looking at the results of \Rfis{p.xi1}{p.xi0}:
(a) What is the physical reason behind the existence of a maximum value of the gravitational number? 
Regarding this question, we definitely encourage the reader to familiarize with the interesting reflections provided and conclusion drawn by Darwin at page 18 of his communication \cite{gd1889ptrs} to the Royal Society of London almost 130 years ago.
(b) The physical variables involved in the definition [\REq{gcn1}] of the gravitational number are \textit{experimentally controllable} and can be selected to produce a value greater than \mbox{$N_{m}$}; what happens in that case?
(c) Are all multiple solutions corresponding to a given gravitational number physical? 
If yes, (c1) what circumstance selects the occurrence of one rather than another? 
If not, (c2) what reason removes the occurrence of the unphysical solutions?
One thing is clear at this point: the answer to question (b) cannot be provided by a gravitofluid-static analysis; the latter simply, and \textit{only}, says that solutions do not exist if $N > N_{m}$.
The answer must, therefore, be sought within a gravitofluid-\textit{dynamics} context.
The quest for answers to the other questions brought us to investigate in details the thermodynamics of the physical system (\Rse{thd}). 

\subsection{Reflections on the Lane-Emden solution} \label{ap-le}
Before turning to thermodynamics, we wish to conclude \Rse{fsgsf} by expressing our point of view, with collegial spirit absolutely free of premeditated criticism, regarding the approach more or less systematically followed in the astrophysical literature that brings to the Lane-Emden solution, particularly for isothermal gas spheres.
A rather accurate and extremely informative historical account is given by Chandrasekhar in the bibliographical notes at page 176 of his textbook \cite{sc1957}.

The conceptual pathways of the literature approach and of our approach basically separate just after \REq{bc.rho.r=0}.
The former approach carries on with the idea that the imposition of the central density
\begin{equation}
    \rhoy(0) = \rhoyc \label{bc.rho.r=0.le} 
\end{equation}
is a legitimate boundary condition; \REq{bc.rho.r=0.le} by itself does not remove the idea of a possible finite extension of the gas sphere but {\color{\newc} certainly} pushes that idea automatically out of the mathematical description's focus. 
The replacement of a gravitational boundary condition [\REq{bc.rho.r=a}] with a fluid-dynamic boundary condition [\REq{bc.rho.r=0.le}] does not affect at all the procedure to cast the problem in nondimensional form.
We introduce again nondimensional variables [\REqq{nd.vars}] and reach once more \REqd{LE-isoT.r.nd.sf}{bc.rho.r=0.nd.sf}; \REq{bc.rho.r=a.nd.sf}, however, must be replaced with the nondimensional version
\begin{equation}
    \xiy(0) = \frac{\rhoyc}{\trhoy} \label{bc.rho.r=0.le.nd} 
\end{equation}
of \REq{bc.rho.r=0.le} and, therefore, the characteristic-number set changes from \REq{cn.sf} to
\begin{equation} \label{cn.sf.le}
    \begin{bmatrix} \; 
        \Piy_{1} = \dfrac{4\piy G {\tilde{r}}^{2} \trhoy}{RT} \quad & 
        \Piy_{3} = \dfrac{\rhoyc}{\trhoy}                               \;  
    \end{bmatrix}
\end{equation}
There are only two independent characteristic numbers this time. 
We resolve \REq{cn.sf.le} for the scale factors 
\begin{equation} \label{sf.cn.le}
    \begin{bmatrix} \; 
        \tilde{r} = \sqrt{\Piy_{1}\Piy_{3}\dfrac{RT}{4\piy G\rhoyc}} \;   &
        \trhoy   = \dfrac{\rhoyc}{\Piy_{3}}  \; 
    \end{bmatrix}
\end{equation}
and set the $\Piy$-type characteristic numbers to the values 
\begin{equation}\label{cn.le.lit}
	\begin{split} 
		 \Piy_{1} =  & 3^{2n} \qquad \begin{cases} n=0 & \text{Emden \cite{re1907}} \\[.5ex] n=1 & \text{King \cite{ik1966aj}} \\[0ex] \end{cases} \\[.5\baselineskip]
		 \Piy_{3} =  & 1 
	\end{split}
\end{equation}
that lead to the scale factors widely used in the literature
\begin{equation} \label{sf.cn.le.lit}
    \begin{bmatrix} \; 
        \rA     = 3^{n}\sqrt{\dfrac{RT}{4\piy G\rhoyc}} \;   &
        \trhoyA = \rhoyc  \; 
    \end{bmatrix}
\end{equation}
In \REq{sf.cn.le.lit}, we have affixed the subscript ``A'' (\underline{A}uthor) to the scale factors in order to avoid confusion with those [\REq{sf.cn.m}] we use in our approach.
A bit of attention should be put at remembering that there is a factor 3 between the radial-distance scale factors of, respectively, Emden (A$\rightarrow$E) and King (A$\rightarrow$K)
\begin{equation}\label{rsf.k.e}
   \rK = 3\rE = 3\sqrt{\dfrac{RT}{4\piy G \rhoyc}}
\end{equation}
that reflects also on the nondimensional radial coordinates 
\begin{equation}\label{ndrc.k.e}
   \etayK = \frac{r}{\rK} = \frac{1}{3}\frac{r}{\rE} = \frac{1}{3} \,\etayE
\end{equation}
The central density is taken as scale factor  
\begin{equation}\label{dsf.k.e}
   \trhoyK = \trhoyE = \rhoyc
\end{equation}
in both cases so that the nondimensional-density variables coincide
\begin{equation}\label{nddv.k.e}
   \xiyK(\etayK) = \xiyE(\etayE) = \frac{\rhoy}{\rhoyc}
\end{equation}
With these premises, the mathematical problem reads 
\begin{subequations} \label{nd.prob.cn.m.le}
\begin{equation}\label{LE-isoT.r.nd.cn.m.le}
    \frac{1}{\etayA^{2}}\pder{}{\etayA}\left( \etayA^{2} \pder{\ln\xiyA}{\etayA} \right) + 3^{2n}\xiyA = 0
\end{equation}
  \begin{align}
    {\left. \pder{\ln\xiyA}{\etayA} \right|}_{\setayA=0} & =  0 \label{bc.rho.r=0.nd.cn.m.le.1} \\[\baselineskip]
    \xiyA(0)                                                    & =  1 \label{bc.rho.r=0.nd.cn.m.le.2} 
  \end{align}    
\end{subequations}
An attractive feature of \REqq{nd.prob.cn.m.le} is the absence in them of any physical characteristic number; their solution, therefore, should be expected in the form of a universal density profile somehow apt to describe \textit{all} isothermal gas spheres.
On the other hand, the scale factors [\REq{sf.cn.le.lit}] contain the central density, a {\color{\newc} \textit{neither} known a priori \textit{nor} controllable} \erase{dimensional} variable \erase{and, therefore,} in demand of being fixed by additional physical information{\color{\newc}. U}ntil the latter is not procured somehow, we believe {\color{\newc} the density profile produced by the numerical integration of \REqq{nd.prob.cn.m.le} remains idle;} \REq{bc.rho.r=0.le} or \REq{bc.rho.r=0.nd.cn.m.le.2} cannot be fully flagged as boundary conditions but are, at most, \textit{scaling} conditions \cite{tp1990pr} and, in turn, the \erase{universal} {\color{\newc} density} profile \erase{produced by the numerical integration of \REqq{nd.prob.cn.m.le}} should be considered of parametric nature mathematically speaking. 
As a matter of fact, before setting in motion numerical machineries, we should dedicate a bit of attention to a question hardly asked: from a \textit{physical} point of view, at which radial distance do we terminate the numerical integration?
Well, if we wish to deal with our test case ($r \leq a$) via \REqq{nd.prob.cn.m.le} rather than \REqq{nd.prob.cn.m} then we should stop the numerical integration at the terminal distance
\begin{equation}\label{termdist}
   \etayAa = \frac{a}{\rA} = \frac{a}{3^{n}}\sqrt{\frac{4\piy G\rhoyc}{RT}}
\end{equation}
obviously.
Operationally, we need an explicit number for $\etayAa$ but we see right away from \REq{termdist} that we do not go very far with the expression on its rightmost-hand side because, although radius $a$ and temperature $T$ are under conceptual control, the central density \erase{seems to escape somehow} {\color{\newc} escapes} our reach.
{\color{\newc} This reflection leads inevitably to think that the formulation based on \REqq{nd.prob.cn.m.le} is not self-contained.}
\erase{We can go one step further to circumvent the} 

{\color{\newc} The} central-density hindrance {\color{\newc} can be circumvented} by calculating the mass of gas contained within the terminal distance
{\color{\newc} 
\begin{equation}\label{m.gas.le}
   m_{g} = 4\piy \int_{0}^{a}\rhoy(r)r^{2} dr = \frac{aRT}{G}\cdot \frac{3^{2n}}{\etayAa}\int_{0}^{\setayAa} \etayA^{2}\,\xiyA d\etayA 
\end{equation}}
and by making resurface from it the gravitational number
\begin{equation}\label{gcn.le}
   N = \frac{G m_{g}}{aRT} = \frac{3^{2n}}{\etayAa}\int_{0}^{\setayAa} \etayA^{2}\,\xiyA d\etayA
\end{equation}
\begin{figure}[h]
  \includegraphics[keepaspectratio=true, trim= 1ex 8ex 2.9ex 20ex , clip , width=\columnwidth]{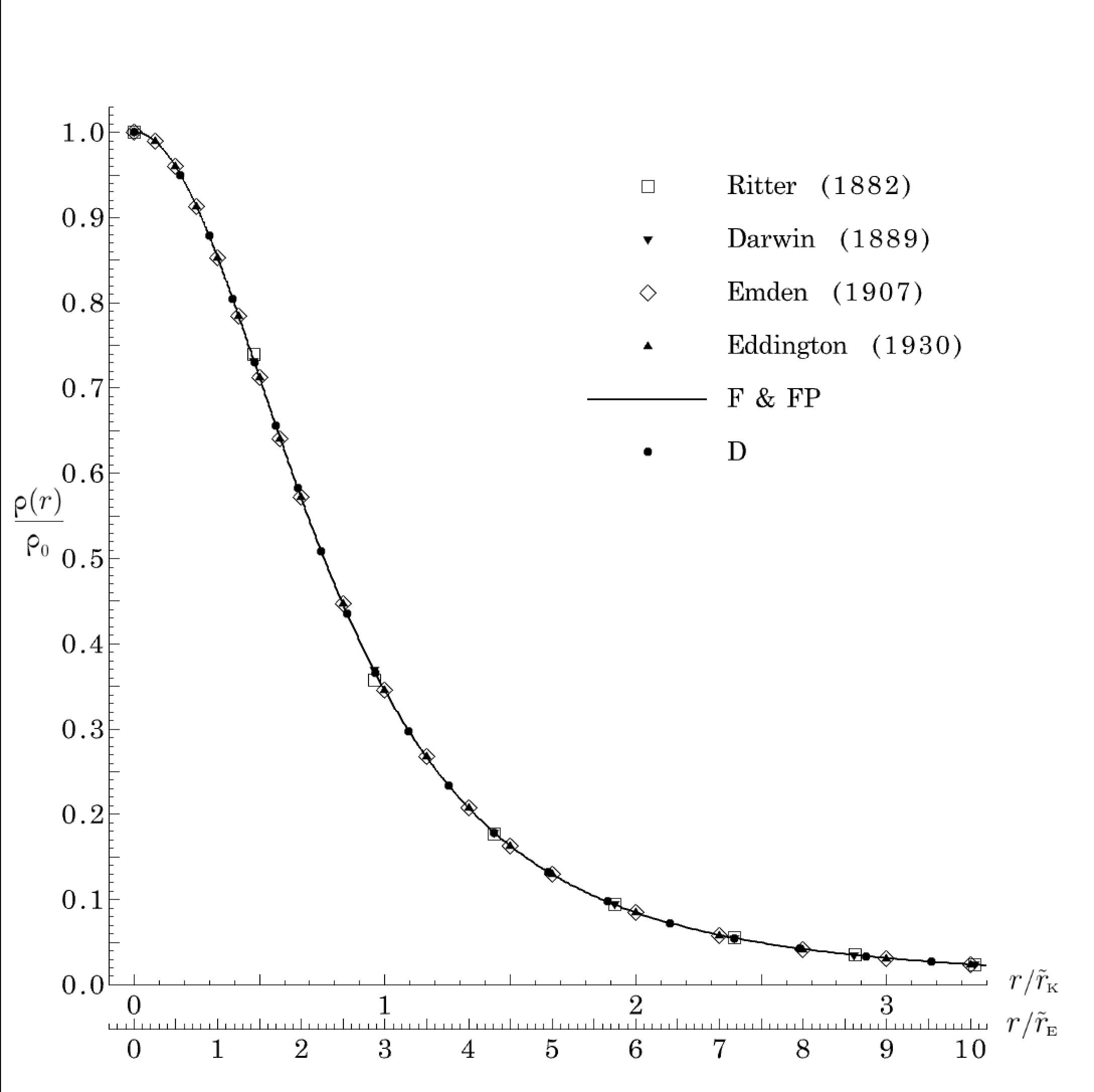}\\
  \caption{Gas-density profile of the Lane-Emden solution according to Ritter \cite{ar1882adp}, Darwin \cite{gd1889ptrs}, Emden \cite{re1907}, Eddington \cite{ae1930}, and our results\hfill\ }
  \label{p.xi-e}
\end{figure}
\REqb{gcn.le} establishes a precise connection between terminal distance and gravitational number. 
Moreover, the connection can even be improved because the integral can be transformed with the help of \REq{LE-isoT.r.nd.cn.m.le} and a bit of integration-by-part jugglery as
\begin{equation}\label{int.le}
    3^{2n}\int_{0}^{\setayAa} \etayA^{2}\,\xiyA d\etayA = - \left[ \etayA^{2} \pder{\ln\xiyA}{\etayA} \right]_{\setayA=\,\setayAa}
\end{equation}
so that \REq{gcn.le} goes into the more convenient form
\begin{equation}\label{gcn.le.1}
   N = \frac{G m_{g}}{aRT} = - \left[ \etayA \pder{\ln\xiyA}{\etayA} \right]_{\setayA=\,\setayAa}
\end{equation}
that spares us the trouble of the numerical evaluation of the integral.
In \REq{gcn.le.1}, we should regard the terminal distance as unknown and the gravitational number as prescribed. 
It comes with no surprise that the latter also fixes nondimensionally the central density
\begin{equation}\label{cd.le}
   \frac{\rhoyc}{\brhoy}  = 3^{2n-1} \frac{\etayAa^{2}}{N}
\end{equation}
\REqb{cd.le} easily follows from the average-density definition [\REq{agd}] taking into account \REq{m.gas.le}, \REq{int.le} and \REq{gcn.le.1}.
In this way of looking at things, \erase{therefore,} the gravitational number{\color{\newc}, an ingredient \textit{foreign}  to the mathematical formulation based on \REqq{nd.prob.cn.m.le}, sanitizes the lack of self-containedness of that formulation and} emerges as the identifier that singles out a specific isothermal gas sphere from the multitude described by the universal solution of \REqq{nd.prob.cn.m.le}.
{\color{\newc} The} path is hence clear.
Numerical algorithms can be smoothly launched to solve \REqq{nd.prob.cn.m.le} by going mathematically as far as wished from the center; for example, Emden went up to \mbox{$\etayE = 2\cdot 10^{3}$}, we reached \mbox{$\etayE=3\etayK=10^{5}$} (F \& FP algorithms only).
This action leads to the \erase{universal} density profile illustrated in \Rfi{p.xi-e} showing our numerical results with superposed data collected in the literature and adequately post-processed. 

With the scale factors of \REq{sf.cn.le.lit}, the nondimensional gravitational field follows straightforwardly from \REq{mom.isoT.r}
\begin{equation}\label{gf.nd.le}
    \frac{g}{\sqrt{4\piy G \rhoyc RT}} = \frac{1}{3^{n}} \pder{\ln\xiyA}{\etayA}
\end{equation}
and its profile, illustrated in \Rfi{p.gf-e}, is thus readily obtained.
\begin{figure}[H]
  {\includegraphics[keepaspectratio=true, trim= 1ex 7ex 5ex 25ex , clip , width=\columnwidth]{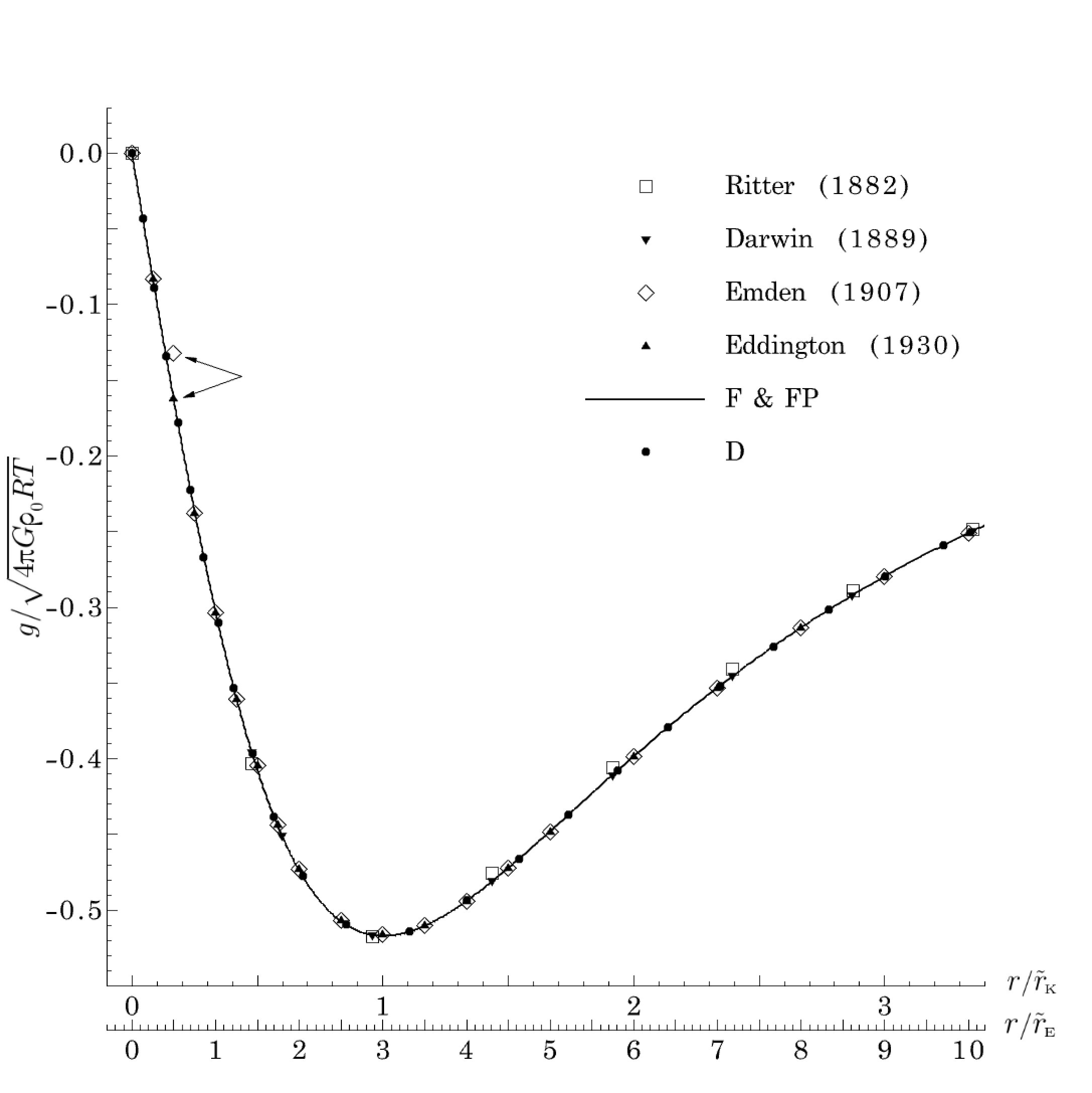}} \\
  \caption{Gravitational-field profile of the Lane-Emden solution according to Ritter \cite{ar1882adp}, Darwin \cite{gd1889ptrs}, Emden \cite{re1907}, Eddington \cite{ae1930}, and our results; Emden's data point indicated by the arrow (-0.13225) is in error (likely a typo in his Table 14) and was corrected (-0.16225) by Eddington\hfill\ }
  \label{p.gf-e}
\end{figure} 
\begin{figure}[h]
  {\includegraphics[keepaspectratio=true, trim= 1ex 7ex 5ex 20ex , clip , width=\columnwidth]{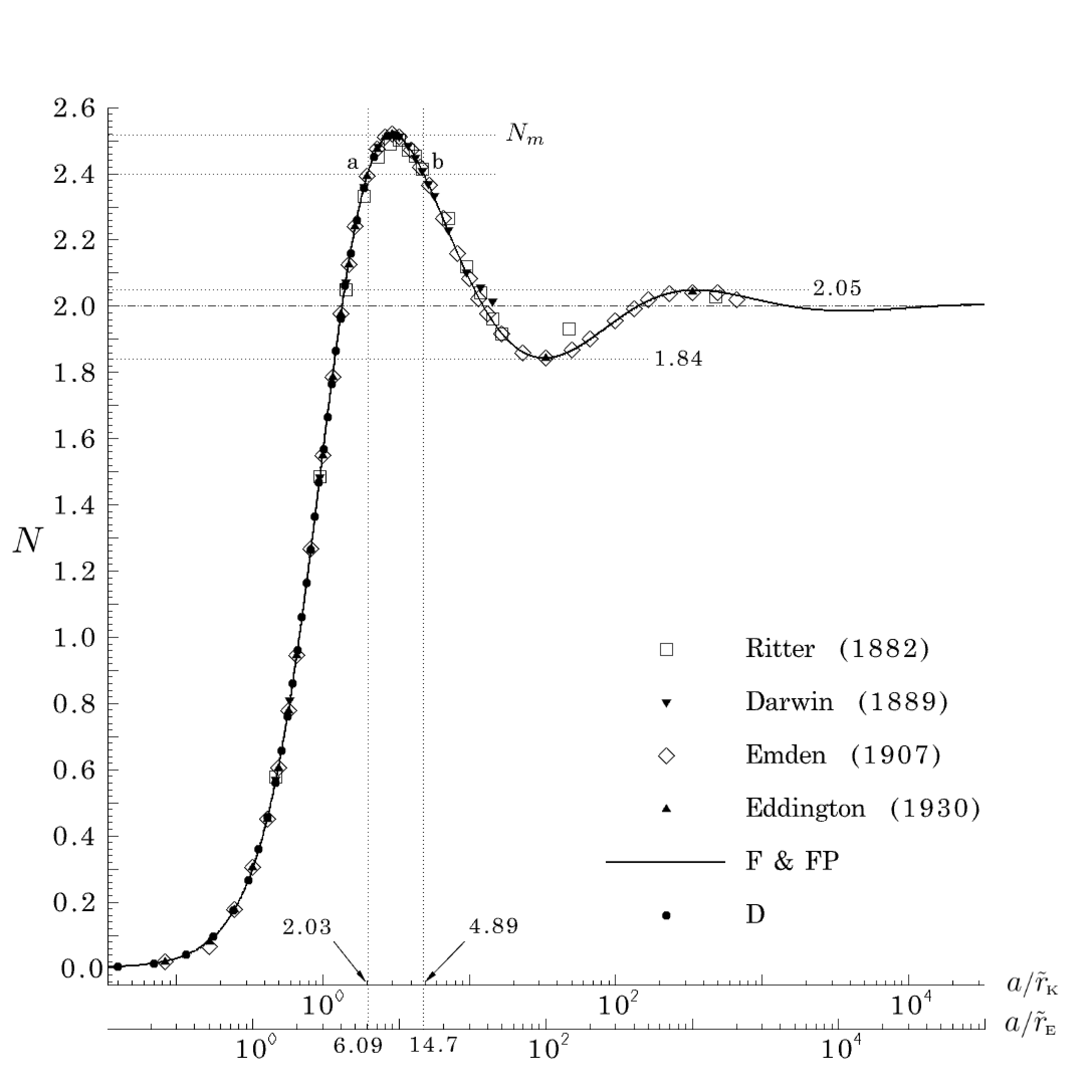}} \\
  \caption{Gravitational-number profile based on the Lane-Emden solution according to Ritter \cite{ar1882adp}, Darwin \cite{gd1889ptrs}, Emden \cite{re1907}, Eddington \cite{ae1930}, and our numerical algorithms\hfill\ }
  \label{p.gn-e}
\end{figure}\noindent
This is also a universal profile encompassing all isothermal gas spheres.
{\color{\newc} The compliance of the peripheral gravitational field
\begin{equation}\label{gf.nd.le.a}
    \frac{g(a)}{\sqrt{4\piy G \rhoyc RT}} = \frac{1}{3^{n}} \left[ \pder{\ln\xiyA}{\etayA} \right]_{\setayA=\,\setayAa} 
\end{equation}
with the Gauss theorem's prescript [\REq{gf.gauss.r=a}] is readily verified with the help of \REqd{termdist}{gcn.le.1}.}

With the solution of \REqq{nd.prob.cn.m.le} in hand, we can \erase{then} proceed to build the graph of \REq{gcn.le.1}, shown in \Rfi{p.gn-e}, which turns out to be rather revealing because it confirms the existence of gravitational-number upper bound $N_{m}$ and of multiple solutions above \mbox{$N\simeq 1.84$}.
The knowledge of the gravitational number allows to enter the graph on the vertical axis and to read off on the horizontal axis the terminal distance that marks the terminus on the profiles in \Rfid{p.xi-e}{p.gf-e} corresponding to the isothermal gas sphere identified by the prescribed gravitational number.
It is this step that basically fixes the central density [\REq{cd.le}] and raises the standing of \REqd{bc.rho.r=0.le}{bc.rho.r=0.nd.cn.m.le.2} to the level of legitimate boundary conditions.
In order to consolidate these thoughts with an explicit example, we have considered again in \Rfi{p.gn-e} the case with $N=2.4$. 
The corresponding horizontal line intersects the profile in the points a and b and selects two values of the terminal distance: \mbox{$\etayEa=3\etayKa  \simeq 6.09$} and \mbox{$\etayEa=3\etayKa  \simeq 14.7$}. 
In turn, these values determine the density-profile portions on the Lane-Emden solution, displayed with thicker lines in \Rfi{p.xi-e-2p4}, that correspond to the curves a and b of \Rfi{p.xi.2.4}, respectively; for the purpose of verification, we have superposed also the results extracted from \Rfi{p.xi.2.4} and  adequately converted.
The situation for the gravitational field is similar and illustrated in \Rfi{p.gf-e-2p4}.
For completeness, we have reproduced in \Rfi{p.xi0-e} the diagram of the central density as function of the gravitational number from \REq{cd.le} for visual comparison with the one shown in \Rfi{p.xi0}: the coincidence is unequivocal. 
\begin{figure}[H]
  \subfigure[\ Intersection a in  \Rfi{p.gn-e}]{\label{p.xi-e-2p4a}\includegraphics[keepaspectratio=true, trim= 3ex 7ex 4ex 20ex , clip , width=\columnwidth]{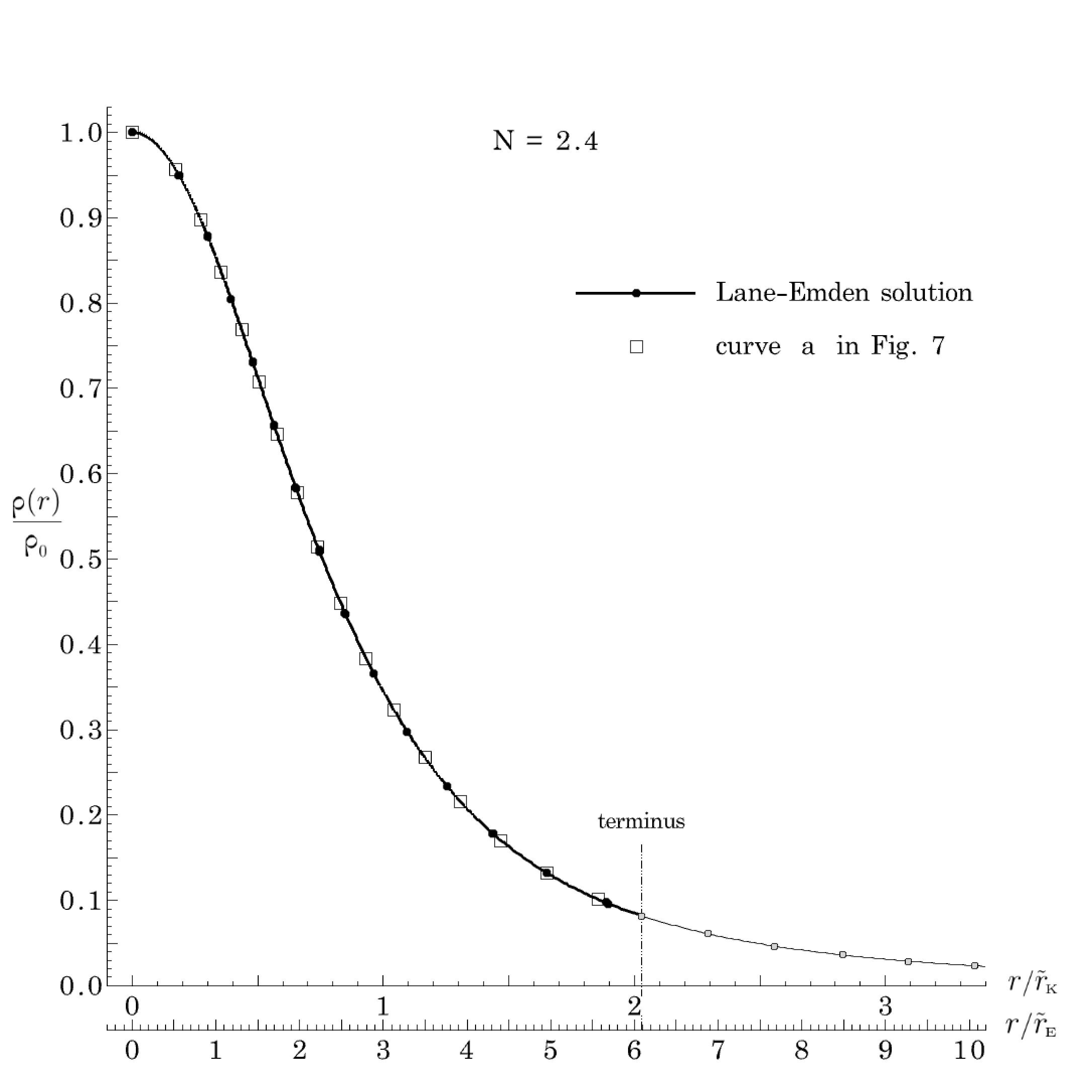}}   \\
  \subfigure[\ Intersection b in  \Rfi{p.gn-e}]{\label{p.xi-e-2p4b}\includegraphics[keepaspectratio=true, trim= 3ex 7ex 4ex 20ex , clip , width=\columnwidth]{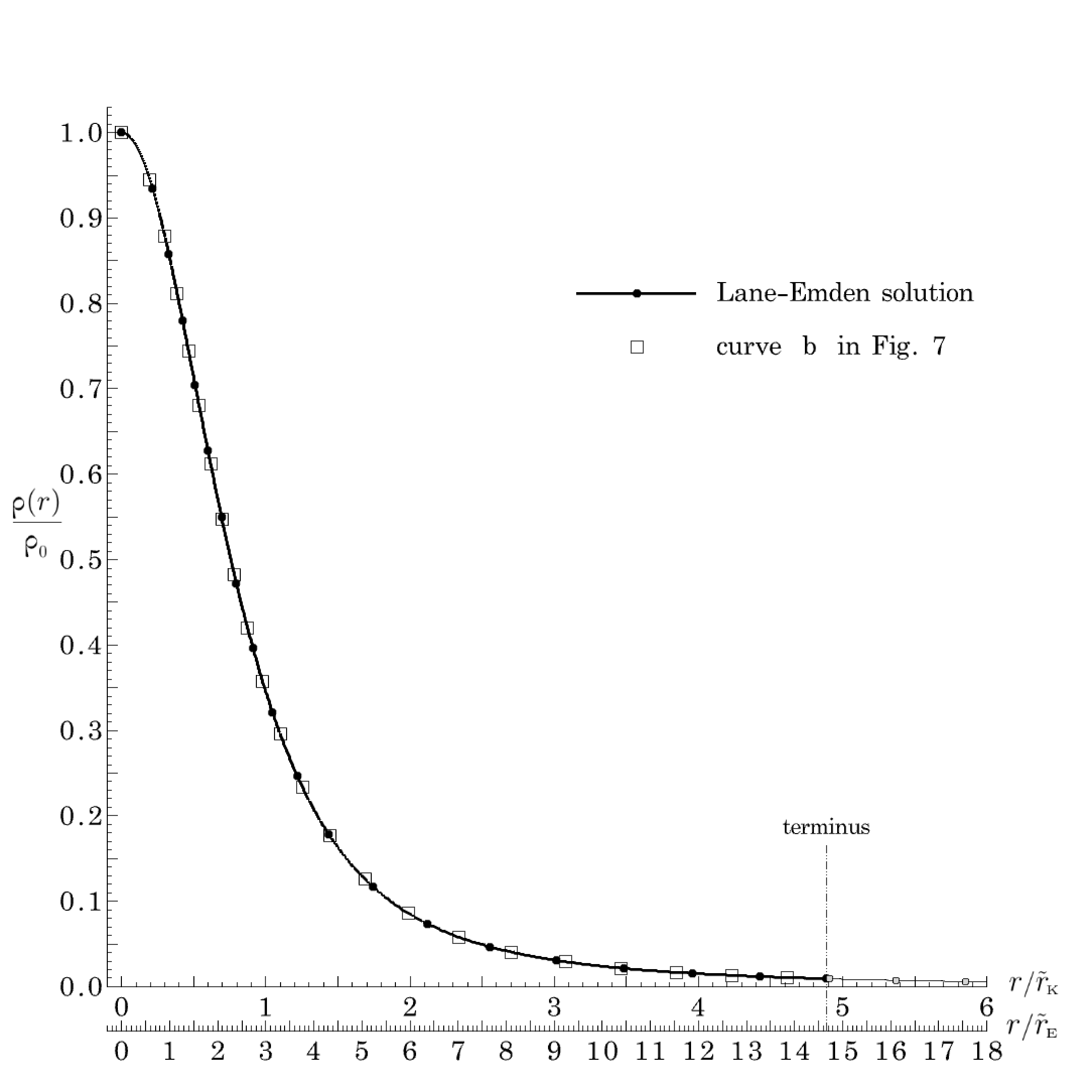}}
  \caption{The two gas-density profiles for \mbox{$N=2.4$} on the Lane-Emden solution\hfill\ }\label{p.xi-e-2p4}
\end{figure}\noindent

{\color{\newc} The graphs in \Rfid{p.xi-e-2p4}{p.gf-e-2p4} may induce in the reader's mind an impression of equivalence between the Lane-Emden solution founded on the idea of the central-density boundary condition [\REq{bc.rho.r=0.nd.cn.m.le.2}] and our solution founded on the peripheral-gravitational-field boundary condition [\REq{bc.rho.r=a.nd.cn.m}] because, after all, they lead to same results although with different parameterization; well, maybe yes from a mathematical point but certainly not from a physical point of view.
In this regard, we wish to stress that the gravitational number is invisible in the formulation based on \REqq{nd.prob.cn.m.le} and the mathematical equivalence is established only after the physical injection of \REq{gcn.le}.}
We are aware that our viewpoint regarding the isothermal Lane-Emden problem and its solution is at variance with the interpretation prevailing in the astrophysical literature for which we quote as representative the statement 
\begin{quote}
  A pure isothermal sphere stretches to infinity and has an infinite mass
\end{quote}
found at page 323 in Saslaw's textbook \cite{ws1987},
{\color{\newc} a seemingly straightforward conclusion when looking at density profile of \Rfi{p.xi-e} without the support of \REq{gcn.le}.}
Nevertheless,
we do not know how to reconcile this quote, and the interpretation it represents, with the graph in \Rfi{p.gn-e} which clearly indicates that there is only one isothermal gas sphere that conforms with the quote: the asymptotic one identified by the rightmost intersection of the horizontal line at \mbox{$N=2$} with the curve, whose intercept corresponds to an infinite terminal distance
\begin{equation}\label{termdist.N=2.last}
    \etayAa =  \frac{1}{3^{n}}\sqrt{\frac{4\piy G}{RT}} \cdot a\sqrt{\rhoyc}\rightarrow\infty
\end{equation}
As a matter of fact, \REq{termdist.N=2.last} implies two possibilities. 
One is the isothermal gas sphere we are looking for that ``stretches to infinity'' (\mbox{$a\rightarrow\infty$}) and ``has an infinite mass''
\begin{equation}\label{m.gas.le.N=2.a}
   m_{g} = 2 \frac{aRT}{G} \; \rightarrow \; \infty 
\end{equation}
Its average gas density vanishes
{\color{\newc}
\begin{equation}\label{m.gas.le.N=2.a.agd}
   \brhoy = \frac{3RT}{2\piy G a^{2}} \; \rightarrow \; 0 
\end{equation}}
but its central density turns into an indeterminate form
\begin{equation}\label{m.gas.le.N=2.a.cd}
   \rhoyc = \left(\frac{3^{n}\etayAa}{a}\right)^{2}\frac{RT}{4\piy G} \; \rightarrow \; \frac{\infty}{\infty} 
\end{equation}
The other one that springs up from \REq{termdist.N=2.last} is the isothermal gas sphere characterized by infinite central density 
\begin{equation}\label{cd.N=2.inf}
   \rhoyc \rightarrow \infty
\end{equation}
with possibly finite but definitely indeterminate size
\begin{equation}\label{m.gas.le.N=2.cd.a}
    a^{2} = \frac{(3^{n}\etayAa)^{2}}{\rhoyc}\frac{RT}{4\piy G} \; \rightarrow \; \frac{\infty}{\infty} 
\end{equation}
mass and average density.
We see at this point how the scale-factor choice [\REq{sf.cn.le.lit}] backfires via the bad repercussions of these peculiarities [\REqd{m.gas.le.N=2.a.cd}{cd.N=2.inf}]:
\REq{sf.cn.le.lit} becomes idle for the infinitely-stretched isothermal sphere because the central density is an indeterminate form [\REq{m.gas.le.N=2.a.cd}] and it goes into the perverse form
\begin{equation} \label{sf.cn.le.lit.cdinf}
    \begin{bmatrix} \; 
        \rA      \rightarrow 0 \quad  &
        \trhoyA \rightarrow \infty  \; 
    \end{bmatrix}
\end{equation}
for the other isothermal sphere because the central density becomes infinite [\REq{cd.N=2.inf}]. 
It is rather easy to convince ourselves that \REq{sf.cn.le.lit.cdinf} molds the \erase{universal} density profile of \Rfi{p.xi-e} into one single physical point, i.e. the center \mbox{$r=0$}, in which the isothermal sphere is concentrated possibly with finite mass but definitely with infinite density.

We conjecture that the root of the interpretation prevailing in the literature can be traced back to Lane's paper \cite{hl1870ajs} whose attention concentrated on adiabatic ideal-gas spheres.
In the text at page 60, Lane introduced the idea that the differential-equation solution, obviously based on \REq{bc.rho.r=0.le}, should also provide a way to determine gas-sphere size and central density.
He began by saying that the adiabatic index, i.e. {\color{\newc} the} specific-heat ratio, influences
\begin{quote}
   ... first the form of the curve that expresses the value of \mbox{$\frac{\rhoy}{\rhoyc}$} for each value of $x$; secondly, the value of the upper limit of $x$ corresponding to \mbox{$\frac{\rhoy}{\rhoyc}=0$}; and thirdly, the corresponding value of $\muy$. 
\end{quote}
In Lane's notation, $x$ is the nondimensional radial coordinate [defined in his \mbox{Eq. (4)}] and $\muy$ is the gas-sphere nondimensional mass [defined in his \mbox{Eqs. (5) and (6)}]. 
Then Lane continued
\begin{quote}
   These limiting, or terminal, values of $x$ and $\muy$ cannot be found except by calculating the curve ... But when these values have been found ... they may be introduced once for all into equations (4) and (5) from which the values of $\rhoyc$ and ... are at once deduced.
\end{quote}
Emden \cite{re1907} extended the analysis to polytropic gas spheres and, according to the results of his calculations, summarized their behavior at page 155 as
\begin{quote}
    Alle polytropen Gaskugeln von \mbox{$n=0$} bis \mbox{$n=\infty$} haben eine Oberfl\"{a}che, an der Druck, Temperatur und Dichte die Werte Null annehmen. F\"{u}r \mbox{$n>5$} liegt dieselbe im Unendlichen. \\[-\baselineskip]
    
    {\small [All polytropic gas spheres from \mbox{$n=0$} to \mbox{$n=\infty$} possess a surface on which pressure, temperature and density vanish.
    For \mbox{$n>5$} the surface lies at infinity]}
\end{quote}
In Emden's notation, $n$ is the polytropic index, not to be confused with the formal exponent we used in \REq{cn.le.lit} (top line). 
There is a very short and straightforward step from the last sentence in the Emden's statement quoted here to the conclusion for isothermal spheres (\mbox{$n=\infty$}) exemplified in Saslaw's statement quoted just before \REq{termdist.N=2.last}.
\begin{figure}[H]
  \subfigure[\ Intersection a in \Rfi{p.gn-e}]{\label{p.gf-e-2p4a}\includegraphics[keepaspectratio=true, trim= 1.5ex 7ex 4ex 20ex , clip , width=\columnwidth]{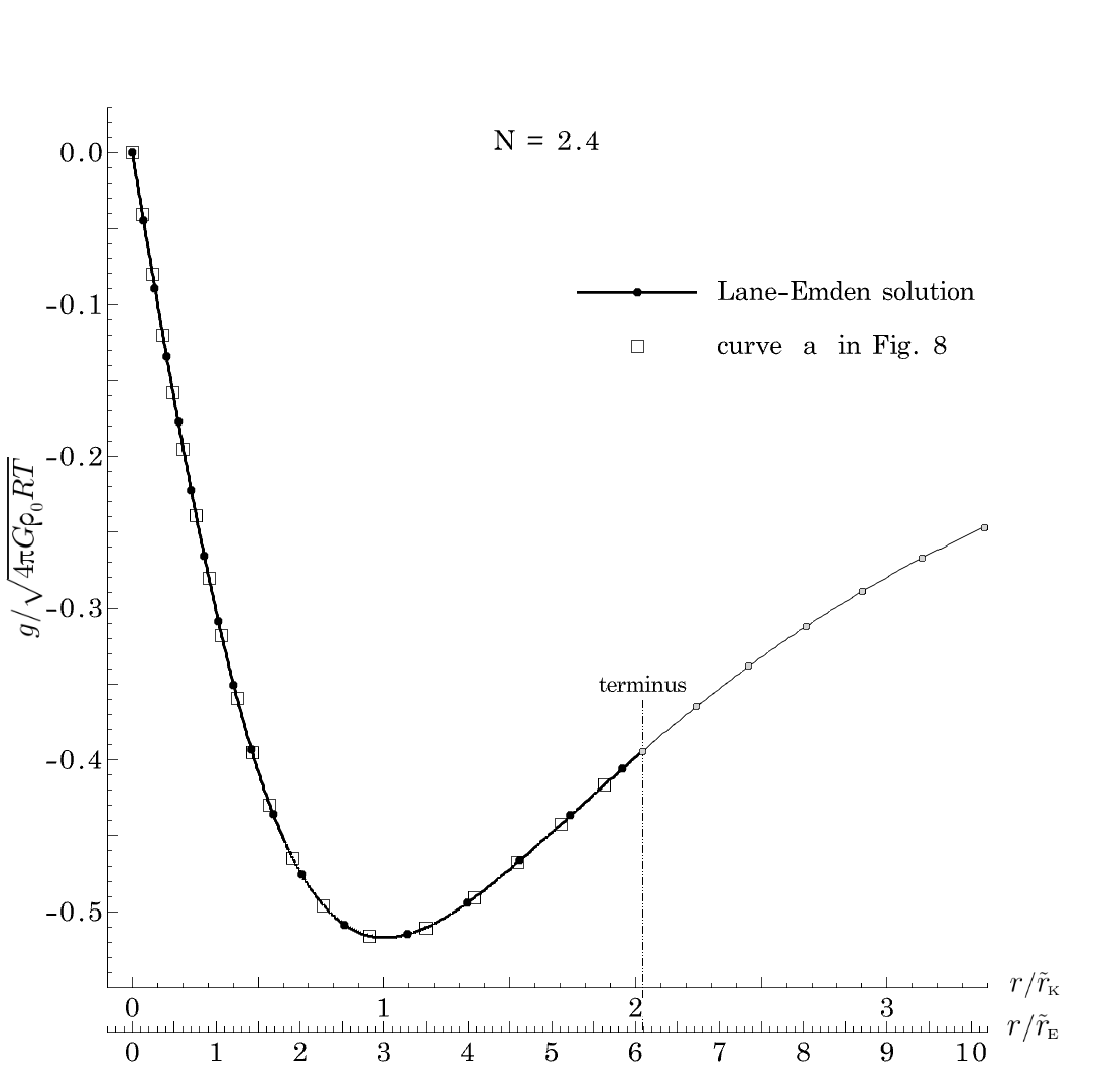}}   \\
  \subfigure[\ Intersection b in \Rfi{p.gn-e}]{\label{p.gf-e-2p4b}\includegraphics[keepaspectratio=true, trim= 1.5ex 7ex 4ex 20ex , clip , width=\columnwidth]{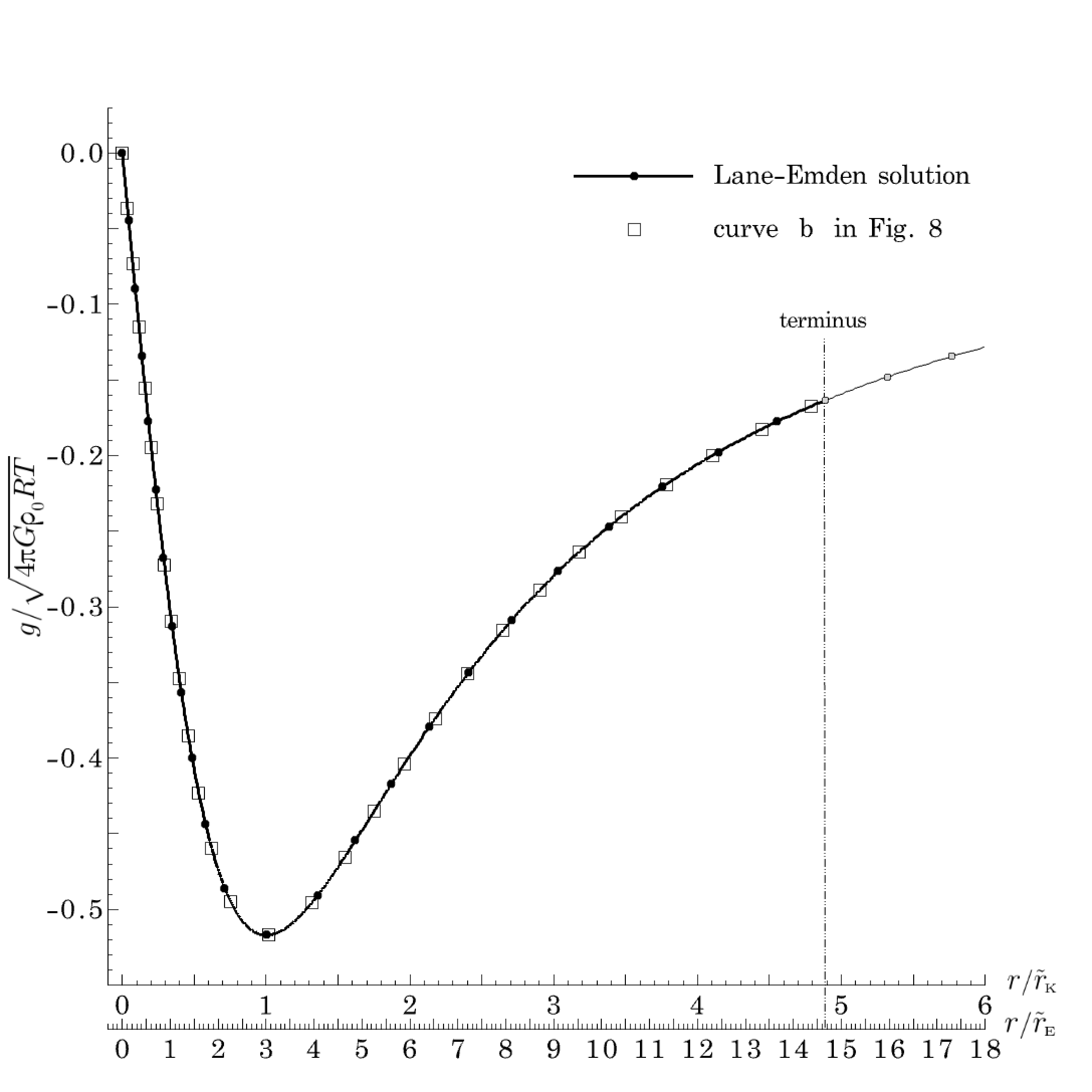}}
  \caption{The two gravitational-field profiles for \mbox{$N=2.4$} on the Lane-Emden solution\hfill\ }\label{p.gf-e-2p4}
\end{figure}
\begin{figure}[h]
  {\includegraphics[keepaspectratio=true, trim= 1ex 7ex 5ex 25ex , clip , width=\columnwidth]{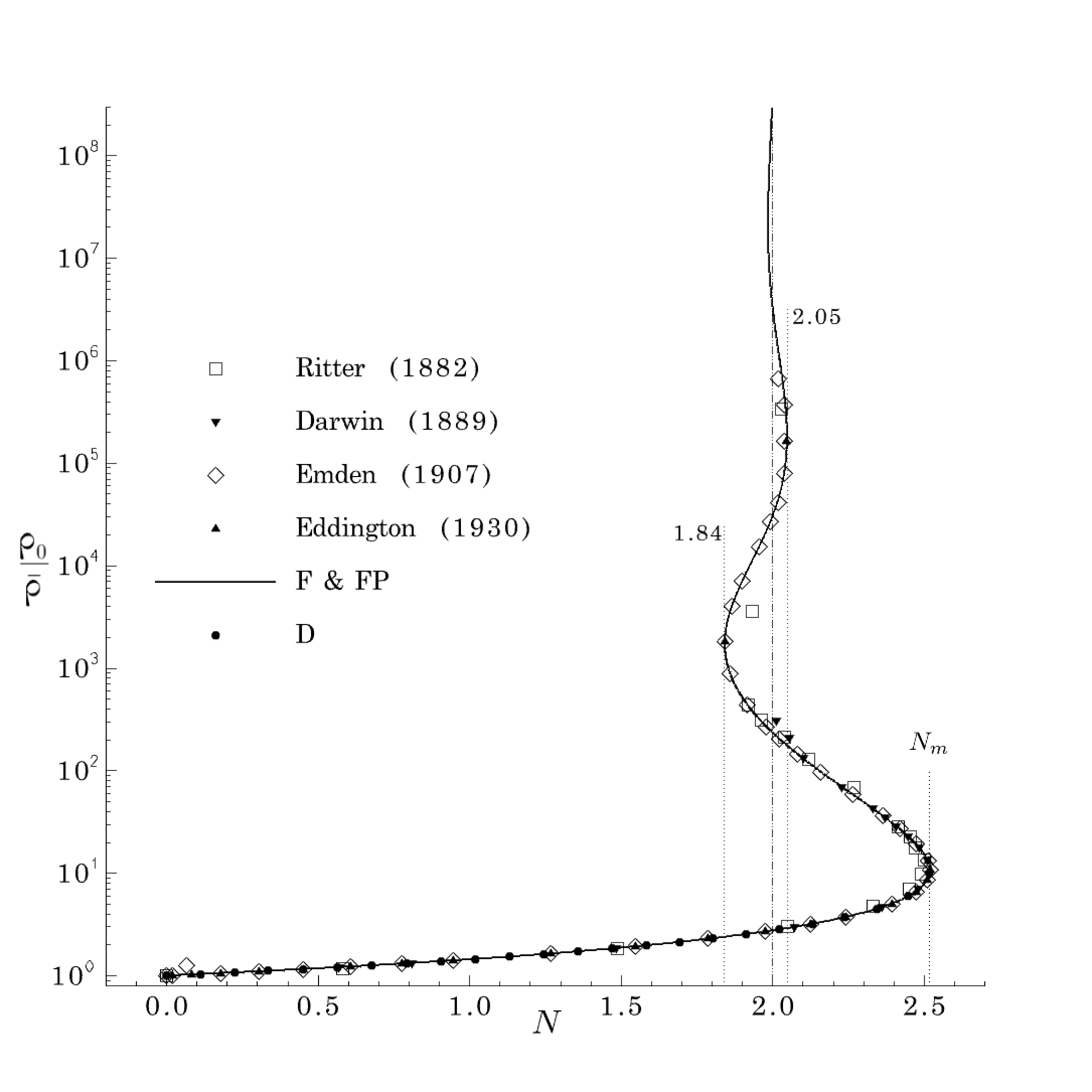}} \\
  \caption{Central density from the Lane-Emden solution as function of the gravitational number (compare with \Rfi{p.xi0})\hfill\ }
  \label{p.xi0-e}
\end{figure}\noindent

Lane and Emden had no hesitation to accept the part of solution to their differential equations at the left of the ``limiting, or terminal, value of $x$'' at which \mbox{$\rhoy/\rhoyc=0$} and to dismiss the part of solution yielding negative densities at the right of that value.
It seems to us that Lane's and Emden's choice, although justified by the necessity of building up a physical model, was simply motivated by convenience without further reflection.
Already 116 years ago, Jeans signaled the conceptual precariousness of a vanishing density at page 3 of \cite{jj1902ptrs}:
\begin{quote}
    Whether we suppose the thermal equilibrium of the gas to be conductive or adiabatic, we are still met by the difficulty that the gas equations break down over the outermost part of the nebula, through the density not being sufficiently great to warrant the statistical methods of the kinetic theory.
\end{quote}
In line with Jeans' appropriate warning, a fluid dynamicist would hesitate to accept and be suspicious about the physical consistency of solutions that contain vanishing mass density, pressure, temperature, because of the incompatibility of such occurrences with the reason of existence of the governing equations that generated them. 
In other words, the applicability of the continuum-medium model breaks down in the left vicinity of the ``limiting, or terminal, value of $x$'' and, therefore, what sense should one make of a result, i.e. \mbox{$\rhoy/\rhoyc=0$}, yielded by a physical model in a place outside of its applicability domain?
Of course, these considerations are absolutely in no way meant to imply rejection of the polytropic-sphere problem as conceived by Lane and Emden but we believe a prudent reflection about whether or not something may be in need of revision in their formulation would not be an idle exercise.

\section{Thermodynamics}\label{thd}
\subsection{Entropy} \label{entr}
The entropy of the physical system is the sum of the entropy of the gas and the entropy of the shell
\begin{equation}\label{ent.t}
   S = S_{g}+S_{s}
\end{equation}
The determination of the shell contribution $S_{s}$ requires the integration on the shell volume of the relative specific entropy
\begin{equation}\label{ent.s}
  S_{s} = \int_{V_{s}} \rhoy_{s} s_{s}(T)dV = m_{s} s_{s}(T)  
\end{equation}
The determination of the gas contribution $S_{g}$ requires the integration over the gas-sphere volume of the perfect-gas specific entropy
\begin{equation}\label{ent.g}
  S_{g} = \int_{V_{g}} \rhoy s_{g}dV = 4\piy \int_{0}^{a} r^{2} \rhoy s_{g} dr
\end{equation}
The specific entropy is composed by two terms
\begin{equation}\label{ent.g.sp}
    s_{g} = s_{g}(T,v)  = s_{gt}(T,v)  + s_{gi}(T)
\end{equation}
The first descends from the translational degrees of freedom of the molecules
\begin{equation}\label{ent.g.sp.t}
    s_{gt}(T,v) = R\left( C + \frac{3}{2} \ln T + \ln v  \right)
\end{equation}
and features a dependence on the specific volume \mbox{$v(r)=1/\rhoy(r)$}; the second term, $s_{gi}(T)$, is associated with the internal molecular structure and does not need to be expanded in explicit details because it depends only on temperature.
In \REq{ent.g.sp.t}, we have set for brevity
\begin{equation}\label{ent.c}
  C = \frac{5}{2} + \frac{3}{2} \ln \frac{km^{5/3}}{2\piy \hbar^{2}}
\end{equation}
with
\begin{tabbing}
	i \= xx \= xxxxxxxxxx \kill
	\> $k$            \> Boltzmann constant, $1.38064852\cdot10^{-23}\,$J$\cdot$K$^{-1}$ \\
	\> $m$           \> gas molecular mass \\
	\> $\hbar$       \> Planck constant over 2\piy, $1.054571800\cdot10^{-34}\,$J$\cdot$s
\end{tabbing}
The integral on the rightmost-hand side of \REq{ent.g} is easily performed and leads to the interesting separation
\begin{subequations}\label{entr.g}
	\begin{equation}\label{entr.g.sep}
	  S_{g} = S_{g,0} + \hat{S}_{g} 
	\end{equation}
	in which
	\begin{align}\label{entr.g.ngf}
	  S_{g,0} & = S_{g,0}(T,V,m_{g}) \notag \\[.5\baselineskip]
	            & =  m_{g} R \left( C + \frac{3}{2} \ln T + \ln \frac{V}{m_{g}} + \frac{s_{gi}(T)}{R}\right)
	\end{align}
	is the entropy the gas would have in the absence of gravitational effects and
	\begin{align}\label{entr.g.gf}
	  \hat{S}_{g} & = \hat{S}_{g}(m_{g},N) \notag \\[.5\baselineskip]
	                &  =  - m_{g} R \int_{0}^{1} 3 \etay^{2} \xiy(\etay,N) \ln \xiy(\etay,N) d\etay
	\end{align}
	is the correction due to the presence of the gravitational field.
\end{subequations}
\REqb{entr.g.gf} clearly emphasizes the importance of the gas-density distribution for the purpose of entropy calculations.
The reformulation of the gravitational number [\REq{gcn1}] in terms of the state variables $T,V,m_{g}$ 
\begin{equation}\label{gcn.t}
   N = \left( \frac{4\piy}{3}\right)^{1/3}\frac{G m_{g}}{RTV^{1/3}}
\end{equation}
gives full evidence of the gravitational correction's lack of first-degree homogeneity; we see through \REqd{ent.t}{entr.g.sep}, therefore, that total entropy and gas entropy are inevitably deprived of the same characteristic.
The integral in \REq{entr.g.gf} can be evaluated numerically after that the solution $\xiy(\etay,N)$ has been obtained, of course; however, with a bit of patience and integration-by-part mastery, it is also possible to obtain the integral in the following analytical form
\begin{align}\label{entr.g.gf.I}
       I_{ent}(N) & = - \int_{0}^{1} 3 \etay^{2} \xiy(\etay,N) \ln \xiy(\etay,N) d\etay  \notag \\[.5\baselineskip] 
                    & = N - \ln \xiy(1,N) - 6 \left( 1 - \xiy(1,N) \rule{0pt}{2.5ex}\right)
\end{align}
that clearly highlights the peripheral density's role in establishing the gravitational correction.
Its nondimensional diagram
is shown in \Rfi{p.Ient}; 
\begin{figure}[h]
  \includegraphics[keepaspectratio=true, trim= 5ex 8ex 4ex 20ex , clip , width=\columnwidth]{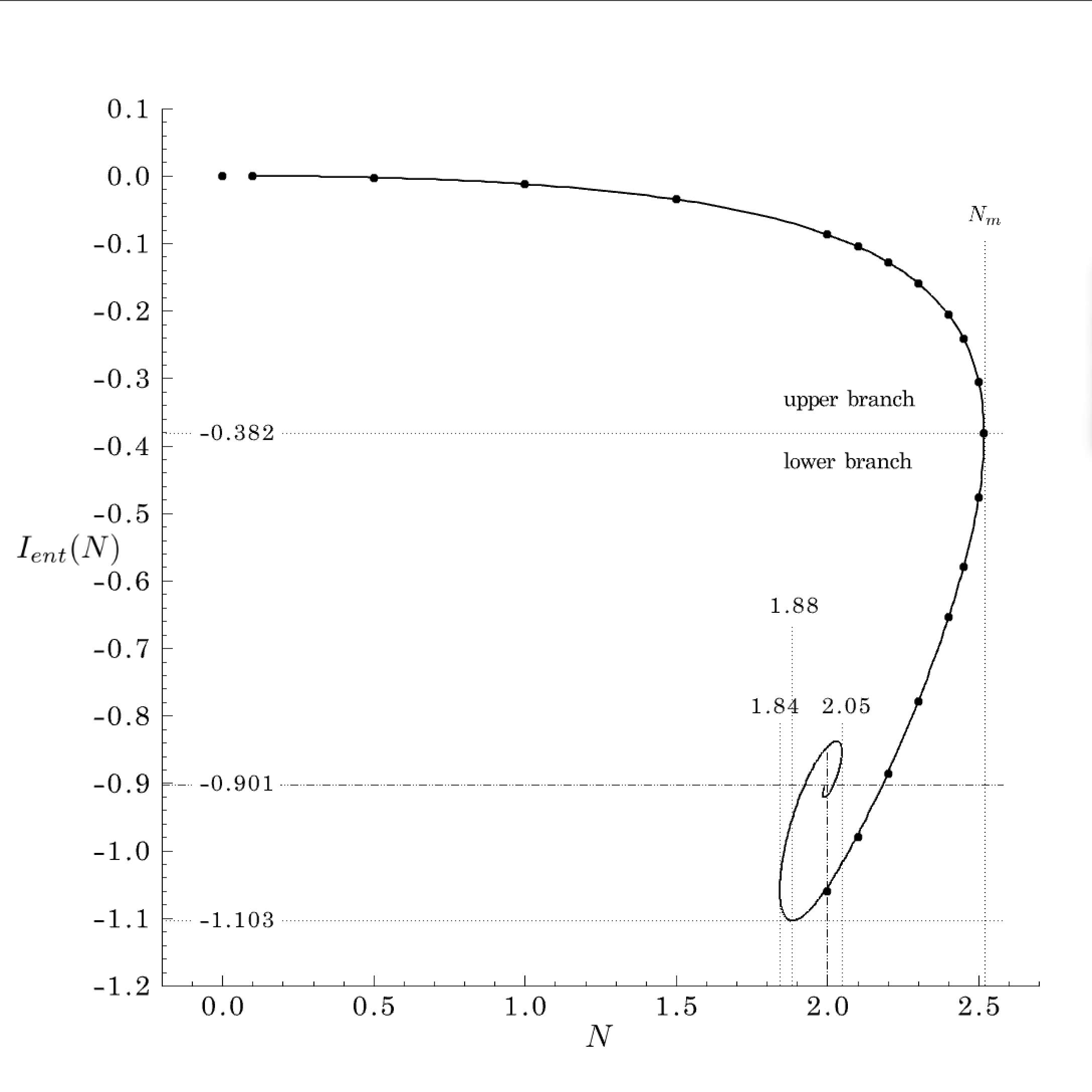}
  \caption{Gravitational correction to the gas entropy}  
  \label{p.Ient}
\end{figure}
the pattern is obviously driven by that of the peripheral density (\Rfi{p.xi1}).
Indeed, it also attains vertical slope at \mbox{$N=N_{m}$}; thereat, it crosses the level 
\begin{equation}\label{branch.b}
  I_{ent}(N_{m}) = N_{m} +\ln3 - 4 \simeq -0.382
\end{equation} 
marking the boundary between upper and lower branches of fluid-static configurations. 
Subsequently, at \mbox{$N\simeq 1.88$}, it goes through the absolute minimum \mbox{$I_{ent}(1.88) = -1.103$} and then spirals around the point \mbox{$\,N=2, I_{ent}(2) = \ln3 - 2\simeq -0.901\,$}.
From \REq{ent.t}, taking into account \REq{ent.s}, \REqq{entr.g} and \REq{entr.g.gf.I}, the explicit expression of the system entropy reads
\begin{align}\label{entr.t.1}
    S  =  m_{g}R \left[ \rule{0pt}{3.5ex}\right. C 
       & + \frac{3}{2}\ln T + \ln\frac{V}{m_{g}}  + \frac{s_{gi}(T)}{R} + \frac{m_{s}}{m_{g}}\frac{s_{s}(T)}{R} \rule{0pt}{3.5ex} \notag \\[.5\baselineskip]
       & \!\!\!\!\!+ N - \ln \xiy(1,N) - 6 \left( 1 - \xiy(1,N) \rule{0pt}{2.5ex}\right)\left.\rule{0pt}{3.5ex}\right] 
\end{align}
Incidentally, it may be worthwhile to point out that this entropy expression depends on the state variables $T,V,m_{g}$ and, therefore, it should not be looked at as a fundamental relation 
\cite{hc1963,lt1966,ln1971,hc1985}.
The gravitational correction is always negative; therefore, the presence of the gravitational field always reduces the entropy with respect to the situation when it is absent.
In case of multiple solutions, the upper-branch configurations have greater entropy than the lower-branch configurations.
This circumstance perhaps gives a somewhat probabilistically privileged status to the former branch but it does not exclude at all the potential realizability of the latter branch.

Some of the lower-branch configurations present physically unusual peculiarities that can be brought to light by the study of the entropy differential. 
But first, a little simplification.
It is clear from \REq{entr.t.1} that the shell mass $m_{s}$ is a legitimate state parameter and should be treated as such in general.
On the other hand, our emphasis is not on the role of the container within the physical problem we are studying; therefore, just to simplify a little bit the forthcoming algebra, we decide at this point to introduce the (irrelevant) constraint 
\begin{equation}\label{mass.s}
     \frac{m_{s}}{m_{g}} = const
\end{equation}
Another important step required in the procedure to obtain the entropy differential is the logarithmic differentiation
\begin{equation}\label{dgcn.t}
   \frac{dN}{N} = \frac{dm_{g}}{m_{g}} - \frac{1}{3}\frac{dV}{V} - \frac{dT}{T}
\end{equation}
of the reformulated gravitational number [\REq{gcn.t}]. 
With that, all the ingredients are in place to expand from \REq{entr.t.1} the entropy differential
\begin{align}\label{diff.entr.t}
    dS     & = \frac{m_{g}R}{T} \left( \frac{3}{2} + \frac{c_{vi}(T)}{R} + \frac{m_{s}}{m_{g}}\frac{c_{s}(T)}{R} - \phient(N) \right) dT \notag \\[.5\baselineskip]
            & + \frac{m_{g}R}{V} \left( 1 - \frac{1}{3} \phient(N)\right) dV  \notag \\[.5\baselineskip]
            & + R \left( C + \frac{3}{2}\ln T + \ln\frac{V}{m_{g}} + \frac{s_{gi}(T)}{R} + \frac{m_{s}}{m_{g}}\frac{s_{s}(T)}{R} \rule{0pt}{3.5ex}\right. \notag \\[.5\baselineskip]
            & \hspace*{11.25em} + \psi_{ent}(N) \left.\rule{0pt}{3.5ex}\right) dm_{g}
\end{align}
in terms of the state-variable differentials $dT, dV, dm_{g}$ and to recognize the corresponding partial derivatives by inspection of \REq{diff.entr.t}.
Therein,
\begin{subequations}\label{sp.heats}
\begin{equation}\label{cvi}
   c_{vi} = T\; \tds{}{s_{gi}}{T}
\end{equation}
is the contribution to the gas constant-volume specific heat due to internal molecular structure and 
\begin{equation}\label{cs}
   c_{s} = T\; \tds{}{s_{s}}{T}
\end{equation}
\end{subequations}
is the shell specific heat; moreover, the following two new functions of the gravitational number appear
\begin{subequations} \label{diff.entr.defs}
\begin{align}
     \phient(N) & = N \tds{}{I_{ent}}{N} \notag \\[.5\baselineskip] 
                        & = N \left[ 1 + \left( 6 - \frac{1}{\xiy(1,N)} \right) \xiy'(1,N)  \right] \label{phient}  \\[.5\baselineskip]
     \psi_{ent}(N)   & = I_{ent}(N) + \phient(N) -1  \label{psient}
\end{align}
In \REq{phient}, $\xiy'(1,N)$ stands for the derivative $d\,\xiy(1,N)/dN$. 
In a first instance, we calculated it numerically via the centered finite-difference stencil
\begin{equation}\label{dxi}
   \xiy'(1,N)=\lim_{\Delta N \rightarrow 0} \frac{\xiy(1,N\! +\!\Delta N) - \xiy(1,N\! -\!\Delta N)}{2\Delta N}
\end{equation}
but later we hit upon a surprising result that allows to express the derivative in analytical form; we will come back to this matter with more details in \Rse{thd.se.T}.
\end{subequations}
The separation structure of \REq{entr.g.sep} is inherited by all partial derivatives in \REq{diff.entr.t}: there is a standard term appropriate when gravitational effects are absent followed by a gravitational correction.
The peculiarities mentioned before descend from the capability of the gravitational corrections to flip the sign of the derivatives and this occurrence is of particular relevance for the derivative \begin{align}
    \pds{}{S}{T}{V,m_{g}}  = \frac{m_{g}R}{T} \left(\rule{0pt}{3.3ex}\right.\frac{3}{2} 
                              & + \frac{c_{vi}(T)}{R} + \frac{m_{s}}{m_{g}}\frac{c_{s}(T)}{R}  \notag \\[.5\baselineskip]
                              & - \phient(N) \left.\rule{0pt}{3.3ex}\right) \label{ent.derivs.T}   
\end{align}
because it is a criterion of thermodynamic stability in the absence of the gravitational field and it may somehow be connected to thermal stability also in the presence of the gravitational field.
The scenario is better brought forward visually by the diagram of the function $\phient(N)$ shown in \Rfi{p.phient}.
The profile is monotonic for the upper-branch configurations (\Rfi{p.phient1}) and goes through a repetitive pattern in between asymptotes for the lower-branch configurations.
The asymptotes correspond to the points of vertical slope in \Rfi{p.xi1} at which the derivative $\xiy'(1,N)$ and, by consequence, all the derivatives in \REq{diff.entr.t} become infinite.
This happens for the first time at \mbox{$N=N_{m}$}.
The function $\phient(N)$ is always negative and the derivative \mbox{$\pdst{}{S}{T}{V,m_{g}}$} is always positive for the upper-branch configurations (\Rfi{p.phient1}); thermodynamically speaking, these are well behaved and all seems in order. 
The sign of the derivative can change for some of the lower-branch configurations according to where they are positioned with respect to the level established by the sum of the specific-heat terms [\REq{ent.derivs.T}, top line] which depends on temperature and, ultimately, on the gravitational number if \REq{gcn.t} is accordingly resolved to read
\begin{equation}\label{gcn.temp}
   T = \left( \frac{4\piy}{3}\right)^{1/3}\frac{G m_{g}}{RV^{1/3}}\frac{1}{N}
\end{equation}
For the mere purpose of visualization, we have considered a diatomic gas at sufficiently low temperature (\mbox{$c_{vi}/R=1$}) and set the shell contribution to a qualitative, although meaningful as order of magnitude, value (\mbox{$c_{s}/R\simeq 1.573$}; mild steel confining molecular nitrogen, for example); in the sequel, we will refer to these values as ``visualization specific-heat settings''.
In this case, the specific-heat level, labeled ``shell (qualitative)'' in \Rfi{p.phient}, does not change with temperature or gravitational number and is obviously horizontal.
So, all lower-branch configurations above the ``shell (qualitative)'' level have negative derivative \mbox{$\pdst{}{S}{T}{V,m_{g}}$}, a peculiarity that we perceive as undeniable omen of thermal instability.
The danger, therefore, exists exclusively for some of the lower-branch configurations among which the one at \mbox{$N=N_{m}$} inaugurates the first series.
The intersections of the several, actually infinite, curves of the function $\phient(N)$ with the ``shell (qualitative)'' level  
identify a series of points F$_{1}$, F$_{2}$, F$_{3}, \ldots$ in correspondence to which the derivative vanishes
\begin{equation}\label{ent.derivs.T.0}
     \pds{}{S}{T}{V,m_{g}}  = 0 
\end{equation}
They represent, therefore, the boundary points between series of well-behaved and peculiar configurations; clearly, the derivative \mbox{$\pdst{}{S}{T}{V,m_{g}}$} changes sign an infinite number of times.

These thermodynamic occurrences point out how much interesting and, at the same time, necessary thermodynamic-stability analyses of self-gravitating fluids are.
The literature offers several \cite{va1985,dl1968mnras,ji1974aj,jk1978mnras,tp1989ajss,tp1990pr,jb2008,phc2002aa} but, in the majority of the references we have explored, the thermodynamic stability is systematically studied within a kinetic-theory framework centered around the Boltzmann's definition of entropy.
We believe it would be beneficial to investigate the thermodynamic stability also within an axiomatic-thermodynamics framework \cite{hc1963,lt1966,ln1971,hc1985} in order to find out whether or not such an alternative conceptual path brings to the same results achieved by kinetic-theory based analyses.
Unfortunately, given the priorities of our research, we have no other choice for the time being than to label this task as future work and carry on.


\subsection{Energy} \label{ener}
The physical system is composed by matter, gas and shell, and gravitational field.
Therefore, we consider two forms of energy whose belongingness is obvious: matter energy and gravitational-field energy.
Their sum gives the total energy.
There is no kinetic energy in static circumstances and matter energy is made up exclusively by the thermodynamic part distributed in space with density  
\begin{equation} \label{m.ener.d}
	  \rhoy u = 
	 \begin{cases}  \; \rhoy \left( \frac{3}{2} R T + u_{gi}(T) \right) & \text{gas} \\[3ex] 
                      \; \rhoy_{s} u_{s}(T)                                    & \text{shell} \\[3ex] 
                      \;  0                                                      & \parbox{2em}{empty \\ space}\end{cases}  \\[0ex]
\end{equation}
\begin{figure}[H]
  \subfigure[\ Upper branch and lower branch between asymptotes (1.84,$N_{m}$)]{\label{p.phient1}\includegraphics[keepaspectratio=true, trim= 4ex 8ex 4ex 20ex , clip , width=\columnwidth]{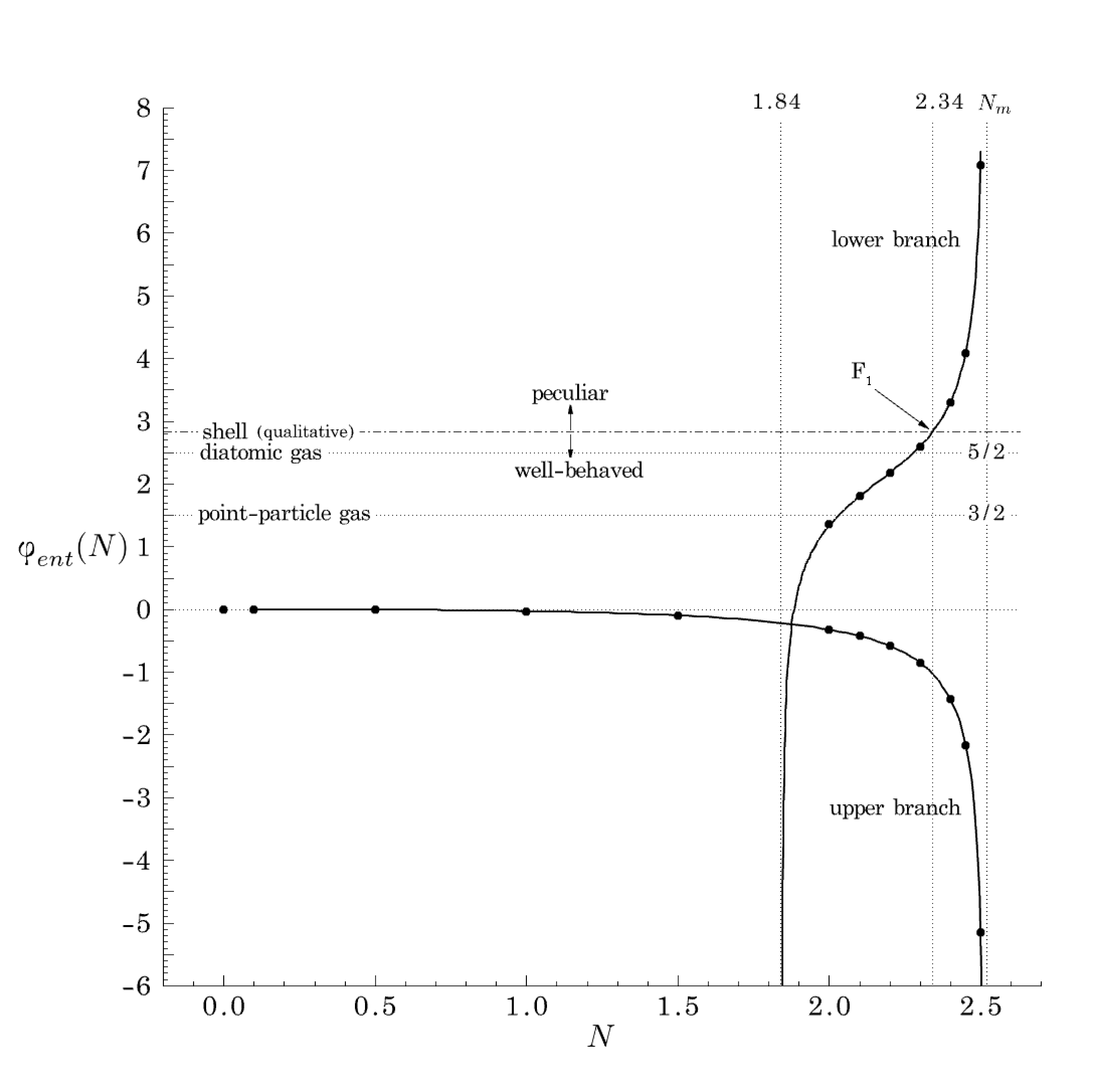}}   \\
  \subfigure[\ Lower branch between asymptotes (1.84,2.05) and (1.98,2.05)]{\label{p.phient2}\includegraphics[keepaspectratio=true, trim= 4ex 8ex 4ex 18ex , clip , width=\columnwidth]{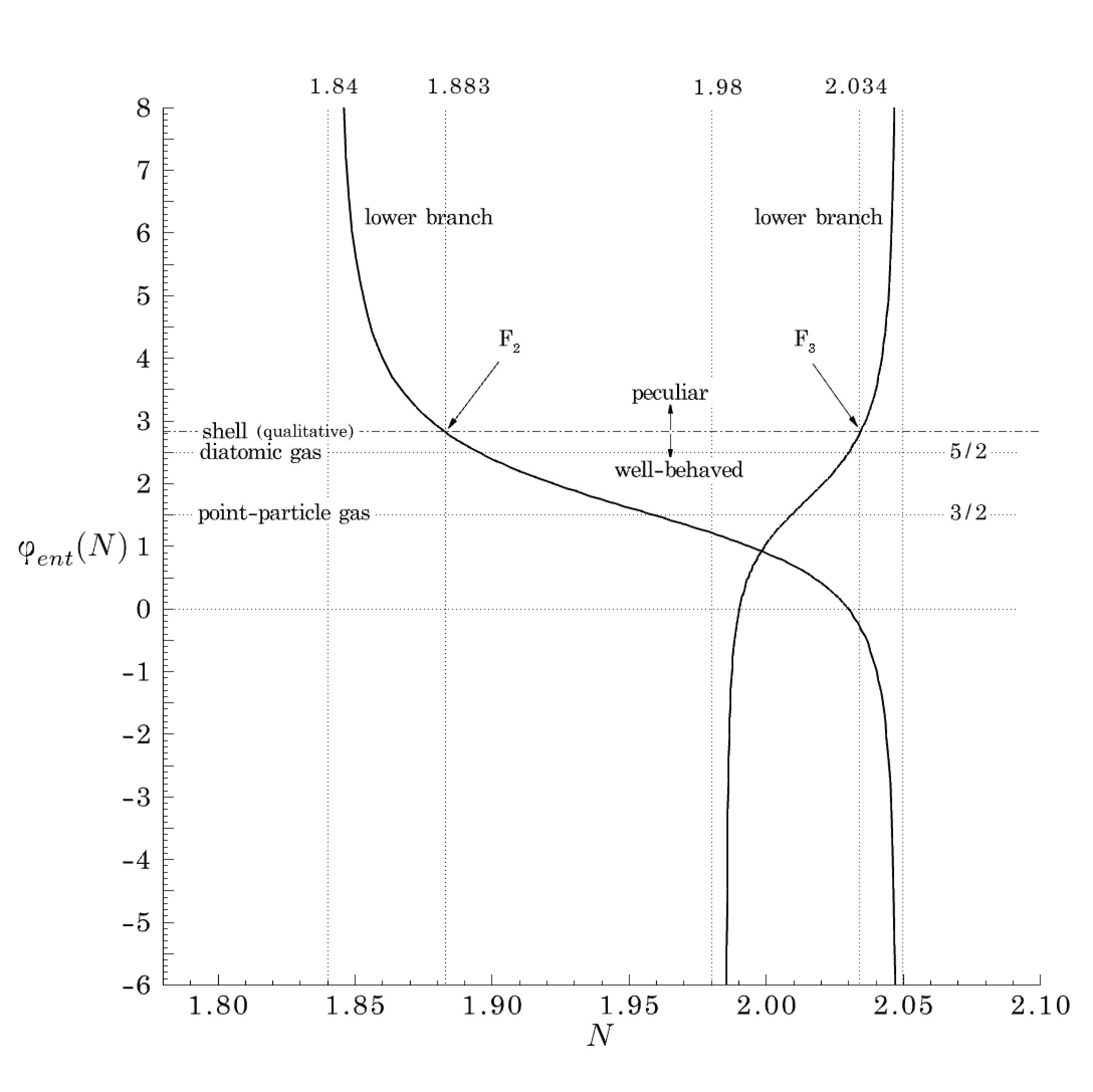}}
  \caption{Function $\protect\phient(N)$}\label{p.phient}
\end{figure}\noindent
In \REqq{m.ener.d}, $u_{gi}(T)$ is the contribution to specific thermodynamic energy of the gas due to internal molecular structure and $u_{s}(T)$ is the shell specific thermodynamic energy.
The gravitational energy is distributed in space with density \cite{ma1915jre,ma2007}
\begin{align} 
	  \varepsilon  & = - \frac{g^{2}}{4\piy G} = \notag \\[.5\baselineskip]
	 & - \frac{1}{4\piy} \frac{G m_{g}^{2}}{2a^{4}}  
	       \begin{cases}  \displaystyle \left( \frac{1}{N} \pder{\ln\xiy}{\etay} \right)^2                                                                             & \!\!\!\text{(g)} \\[4ex] 
                            \displaystyle \frac{1}{\etay^{4}} \left( 1\! +\! \frac{m_{s}}{m_{g}} \frac{\etay^{3} - 1}{(1 + \deltay/a)^{3}- 1}\right)^{2}    & \!\!\!\text{(s)} \\[4ex] 
                            \displaystyle \frac{1}{\etay^{4}} \left( 1 + \frac{m_{s}}{m_{g}} \right)^{2}                                                             & \!\!\!\text{(es)}
           \end{cases}  
           \label{g.ener.d} 
\end{align}
The expressions tagged (g), (s) and (es) in \REq{g.ener.d} are obtained by direct substitution of the gravitational-field solution [\REq{gf.nd}; \REq{gf.gauss.nd} (shell) and (empty space)] and apply locally in the space inside the gas sphere, inside the shell and outside the shell in empty space, respectively.
Our position is conceptually different from the one usually taken in the literature because we do not make use of the concept ``potential energy of the gas''.
The equivalence between the concepts ``matter potential energy'' and ``field energy'' holds \textit{only} under stationary, a fortiori static, circumstances, as we learn from Feynman's chapters 8 and 27 in \cite{rf1964v2}.
The definition on the top line of \REq{g.ener.d} always holds, within Newton's theory of gravity, regardless whether a dynamic or a stationary or a static situation is prevailing.
The justification for our opting in favor of the ``field energy'' interpretation even under static circumstances is not as drastically severe as the devastating motivation for rejection expressed by Heaviside \cite{oh1893te1,oh1971v1}
\begin{quote}
   Potential energy, when regarded merely as expressive of the work that can be done by forces depending upon configuration, does not admit of much argument. It is little more than a mathematical idea, for there is scarcely any physics in it. It explains nothing. 
\end{quote}
It simply resides in the wish to be self-consistent all the way along our study path 
because, as explained in the introduction (\Rse{intro}), our research objectives foresee also gravitofluid-\textit{dynamic} analyses in which the equivalence does not hold.

The total-energy density is given by the sum
\begin{equation}\label{ener.d}
   e = \rhoy u + \varepsilon
\end{equation}
and its integral extended to all space provides the total energy of the physical system
\begin{equation}\label{t.ener.int}
   E = \int_{\parbox{1.5em}{\tiny \centering all \\ space}} e dV = 4\piy \int_{0}^{\infty} r^{2} e dr
\end{equation}
\begin{subequations}\label{energies}%
With the availability of \REqd{m.ener.d}{g.ener.d}, plus a bit of care, the integral in \REq{t.ener.int} can be carried out smoothly and leads to the expectedly obvious separation
\begin{equation}\label{t.ener}
   E = E_{m} + E_{gf}
\end{equation}
The first term on the right-hand side of \REq{t.ener} is the matter energy in the thermodynamic form 
\begin{equation}\label{m.ener}
    E_{m} = m_{g} R T \left(\frac{3}{2} + \frac{u_{gi}(T)}{RT} + \frac{m_{s}}{m_{g}}\frac{u_{s}(T)}{RT} \right)
\end{equation}
comprising the gas contributions, due to translational degrees of freedom and internal molecular structure, and the shell contribution; its order of magnitude is set by \mbox{$m_{g} R T$}.
The other term in \REq{t.ener} is the gravitational energy
\begin{equation}\label{gf.ener}
   E_{gf} = - \frac{G m_{g}^{2}}{2a} \left( 6 \frac{1-\xiy(1,N)}{N} + \Phi \right)
\end{equation}
It depends on the state variables via the gravitational number [\REq{gcn.t}], is not first-degree homogeneous, and its order of magnitude is set by $G m_{g}^{2}/a$.
\end{subequations}
\begin{subequations}
In \REq{gf.ener}, the function
\begin{align}
   \Phi & = 3 \; \frac{\rhoy_{s}}{\brhoy} \left(1-\frac{\rhoy_{s}}{\brhoy}\right) \left[ \left( 1 + \frac{\deltay}{a} \right)^{2} - 1 \right] \notag \\[.5\baselineskip]
         & + \frac{6}{5}\left(\frac{\rhoy_{s}}{\brhoy}\right)^{2} \left[ \left( 1 + \frac{\deltay}{a} \right)^{5} - 1 \right]\label{g.ener.shell}
\end{align}
with [\REqd{m.shell}{agd}]
\begin{equation}
   \frac{\rhoy_{s}}{\brhoy} = \frac{m_{s}}{m_{g}} \left[ \left(1 + \frac{\deltay}{a}\right)^{3} - 1\right]^{-1} \label{m.ratio} 
\end{equation}    
represents the contribution to the gravitational energy due to the field inside the shell; in our case, $\deltay/a=0.1$ and $m_{s}/m_{g}=0.21$ so that
\begin{equation}\label{g.ener.shell.our.case}
   \Phi  \simeq 0.441
\end{equation}    
\end{subequations}
We draw the reader's attention to the fact that the presence of the shell, always systematically ignored in the literature, affects not only the matter energy via the contribution \mbox{$u_{s}(T)$} but also the gravitational energy via the function $\Phi$.

There are two crosschecks that we can carry out on \REq{gf.ener} to verify if we are on the right track.
If \mbox{$N \rightarrow \epsilon \ll 1$} then \REq{linear} is applicable and we can substitute it in \REq{gf.ener} to obtain
\begin{equation}\label{gf.ener. Nsmall}
   E_{gf} \simeq - \frac{G m_{g}^{2}}{2a} \left(  \frac{6}{5} + \Phi \right)
\end{equation}
If the shell contribution also turns out to be negligible ($\Phi \ll 6/5 $) then the gravitational energy reduces further to
\begin{equation}\label{g.ener.Nsmall.noshell}
   E_{gf} \simeq - \frac{3}{5}\frac{G m_{g}^{2}}{a} 
\end{equation}
a well known textbook result, although the limitations \mbox{($N \rightarrow \epsilon \ll 1$ and $\Phi \ll 6/5$)} of its applicability are hardly emphasized.
The second crosscheck is slightly more elaborate.
We make the total energy [\REq{t.ener}] nondimensional by dividing it with the gravitational-energy order of magnitude $G m_{g}^{2}/a$. 
Moreover, we conform to the ``literature settings'' by assuming a point-particle gas \mbox{[$u_{gi}(T)=0$]} and by neglecting the shell contributions to both matter energy \mbox{[$u_{s}(T)\ll RT$]} and gravitational energy \mbox{($m_{s} \ll m_{g} \Rightarrow \Phi \ll 1$)}.
In this way, \REq{t.ener} reduces to the simplified nondimensional form
\begin{equation}
   \frac{aE}{G m_{g}^{2}} = \frac{3}{2}\frac{1}{N} -  3 \frac{1-\xiy(1,N)}{N} \label{t.ener.lit}
\end{equation}
that corresponds numerically to the expressions deduced within the Lane-Emden's approach and invariably used and commented throughout the literature.
As visual proof, we show the diagram of \REq{t.ener.lit} in \Rfi{p.t.ener.lit.cha} styled according to \mbox{Fig. 2} at page 344 of \Ref{phc2002aa} and in \Rfi{p.t.ener.lit.pad} styled according to \mbox{Fig. 4.4} at page 317 of \Ref{tp1990pr}.
The graphs in \Rfi{p.t.ener.lit} clearly illustrate that the profile spirals around the point \mbox{$N=2\;\mbox{or}\;1/N=0.5,aE/G m_{g}^{2}=-0.25$}, and, at \mbox{$N\simeq 2.03$ or $1/N\simeq 0.492$}, it goes through the notorious minimum \mbox{$aE/G m_{g}^{2}\simeq-0.335$} discovered by Antonov \cite{va1985} in 1962 and declared verge of the gravothermal catastrophe by Lynden-Bell and Wood \cite{dl1968mnras}.
\REqb{t.ener.lit} has an undesirable feature: it diverges if \mbox{$N\rightarrow 0$}. 
This occurrence would persist even if we repristinate the neglected terms.
The reader should rest assured that this divergence is not the hallmark of an additional catastrophe but is, simply, a warning that the order of magnitude $G m_{g}^{2}/a$ is not an adequate, or perhaps convenient, scale factor of the total energy [\REq{t.ener}] for small values of the gravitational number.
From our point of view, the matter-energy order of magnitude \mbox{$m_{g} R T$} is a more appropriate scale factor
because the ensuing gravitational energy 
\begin{equation}\label{g.ener.2}
   E_{gf} = -  m_{g} R T\left[  3 \left( 1-\xiy(1,N) \rule{0pt}{2.5ex}\right) + \frac{\Phi}{2} N\right] 
\end{equation}
is better behaved under all circumstances.
The diagram of \REq{g.ener.2} in nondimensional form is shown in \Rfi{p.t.ener.us}. 
Before anything else, we notice that the gravitational field in the shell affects the profile and its characteristic points via the function $\Phi$.
The profile is basically linear
\begin{equation}\label{g.ener.2.nd.lin}
    \frac{E_{gf}}{m_{g} R T} \simeq - \left( \frac{3}{5} + \frac{\Phi}{2}\right) N 
\end{equation}
for \mbox{$N\le 1$}, then it slopes down until \mbox{$N=N_{m}$} where it achieves the gravitational-energy level
\begin{equation}\label{g.ener.2.nd.Nm}
    \left.\frac{E_{gf}}{m_{g} R T}\right|_{N=N_{m}} = - \left( 2 + \frac{\Phi}{2}N_{m} \right) 
\end{equation}
The center of the spiral is located at $N=2$ and at the gravitational-energy level
\begin{equation}\label{g.ener.2.nd.cs}
    \left.\frac{E_{gf}}{m_{g} R T}\right|_{N=2} = - \left( 2 + \Phi \right) 
\end{equation}
\begin{figure}[H]
  \subfigure[\ Reproduction of Fig. 2 at page 344 of \Ref{phc2002aa}]{\label{p.t.ener.lit.cha}\includegraphics[keepaspectratio=true, trim= 5ex 2ex 2ex 20ex , clip , width=\columnwidth]{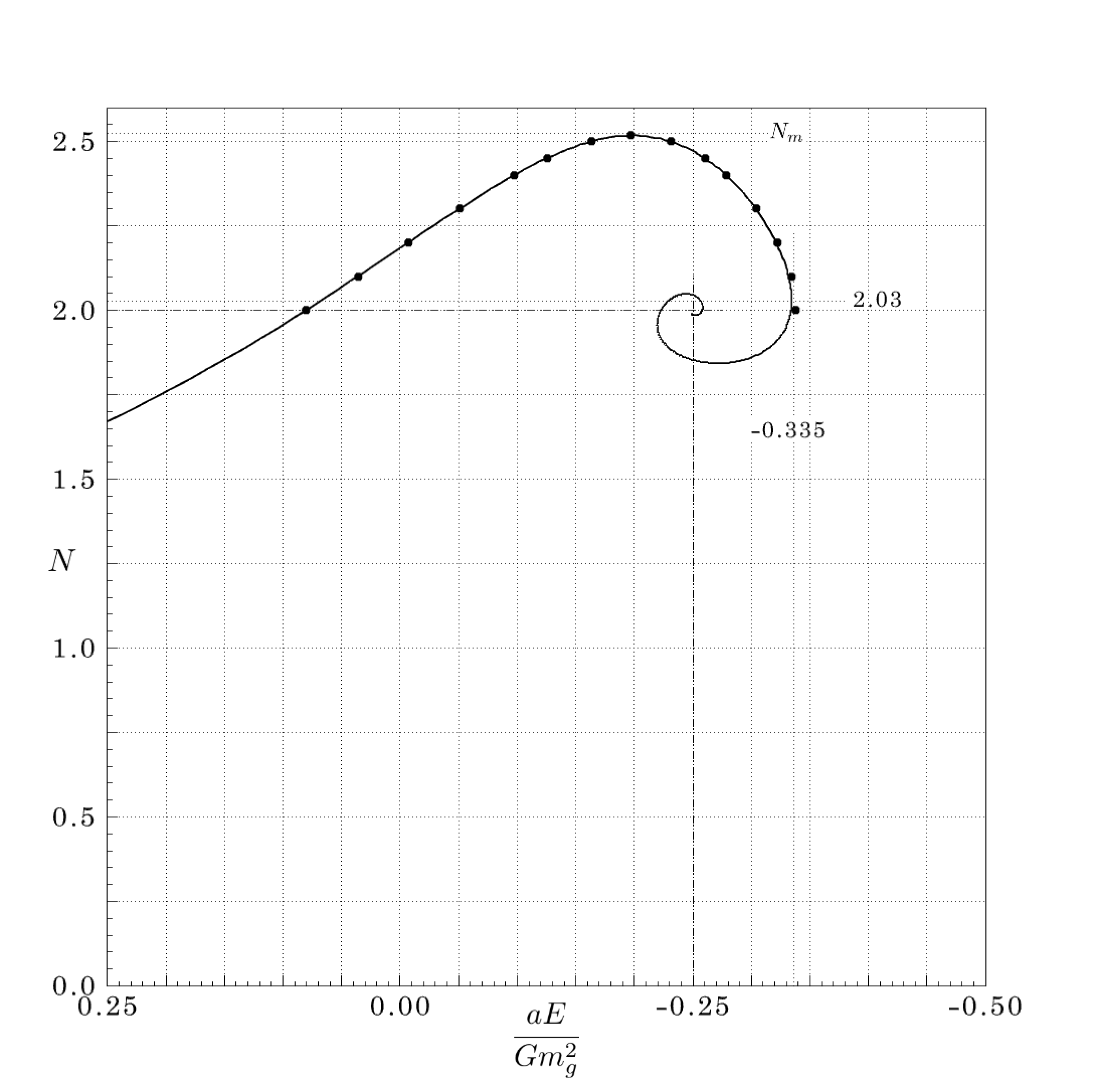}}   \\
  \subfigure[\ Reproduction of Fig. 4.4 at page 317 of \Ref{tp1990pr}]
  {\label{p.t.ener.lit.pad}\includegraphics[keepaspectratio=true, trim= 5ex 2ex 2ex 20ex , clip , width=\columnwidth]{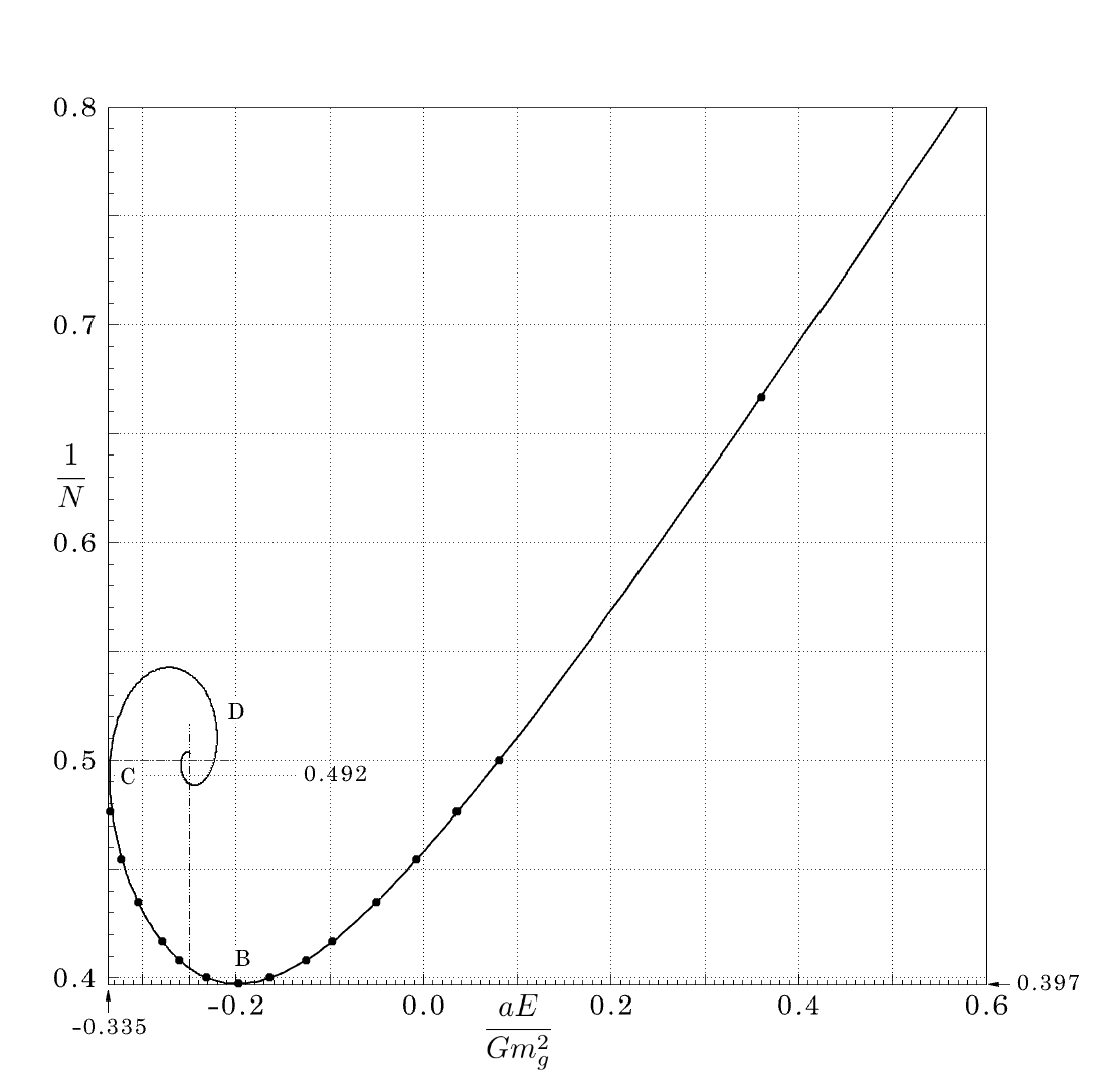}}
  \caption{Nondimensional total energy of the physical system according to \REq{t.ener.lit}\hfill\ }\label{p.t.ener.lit}
\end{figure}\noindent
\REqb{g.ener.2} also presents an absolute minimum whose position depends on the function $\Phi$. 
The extrema are fixed by the vanishing of the derivative with respect to the gravitational number which leads to a condition
\begin{equation}\label{g.ener.2.nd.am}
    \xiy'(1,N) = \frac{\Phi}{6} 
\end{equation}
that unfortunately is not exploitable analytically because it is not resolvable explicitly.
In our calculations, \mbox{$\Phi\simeq 0.441$} [\REq{g.ener.shell.our.case}] and we find numerically the absolute-minimum location at \mbox{$N\simeq2.32$} with minimal gravitational energy 
\begin{equation}\label{g.ener.2.nd.am.e}
    \left.\frac{E_{gf}}{m_{g} R T}\right|_{N=2.32} \simeq  -2.707  
\end{equation}
According to \REq{t.ener}, the explicit expression of the total energy is obtained by adding matter [\REq{m.ener}] and gravitational-field [\REq{g.ener.2}] contributions
\begin{figure}[h]
  \includegraphics[keepaspectratio=true, trim= 5ex 8ex 4ex 20ex , clip , width=\columnwidth]{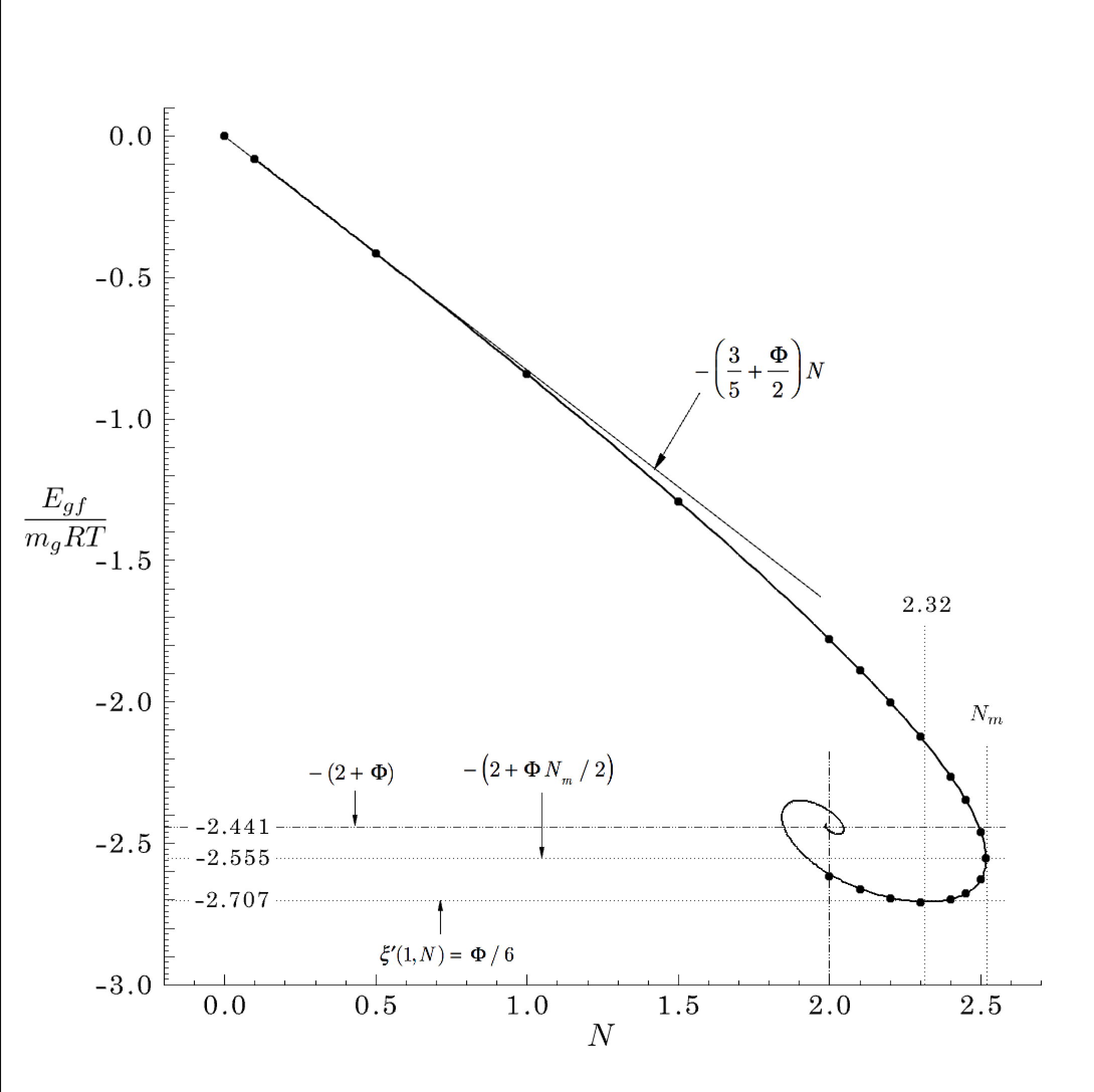}
  \caption{Nondimensional gravitational energy according to \REq{g.ener.2}; $\Phi\simeq 0.441$ in our calculations [\REq{g.ener.shell.our.case}]\hfill\ }
  \label{p.t.ener.us}
\end{figure}
\begin{align}\label{t.ener.1}
    E & = m_{g} R T \left(\frac{3}{2} + \frac{u_{gi}(T)}{RT} + \frac{m_{s}}{m_{g}}\frac{u_{s}(T)}{RT} \right)  \notag \\[.5\baselineskip]
            &  -  m_{g} R T\left[  3 \left( 1-\xiy(1,N) \rule{0pt}{2.5ex}\right) + \frac{\Phi}{2} N\right] 
\end{align}

The discovery of thermodynamic peculiarities connected with the study of the entropy differential [\REq{diff.entr.t}] discussed in \Rse{entr} auto-suggests analogous action with respect to the total-energy differential.
The latter can be obtained straightforwardly from \REq{t.ener.1} by taking into account the same ingredients [\REq{gcn.t}, \REqd{mass.s}{dgcn.t}] used to derive the entropy differential plus the additional (irrelevant) constraint
\begin{equation}\label{thick.s}
     \frac{\deltay}{a} = const
\end{equation}
that we introduce by invoking the same justification we adduced for adopting \REq{mass.s}.
The final result reads
\begin{align}\label{diff.ener.t}
  dE  & = m_{g}R \left(\rule{0pt}{3.3ex}\right. \frac{3}{2} + \frac{c_{vi}(T)}{R} + \frac{m_{s}}{m_{g}}\frac{c_{s}(T)}{R} - \phienep(N)\left.\rule{0pt}{3.3ex}\right) dT   \notag \\[.5\baselineskip]
       & - \frac{m_{g}RT}{V} N \left( \xiy'(1,N) - \frac{\Phi}{6} \right) dV \notag \\[.5\baselineskip]
       & + R T \left(\rule{0pt}{3.3ex}\right.\frac{3}{2} + \frac{u_{gi}(T)}{RT} + \frac{m_{s}}{m_{g}}\frac{u_{s}(T)}{RT}  - \phienem(N) \notag \\[.5\baselineskip]
       & \hspace*{9em} - \Phi N \left.\rule{0pt}{3.3ex}\right) dm_{g}
\end{align}
and contains two new functions 
\begin{subequations} \label{diff.ener.defs}
\begin{align}
     \phienep(N)  &  = 3 \left(\rule{0pt}{2.5ex} 1 - \xiy(1,N) + N \xiy'(1,N)  \right) \label{phienep}  \\[.5\baselineskip]
     \phienem(N) &  = 3 \left(\rule{0pt}{2.5ex} 1 - \xiy(1,N) - N \xiy'(1,N)  \right)  \label{phienem}
\end{align}
of the gravitational number.
\end{subequations}
The specific heats on the top line of \REq{diff.ener.t}, given in \REqq{sp.heats} in terms of the specific entropies, can be expressed also in terms of the corresponding specific energies
\begin{subequations}\label{sp.heats.e}
\begin{equation}\label{cvi.e}
   c_{vi} =  \tds{}{u_{gi}}{T} 
\end{equation}
\begin{equation}\label{cs.e}
   c_{s} =  \tds{}{u_{s}}{T}
\end{equation}
\end{subequations}
\begin{figure}[H]
  \includegraphics[keepaspectratio=true, trim= 5ex 5ex 4ex 20ex , clip , width=\columnwidth]{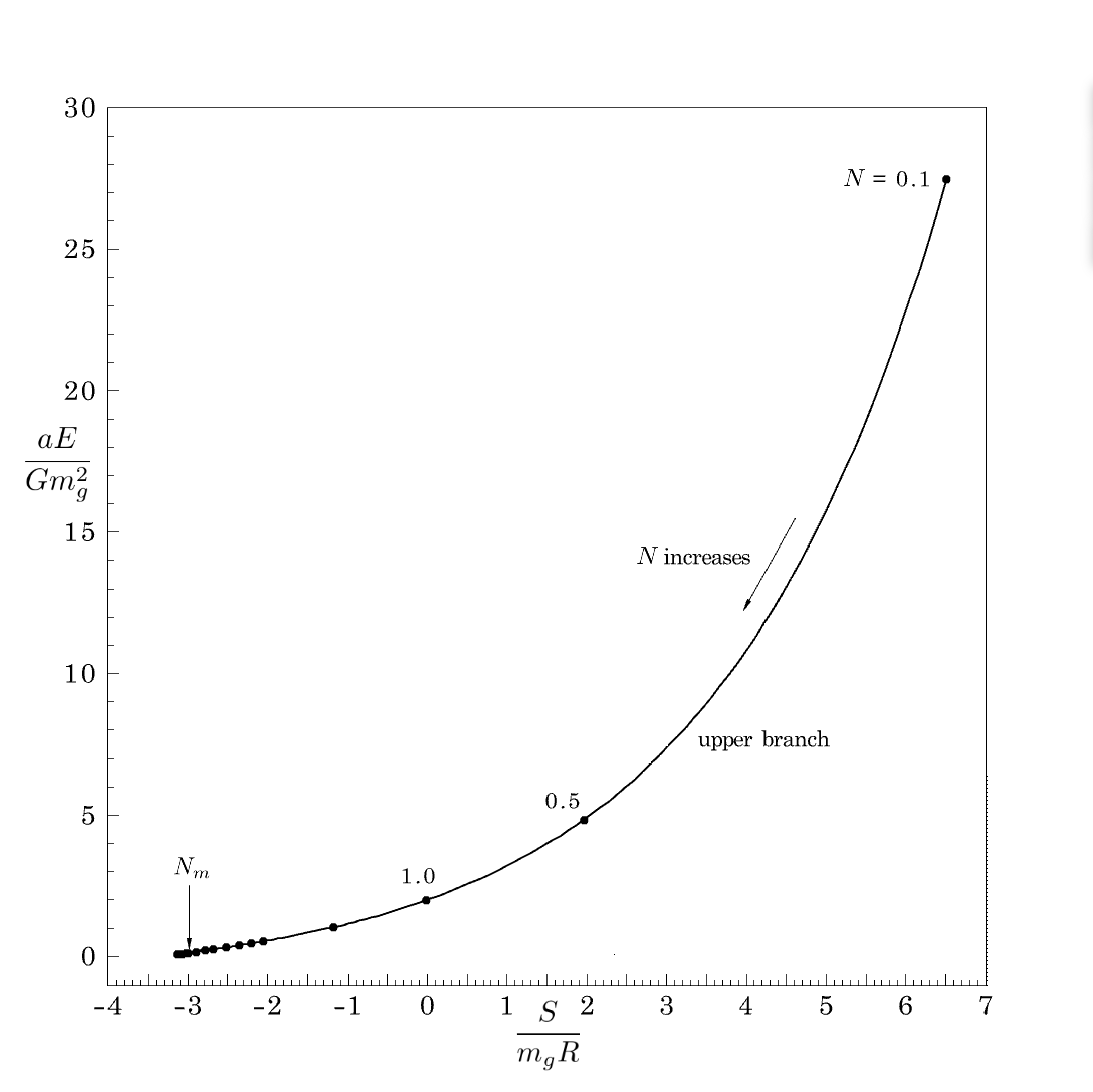}
  \caption{Fundamental relation energy versus entropy for prescribed volume and gas mass, mainly for the upper-branch configurations; visualization specific-heat settings are enforced.\hfill\ }
  \label{p.fr.ener}
\end{figure}\noindent 
The partial derivatives of the total energy with respect to the state variables are easily identified by inspection of \REq{diff.ener.t} and once again we recognize in them the separation structure already encountered when dealing with the entropy partial derivatives [\REq{diff.entr.t}].
We put on hold discussing them here, however, and just save the result in view of the thermodynamic considerations of the next section because, we believe, the discussion seems more appropriate within that context.

\subsection{Axiomatic-thermodynamics perspective}\label{thd.con}
\subsubsection{Fundamental relation}
The astrophysical literature systematically regards total energy as entitled to play the role of thermodynamic fundamental relation in the energetic representation: \mbox{$E=E(S,V,m_{g})$}; its inverse \mbox{$S=S(E,V,m_{g})$} provides the fundamental relation in the equivalent entropic representation \cite{hc1963,lt1966,ln1971,hc1985}.
If we conform to this idea then both fundamental relations are available in parametric form via \REq{gcn.t}, \REq{entr.t.1} and \REq{t.ener.1}, the parameter being the temperature.
Regrettably, the latter variable is not eliminable analytically within the set of the indicated equations in general and, therefore, the fundamental relation may not be obtainable in explicit form save for simplified circumstances such as that corresponding to the visualization specific-heat settings introduced just after \REq{gcn.temp}.
Nevertheless, we believe it is worthwhile to dig up the details for this simplified case because of the insight into the physics of the problem that can be gained.
The integration of \REqq{sp.heats} and \REqq{sp.heats.e} is straightforward; the integration constants can be conveniently brought to the left-hand side of \REqd{entr.t.1}{t.ener.1} and incorporated as harmless reference levels of total energy and entropy. 
After that, the elimination of the temperature follows smoothly. 
The diagram illustrating the global view of energy versus entropy for given volume and gas mass is shown in \Rfi{p.fr.ener}. 
The profile has a typical monotonic trend and all seems in order and as expected for the upper-branch configurations.
The peculiar part of the profile lies, of course, in the neighborhood of $N_{m}$ but it is invisible in the scale of \Rfi{p.fr.ener}. 
Before zooming in, however, it is appropriate first to investigate the partial derivative \mbox{$\pdst{}{E}{S}{V,m_{g}}$}.

\subsubsection{State equation \mbox{$\pdst{}{E}{S}{V,m_{g}}$} \label{thd.se.T}}
The state equations are provided by the fundamental relation's partial derivatives and the latter can be obtained in a straightforward manner from the entropy and total-energy differentials 
[\REqd{diff.entr.t}{diff.ener.t}] without specifying any simplification regarding the specific-heat terms.
The recipe is rather simple.
Suppose we wish to obtain \mbox{$\pdst{}{E}{S}{V,m_{g}}$}: then first we set \mbox{$dV=dm_{g}=0$} in \REqd{diff.entr.t}{diff.ener.t} and after eliminate $dT$ from the reduced expressions of the differentials.
This algebraic manipulation leads to a result
\begin{align}\label{se.T}
    \frak{T} & = \pds{}{E}{S}{V,m_{g}} \notag \\[.5\baselineskip]
               & = T\;\frac{\dfrac{3}{2} + \dfrac{c_{vi}(T)}{R} + \dfrac{m_{s}}{m_{g}}\dfrac{c_{s}(T)}{R} - \phienep(N)}
                              {\dfrac{3}{2} + \dfrac{c_{vi}(T)}{R} + \dfrac{m_{s}}{m_{g}}\dfrac{c_{s}(T)}{R} - \phient(N)}
\end{align}
that provokes some anxiety because, as it stands, it seems to deny the variables $(T, S)$ the status of couple of conjugated variables in the energetic representation.
Actually, \REq{se.T} teaches a good lesson: in the presence of the gravitational field, we should not make the mistake of taking for granted thermodynamic notions we are accustomed to in the absence of the gravitational field.
Incidentally, this is the reason why we refrained to put instinctively the stamp of thermal-stability criterion on \REq{ent.derivs.T}, as we can do in the absence of the gravitational field, and awaited, instead, to follow through with the standard analysis of the fundamental relation and its first/second derivatives.
Of course, the first thought that comes to mind when looking at \REq{se.T} is to compare the functions $\phient(N)$ and $\phienep(N)$ [\REqd{phient}{phienep}], although they popped in rather independently.
To this aim, we have superposed the function $\phienep(N)$ on the profiles of \Rfi{p.phient}. 
The outcome is shown in \Rfi{p.phient.phienep} and reveals an unexpected surprise: the two functions coincide
\begin{equation}\label{phient=phienep}
   \phient(N) = \phienep(N)
\end{equation}
The immediate consequence is the reduction of \REq{se.T} to the more reassuring form
\begin{equation}\label{se.T.1}
       \frak{T} = \pds{}{E}{S}{V,m_{g}} = T
\end{equation}
that sanctions the status of $(T,S)$ as couple of conjugated variables and, at the same time, dissipates the anxiety provoked by \REq{se.T}.
With reassurance [\REq{se.T.1}] that the slope of the profile in \Rfi{p.fr.ener} \textit{is} the temperature, we can swiftly determine also the curvature
\begin{equation}\label{p.fr.ener.c}
   \pds{2}{E}{S}{V,m_{g}} = \pds{}{T}{S}{V,m_{g}} = \pds{}{S}{T}{V,m_{g}}^{-1}
\end{equation}
and look back \textit{now} at \REq{ent.derivs.T} as legitimate criterion of thermal stability also in the presence of the gravitational field.
On the basis of \REq{phient=phienep}, with a view to the differentials of entropy [\REq{diff.entr.t}] and total energy [\REq{diff.ener.t}], we can affirm that
\begin{figure}[H]
  \subfigure[\ Upper branch and lower branch between asymptotes (1.84,$N_{m}$); solid circles and hollow squares correspond to D algorithm]{\label{p.phient.phienep.1}\includegraphics[keepaspectratio=true, trim= 5ex 8ex 4ex 20ex , clip , width=\columnwidth]{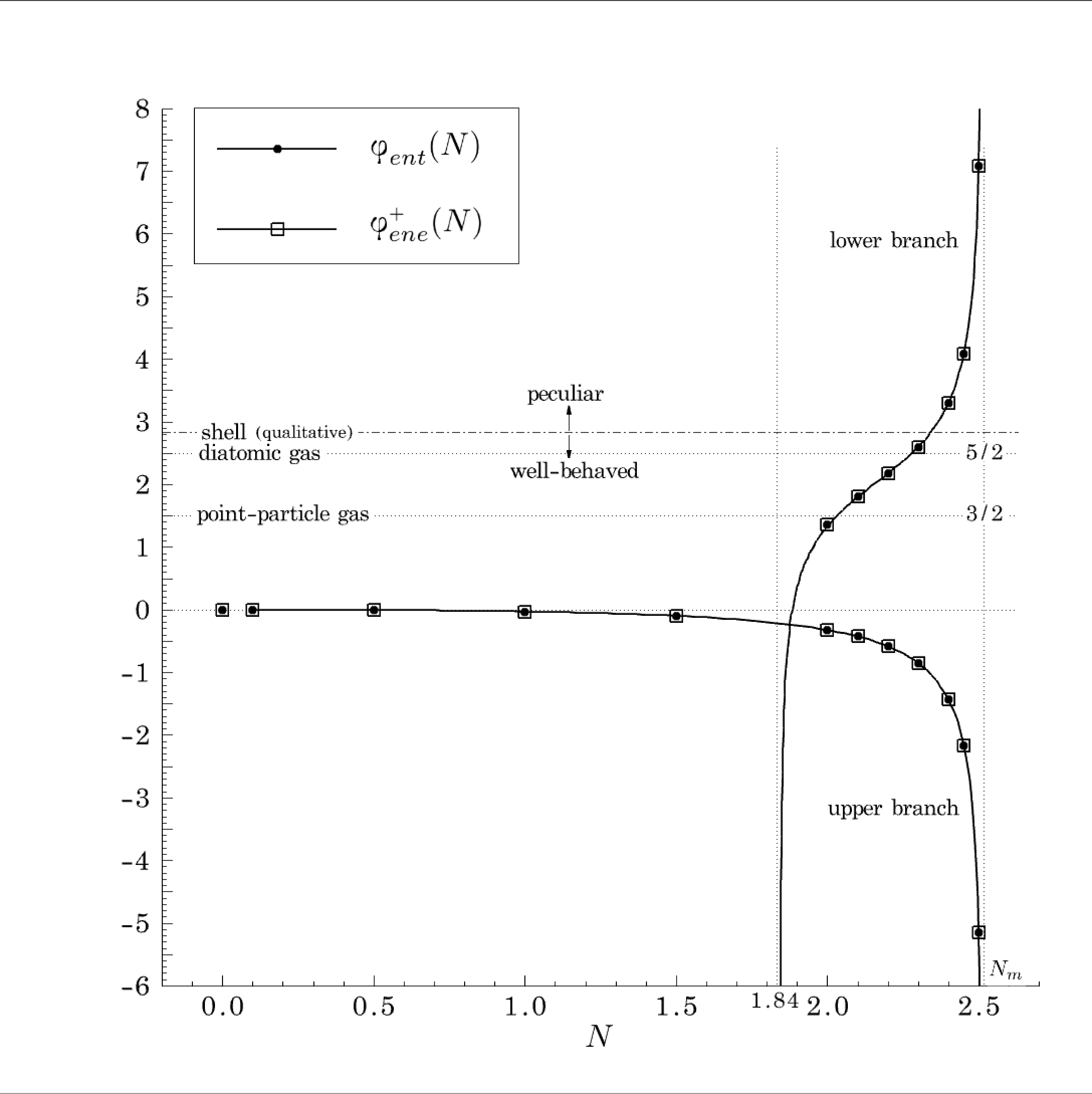}}   \\[3\baselineskip]
  \subfigure[\ Lower branches between asymptotes (1.84,2.05) and (1.98,2.05); F \& FP algorithms only]{\label{p.phient.phienep.2}\includegraphics[keepaspectratio=true, trim= 5ex 8ex 4ex 20ex , clip , width=\columnwidth]{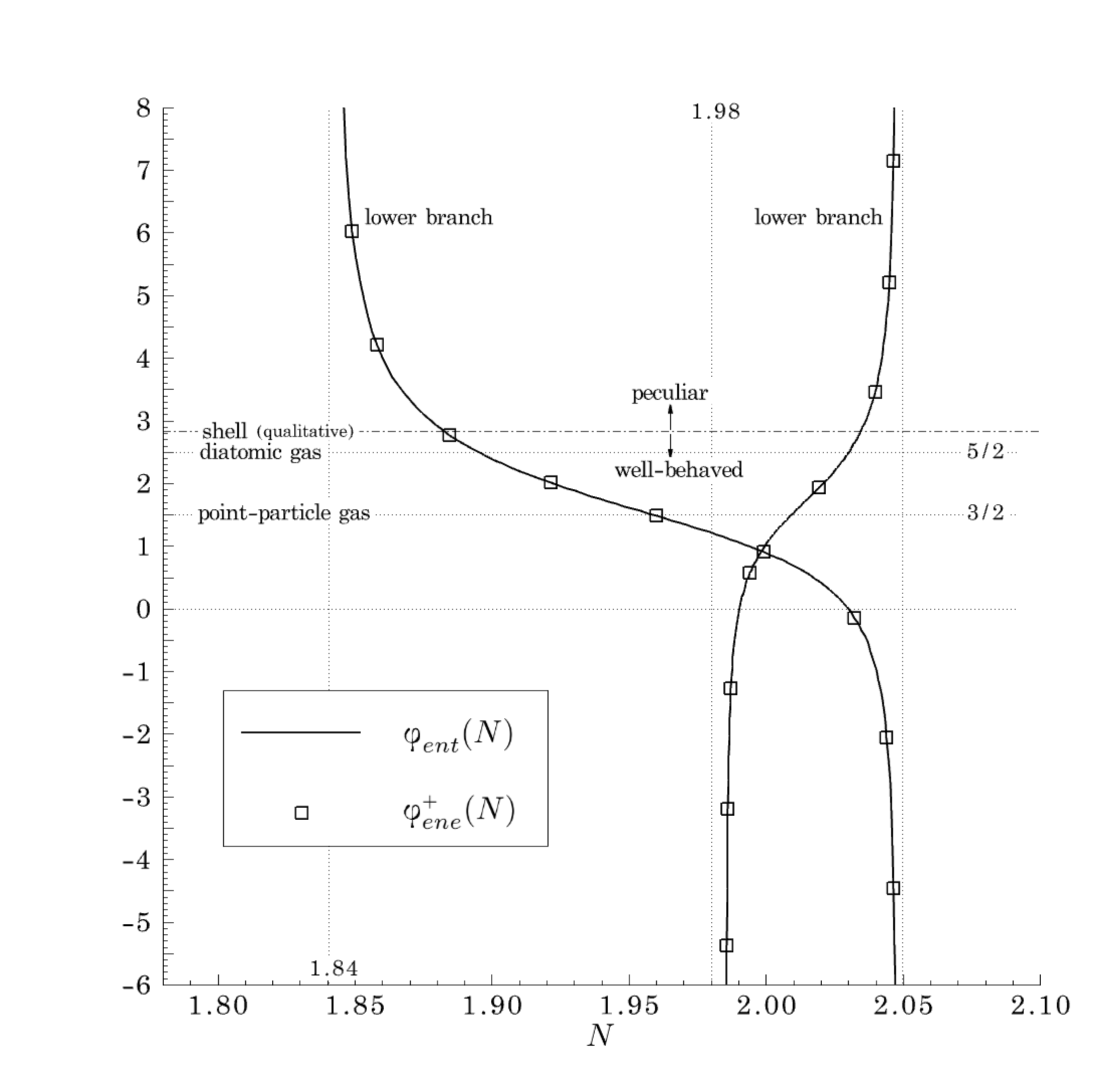}}
  \caption{Functions $\protect\phient(N)$ and $\protect\phienep(N)$}\label{p.phient.phienep}
\end{figure}\noindent%
\begin{equation}\label{ener.deriv.T}
   \pds{}{E}{T}{V,m_{g}} = T \pds{}{S}{T}{V,m_{g}}
\end{equation}
and recognize that 
\begin{equation}\label{ener.deriv.T.neg}
    \pds{}{E}{T}{V,m_{g}} < 0
\end{equation}
for the thermally unstable configurations.
{\color{\newc} With regard to \REq{ener.deriv.T.neg}, we} \erase{We} are aware of Lynden-Bell and coauthors' crusade \cite{dl1977mnras, dl1968mnras,dl1999p} \erase{based on \REq{ener.deriv.T.neg}} in support of existence and physical plausibility of a negative heat capacity of the gas and the controversy it spawned in the literature {\color{\newc}\cite{wt1970zfp,ih1978ptp,wt2003prl,phc2006ijmp,lv2016jsm}}.
However, we refrain from {\color{\newc} discussing virial-theorem applications and statistical-ensemble nonequivalence because it would take us far away from the mainstream of the present discourse and postpone the exposition of our viewpoint to future communications.
Having taken that commitment, here we just state that we do not interpret} \erase{interpreting} \mbox{$\pdst{}{E}{T}{V,m_{g}}$} as heat capacity, in general, and of the gas, in particular, because its more explicit expression 
\begin{align}\label{ener.deriv.T.dim}
    \pds{}{E}{T}{V,m_{g}} & = m_{g}  \left( \frac{3}{2} R + c_{vi}(T)  \right) \notag \\[.5\baselineskip]
                                & + m_{s} c_{s}(T) \notag \\[.5\baselineskip]
                                & - m_{g} R \;\phienep(N)
\end{align}
adapted from \REq{diff.ener.t} includes not only terms belonging to the gas [\REq{ener.deriv.T.dim}, top line] but also properties belonging to the shell [\REq{ener.deriv.T.dim}, middle line] and a \textit{correction} belonging to the gravitational field [\REq{ener.deriv.T.dim}, bottom line]; therefore, what sense would it make to regard \mbox{$\pdst{}{E}{T}{V,m_{g}}$} as a physical property of the gas?
\erase{The narrower interpretation of Lynden-Bell and coauthors descends maybe from the fact that they never considered presence and influence of the shell and always saw the gravitational-field energy as potential energy of a point-particle gas.}
\begin{figure}[h]
  \includegraphics[keepaspectratio=true, trim= 5ex 8ex 4ex 20ex , clip , width=\columnwidth]{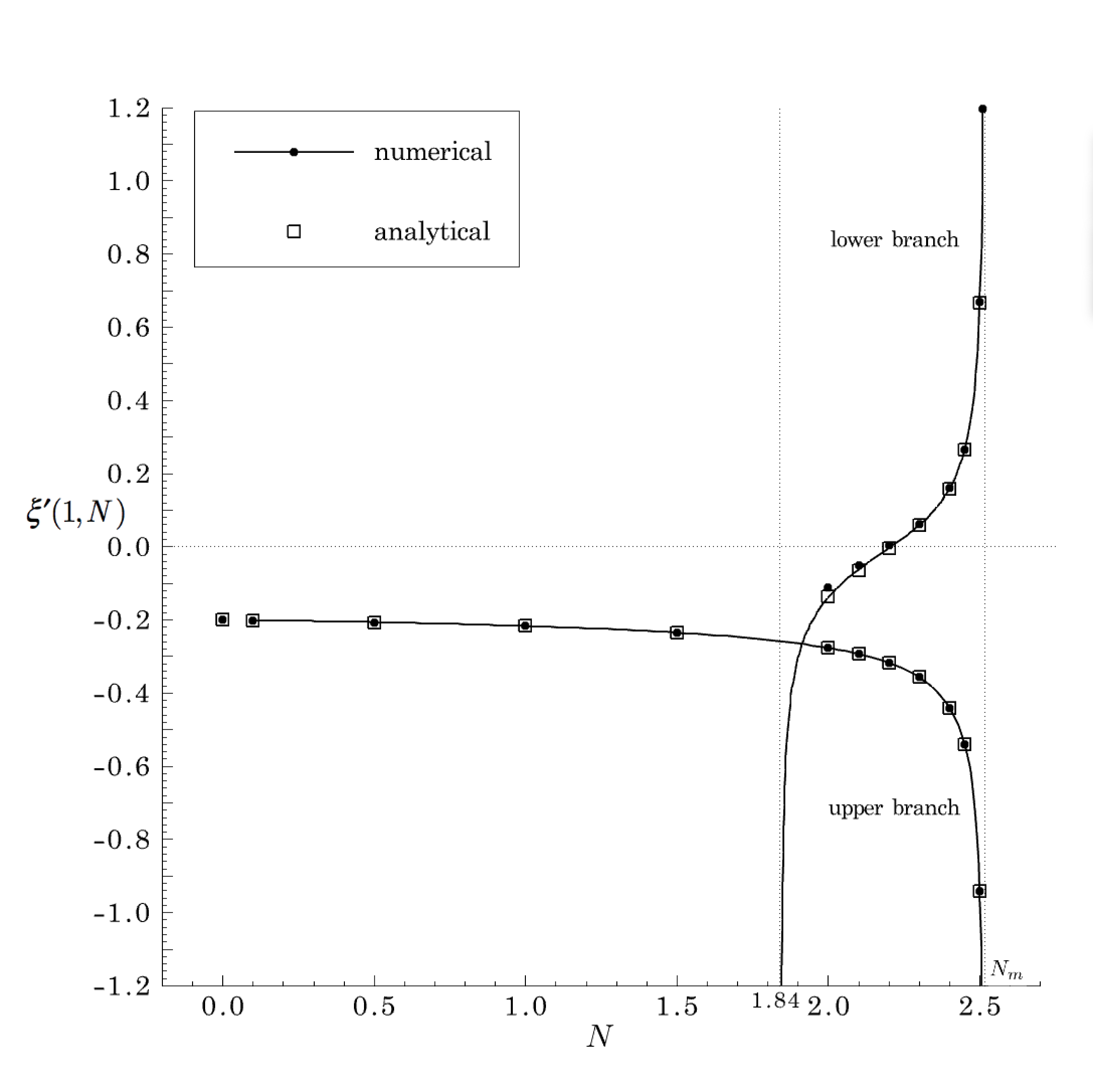}
  \caption{Peripheral-density derivative with respect to gravitational number; upper branch and lower branch between asymptotes (1.84,$N_{m}$)\hfill\ }
  \label{p.xi1p}
\end{figure}
From our standpoint, we view the mentioned terms on top and middle lines of \REq{ener.deriv.T.dim} as, respectively, gas and shell heat capacities; they are always positive, by definition.
It is the gravitational correction on the bottom line of \REq{ener.deriv.T.dim} that can flip the sign of \mbox{$\pdst{}{E}{T}{V,m_{g}}$} and introduce thermal instability.
\erase{At any rate, disputing interpretations is a moot debate because, in the end, it is physical content and consequences of \REq{ener.deriv.T.neg} that really matter, regardless of whatever interpretation is attached to the derivative on its left-hand side.}

Another amazing consequence of \REq{phient=phienep} is the possibility to obtain an analytical expression for the derivative $\xiy'(1,N)$, as anticipated in \Rse{entr} near \REq{dxi}.
Substituting \REqd{phient}{phienep} into \REq{phient=phienep} and solving for the derivative yields
\begin{equation}\label{dxi.a}
   \xiy'(1,N) = \frac{\xiy(1,N)}{N} \frac{3\left(1 - \xiy(1,N)\right) - N}{3\,\xiy(1,N) - 1}
\end{equation}
We see in \REq{dxi.a} the analytical confirmation of a numerical result displayed in \Rfi{p.xi1}: all vertical-slope configurations share the peripheral-density level \mbox{$\xiy(1,N)=1/3$}.
As further verification, we show the comparison of numerical [\REq{dxi}] versus analytical [\REq{dxi.a}] values of \mbox{$\xiy'(1,N)$} in \Rfi{p.xi1p}; the agreement is satisfactory notwithstanding a slight deterioration in accuracy of the D-algorithm (solid circle) for the lower-branch configurations in the vicinity of the asymptote \mbox{$N=1.84$}.

\subsubsection{Fundamental relation (reprise).\label{thd.fr.r}}
We return now to the profile in \Rfi{p.fr.ener} and resume the discussion regarding the fundamental relation put on hold in the paragraph ending \Rse{thd.se.T}. 
Having acquired reliable knowledge about slope [\REq{se.T.1}] and curvature [\REq{p.fr.ener.c}], we proceed to expand the axes' scales to bring the situation in the vicinity of $N_{m}$ into focus. 
The zoomed-in view illustrated in \Rfi{p.fr.ener.z.us} reveals a singular zigzag pattern whose cusps correspond to the characteristic points F$_{1}$, F$_{2}$, F$_{3}, \ldots$ we have encountered in \Rfi{p.phient}.
At the right of $N_{m}$, as we already know from \Rfi{p.fr.ener}, the configurations are thermally stable [\mbox{$\pdst{}{S}{T}{V,m_{g}}>0$}] and the profile has positive curvature 
\begin{equation}\label{p.fr.ener.c.ts}
   \pds{2}{E}{S}{V,m_{g}} > 0
\end{equation}
At $N_{m}$, the derivative \mbox{$\pdst{}{S}{T}{V,m_{g}}$} diverges to \mbox{$\pm\infty$} [\REq{ent.derivs.T} and \Rfi{p.phient1}] and the profile goes through an inflection point 
\begin{equation}\label{p.fr.ener.c.ip}
   \pds{2}{E}{S}{V,m_{g}} = 0
\end{equation}
where the curvature transitions from positive to negative.
At the left of $N_{m}$, there is the first series of thermally unstable configurations 
[\mbox{$\pdst{}{S}{T}{V,m_{g}}<0$}] with negative curvature
\begin{equation}\label{p.fr.ener.c.tus}
   \pds{2}{E}{S}{V,m_{g}} < 0
\end{equation}
\begin{figure}[H]
  \subfigure[\ With visualization specific-heat settings; $\Phi\simeq 0.441$ in our calculations \mbox{[\REq{g.ener.shell.our.case}]}]{\label{p.fr.ener.z.us}\includegraphics[keepaspectratio=true, trim= 5ex 6ex 4ex 20ex , clip , width=\columnwidth]{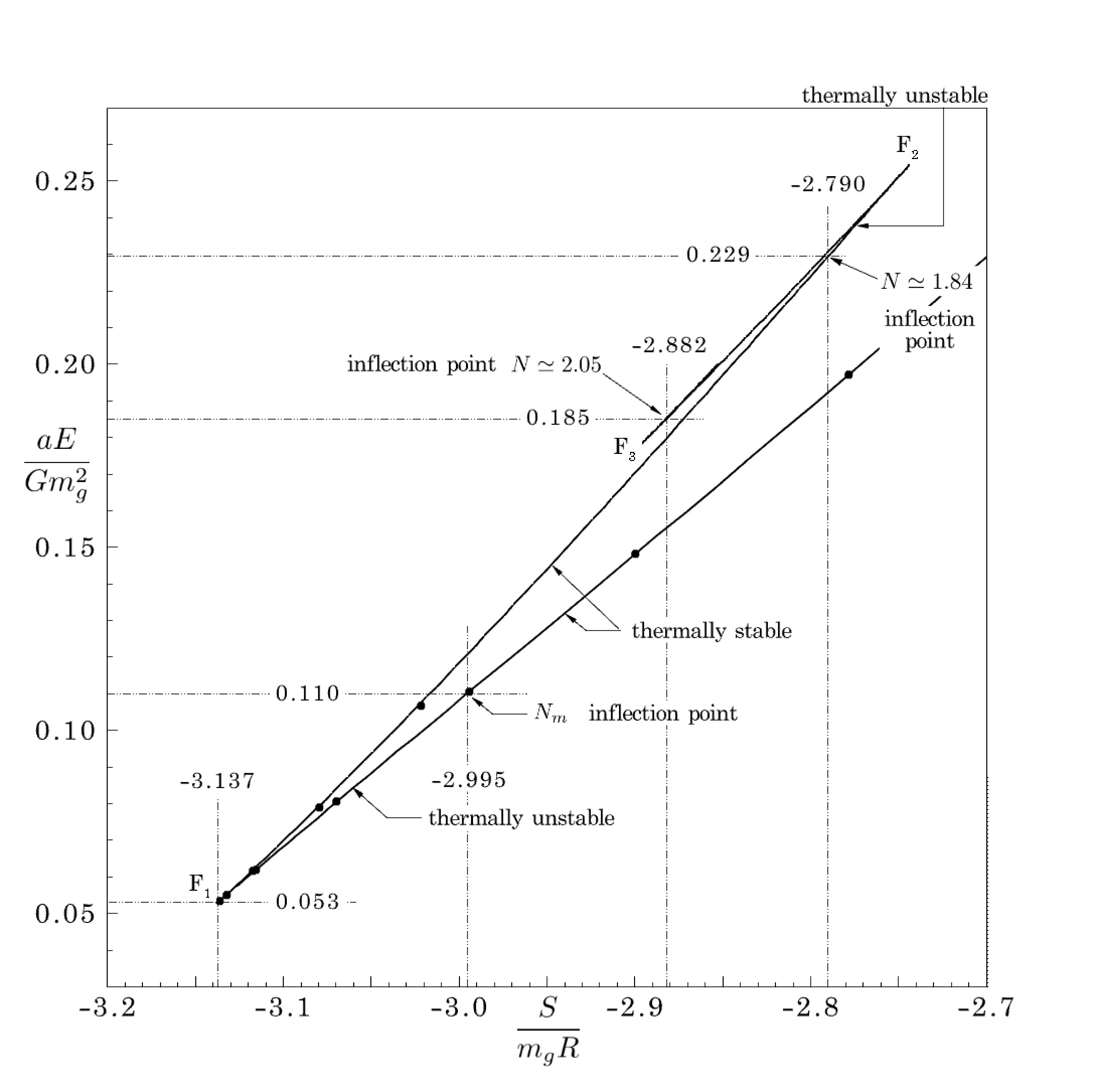}}
  \subfigure[\ With literature settings]{\label{p.fr.ener.z.lit}\includegraphics[keepaspectratio=true, trim= 5ex 6ex 4ex 20ex , clip , width=\columnwidth]{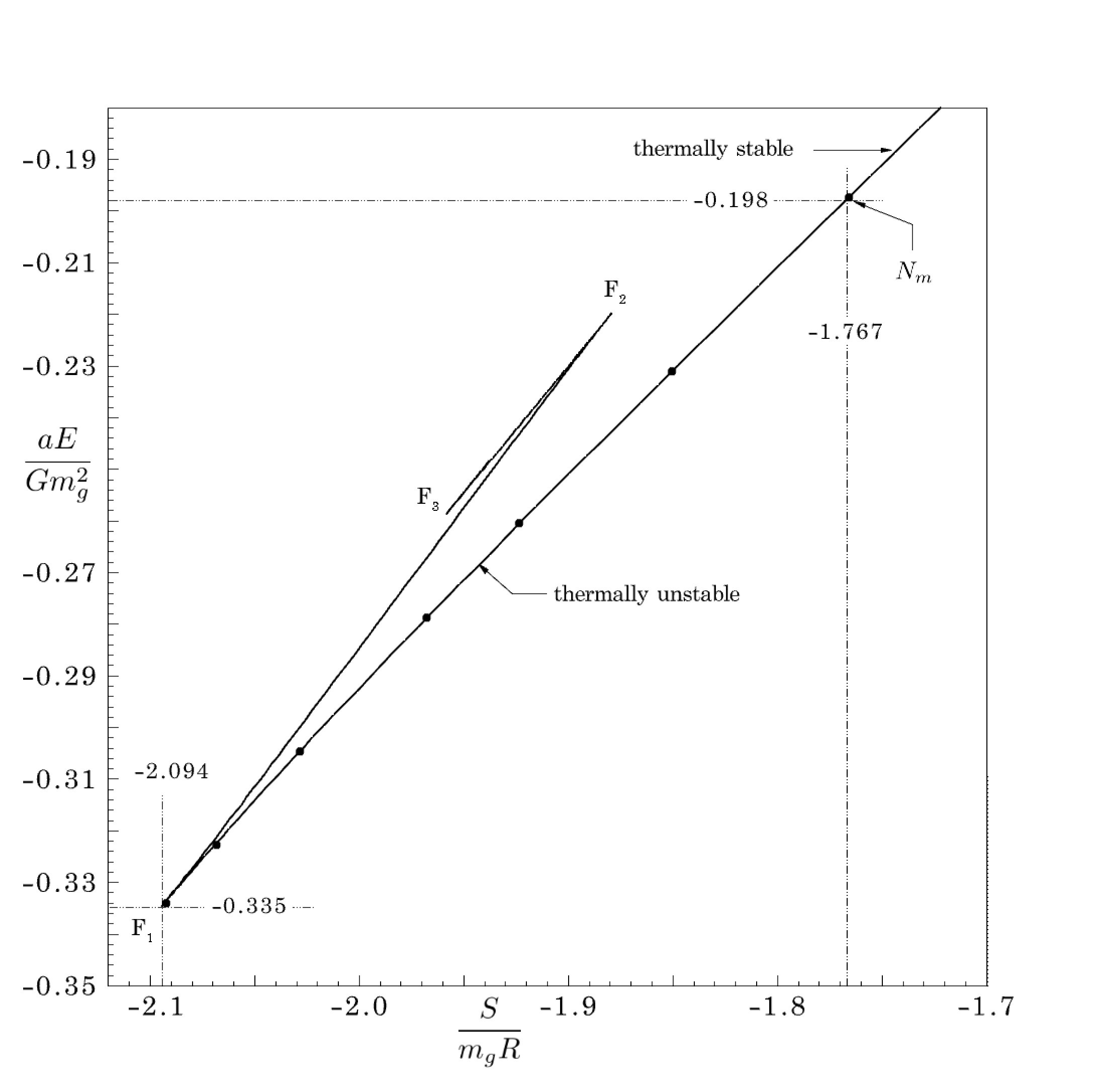}}
  \caption{Fundamental relation energy versus entropy for prescribed volume and gas mass. Enlarged view of the zone relative to lower-branch configurations\hfill\ }\label{p.fr.ener.z}
\end{figure}\noindent
which terminates in F$_{1}$ where the slope, i.e. the temperature, is continuous but the curvature 
flips from $-\infty$ to $+\infty$ reintroducing thermal stability.
The sequence ``thermally stable configurations - inflection point - thermally unstable configurations'' repeats in between each couple of cusps endlessly, in consistent agreement with the picture conveyed by \Rfi{p.phient}.
Each thermally unstable configuration, conceded though that there is a feasible way \textit{to prepare} the physical system in such desired configuration, is clearly the brink of a potential gravothermal catastrophe because any thermal disturbance, however small, will knock the physical system off and make it drift away from the initial configuration: if heated up/cooled down then the physical system will demand/yield more energy.
The location of the cusp F$_{1}$ is important because it signals the existence of bounding values of entropy and total energy: no fluid-static configuration exists to the left of and/or below F$_{1}$.
Therefore, if the physical system turns out to be in a thermodynamic state with entropy and/or total energy beyond the bounds indicated by F$_{1}$ then it will necessarily be in a gravitofluid-\textit{dynamic} condition.
In order to emphasize more incisively this aspect, we show in \Rfi{p.fr.ener.z.lit} the zoomed-in view relative to the calculation of the fundamental relation with the literature settings because, in it, the reader will feel comfortable to recognize the Antonov's well-publicized total-energy minimum \mbox{(-0.335)} and more receptive about the existence of a, perhaps less-publicized, entropy minimum \mbox{(-2.094)}.

We ought to point out that the thermal-stability conclusions we reached along the axiomatic-thermodynamics guidelines differ to some extent from what reported in the literature.
For example, Padmanabhan \cite{tp1990pr} referred to and extended Antonov's analysis \cite{va1985}; with reference to his Fig. 4.4, reproduced here in \Rfi{p.t.ener.lit.pad}, he wrote at page 317
\begin{quote}
   As we shall see later, the branch CD is unstable and is not physically realisable. (Point C corresponds to a density contrast of 709; thus isothermal spheres with a density contrast in the range (32,709) have a negative specific heat.)
\end{quote} 
We share Padmanabhan's opinion that thermally unstable configurations may not be physically realizable.
On the other hand, our findings indicate that not all the configurations on his CD branch, which corresponds to the zigzagging arcs F$_{1}$F$_{2}$, F$_{2}$F$_{3},\ldots$ in \Rfi{p.fr.ener.z.lit}, are thermally unstable. 
Moreover, the quoted statement seems to exclude his BC branch, which represents the ``isothermal spheres with a density contrast in the range (32,709)'' and corresponds to the arc $N_{m}$F$_{1}$ in \Rfi{p.fr.ener.z.lit}, from the set of thermally unstable configurations.
This impression is corroborated by statement (iii) at page 323 of \cite{tp1990pr}
\begin{quote}
   Systems with \mbox{$RE/GM^{2}>-0.335$} and $\rhoy_{c}<709\,\rhoy_{e}(R)$ can form isothermal sphere which are local maxima of entropy.
\end{quote}
In Padmanabhan's notation, $R$ is the sphere radius, corresponding to our $a$, $\rhoy_{c}$ and $\rhoy_{e}(R)$ are central and peripheral densities respectively.
Padmanabhan's latter quote is aligned with the more explicit conclusion drawn by Katz at page 768 of \cite{jk1978mnras} 
\begin{quote}
   The branch of the curve between $h=1$ (at infinity) and $h=709$ is a branch of stable configurations.
\end{quote}
while referring to his \mbox{Fig. 3} which illustrates the curve $1/T$ against $E$; in Katz's notation, $h$ is the density contrast.
However, the text in parentheses after the semicolon in Padmanabhan's former quote explicitly acknowledges the applicability of \REq{ener.deriv.T.neg}, unambiguous mark of thermal instability according to axiomatic thermodynamics, to his BC branch.
We perceive irreconcilable contradictions in the quoted statements.

\subsubsection{State equation \mbox{$\pdst{}{E}{V}{S,m_{g}}$}}
We move on now to next partial derivative \mbox{$\pdst{}{E}{V}{S,m_{g}}$}.
The application of the recipe described in the paragraph just before \REq{se.T} and adapted to the present case yields
\begin{equation}\label{se.p}
    -\frak{p} = \pds{}{E}{V}{S,m_{g}} = - \bpres \left( \xiy(1,N) - \dfrac{\Phi}{6} N \right)
\end{equation}
Apart the minus sign, the right-hand side of \REq{se.p} shows the difference between the peripheral pressure and a term due to the shell's presence.
This state equation, with proper consideration of \REqq{agdp} and \REq{gcn.t}, expands into the form   
\begin{equation}\label{se.p.terletsky}
   \frak{p}V = m_{g} R T \cdot \xiy(1,N) - \left(\frac{4\piy}{3}\right)^{1/3} \frac{\Phi}{6} G m_{g}^{2} V^{-1/3}
\end{equation}
bearing a strong resemblance to the state equation that Bonnor \cite{wb1956mnras} displayed as \mbox{Eq. (1.2)} and claimed to Terletsky (Ref. 7 in Bonnor's list of references); the gravitational effects reside in the presence of $\xiy(1,N)$, which Terletsky did not have because he considered a gas of uniform density, and of the second term on the right-hand side whose structure indeed conforms with the correction suggested by Terletsky and whose origin is clearly attributable to the shell.

Taking into account the definition of the gravitational energy [\REq{g.ener.2}], the right-hand side of \REq{se.p} can be easily manipulated in order to rephrase the same equation into the, perhaps more meaningful, form
\begin{equation}\label{se.p.Egf}
    \frac{\frak{p}}{\bpres} =  1 + \frac{1}{3} \frac{E_{gf}}{m_{g} R T}
\end{equation}
\REqb{se.p.Egf} informs that the profile of $\frak{p}/\bpres$ is simply a rescaling of the gravitational-energy profile (\Rfi{p.t.ener.us}) and, apparently, does not add any novel information.
However, a curious thermodynamicist would definitely wonder how the isotherms in the \mbox{$\frak{p},V$} plane look like and for a good reason: the isotherm's slope \mbox{$\pdst{}{\frak{p}}{V}{T,m_{g}}$} constitutes a criterion of thermodynamic stability \cite{wb1956mnras,wb1958mnras,ws1987}. 
We \textit{do} know this, by the way, strong of the reassuring \REq{se.T.1}.
\REqb{se.p} requires a bit of adaptation in order to satisfy the thermodynamicist's curiosity.
The step sequence consists in resolving the volume in terms of the gravitational number from \REq{gcn.t}
\begin{equation}\label{gcn.V}
   V = \tfrac{4}{3}\piy   \left( \frac{G m_{g}}{RT} \right)^{3}    \frac{1}{N^{3}}
\end{equation}
\begin{figure}[H]
  \subfigure[\ With the original's CGI units]{\label{p.pwall.b.d}
  \includegraphics[keepaspectratio=true, trim= 1ex 8ex 4ex 20ex , clip , width=\columnwidth]{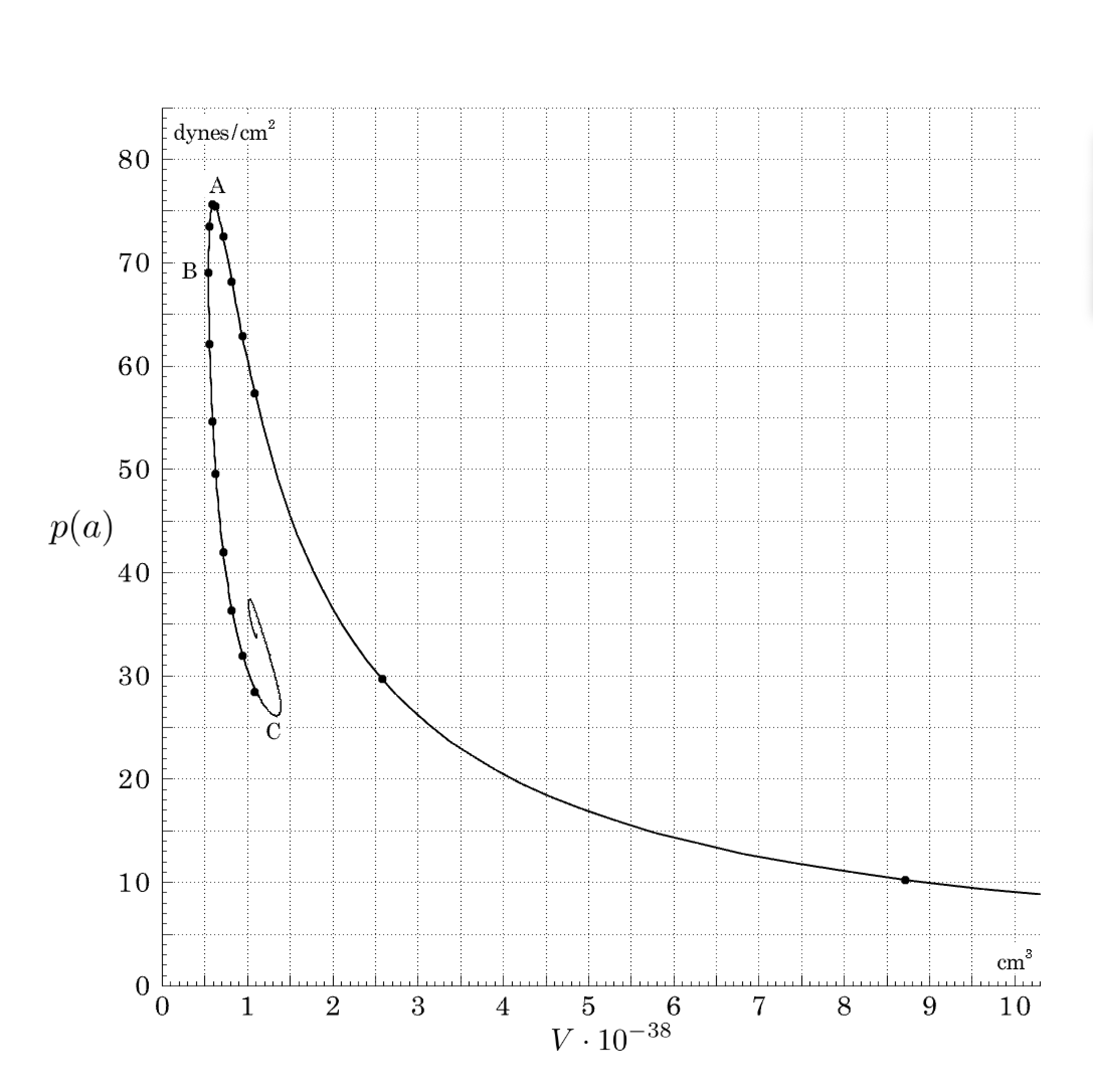}}
  \subfigure[\ In the corresponding nondimensional form]{\label{p.pwall.b.nd}
  \includegraphics[keepaspectratio=true, trim= 1ex 8ex 4ex 20ex , clip , width=\columnwidth]{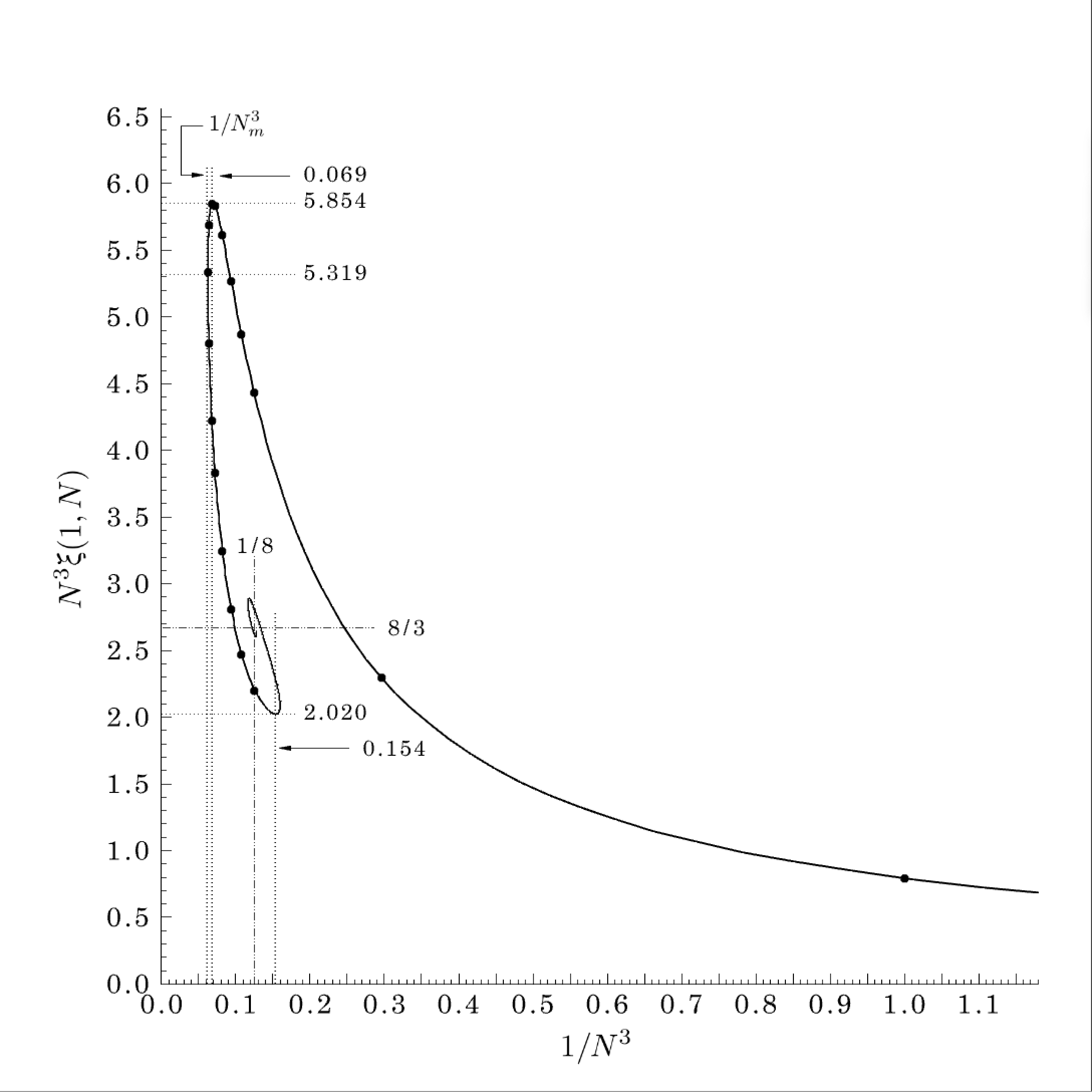}}
  \caption{Isotherm \mbox{$T=273.15\,$}K in the $p(a),V$ plane; reproduction of Fig. 1 at page 355 of \Ref{wb1956mnras}\hfill\ }\label{p.pwall.b}
\end{figure}\noindent
then making the average pressure [\REqq{agdp}] explicit in \REq{se.p}
\begin{equation}\label{se.p.s2}
   \frak{p} = \frac{m_{g}RT}{V}\left( \xiy(1,N) - \dfrac{\Phi}{6} N \right) 
\end{equation}
and finally replacing the volume in \REq{se.p.s2} with \REq{gcn.V} to obtain
\begin{subequations}
	\begin{equation}\label{se.p.s3}
	   \frak{p} = \frac{(RT)^{4}}{\tfrac{4}{3} \piy G^{3} m_{g}^{2}} N^{3} \left( \xiy(1,N) - \dfrac{\Phi}{6} N \right) 
	\end{equation}
	\REqdb{gcn.V}{se.p.s3} are the isotherms' parametric equations and we can conveniently plot them in nondimensional form.
	But first a crosscheck to validate them.
	Bonnor \cite{wb1956mnras} provided isotherms' parametric equations, just below his Eq.~(2.17) at page 355, constructed on and numerically processed with the data of the isothermal-sphere solution of Emden \cite{re1907}.
	He applied his equations to a sphere of \mbox{$m_{g}=10^{30}$} gm ($10^{27}$ kg) of molecular hydrogen at \mbox{$T=0\,^{\circ}$}C (273.15\,K) but did not consider the presence of a shell; therefore, we have to set $\Phi=0$ in \REq{se.p.s3} and use
	\begin{equation}\label{se.p.s3.bonnor}
	   \frak{p}_{\mbox{\tiny\cite{wb1956mnras}}} = p(a) = \frac{(RT)^{4}}{\tfrac{4}{3} \piy G^{3} m_{g}^{2}} N^{3}\xiy(1,N)
	\end{equation}
	to compare results.
\end{subequations}
As visual proof of the equivalence between Bonnor's parametric equations and our \REqd{gcn.V}{se.p.s3.bonnor}, we show the reproduction of Bonnor's Fig.~1 styled according to the original's CGI units in \Rfi{p.pwall.b.d} and in the corresponding nondimensional form in \Rfi{p.pwall.b.nd}.
\begin{figure}[h]
  \includegraphics[keepaspectratio=true, trim= 1ex 8ex 4ex 18ex , clip , width=\columnwidth]{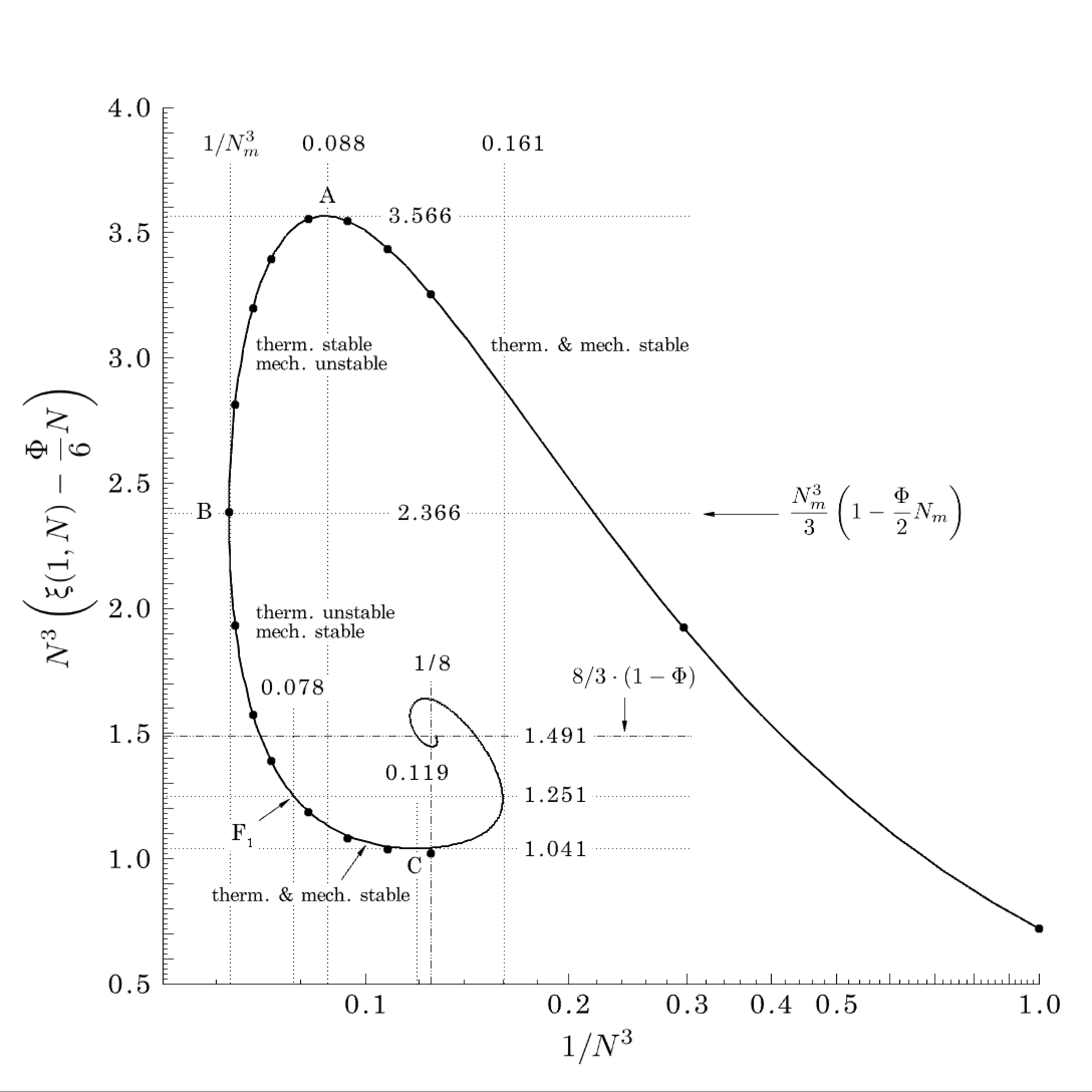}
  \caption{Isotherms' nondimensional profile [\REqd{gcn.V}{se.p.s3}]; $\Phi\simeq 0.441$ in our calculations [\REq{g.ener.shell.our.case}]\hfill\ }
  \label{p.pwall}
\end{figure}
Returning now to the more complete case with inclusion of the shell's term [\REq{se.p.s3}], we show the isotherms' nondimensional profile in \Rfi{p.pwall}, in logarithmic scale for clearer visualization of the characteristic points labeled A,B,C in compliance with Bonnor's notation; for convenience, we have also included the cusp F$_{1}$.
Their locations depend, in general, on the function $\Phi$.
The spiral center is positioned at \mbox{$N=2\;\mbox{or}\;1/N^{3}=1/8$} and at the level $8/3(1-\Phi)$; if \mbox{$\Phi=0$}, the level reduces to $8/3$ in full agreement with Bonnor's results.
The left bound B is located at $N=N_{m}\;\mbox{or}\;1/N_{m}^{3}\simeq 0.063$ and at the level $N_{m}^{3}(1-\Phi N_{m}/2)/3$.
The determination of the extrema A, C, etc., requires the vanishing of the derivative
\begin{multline}\label{der.pfrak}
    \pds{}{\frak{p}}{V}{T,m_{g}} =  \\[.25\baselineskip]
              - \frac{\bpres}{V}\left[ \xiy(1,N) + \frac{N}{3} \left( \xiy'(1,N) - \frac{2}{3}\Phi \right)   \right]
\end{multline}
which follows smoothly from \REqd{gcn.V}{se.p.s3} by logarithmic differentiation; the nondimensional profile of the negative isotherm's slope is shown in \Rfi{p.pwall.slope} to facilitate thermodynamic-stability considerations.
\begin{figure}[h]
  \includegraphics[keepaspectratio=true, trim= 1ex 8ex 4ex 18ex , clip , width=\columnwidth]{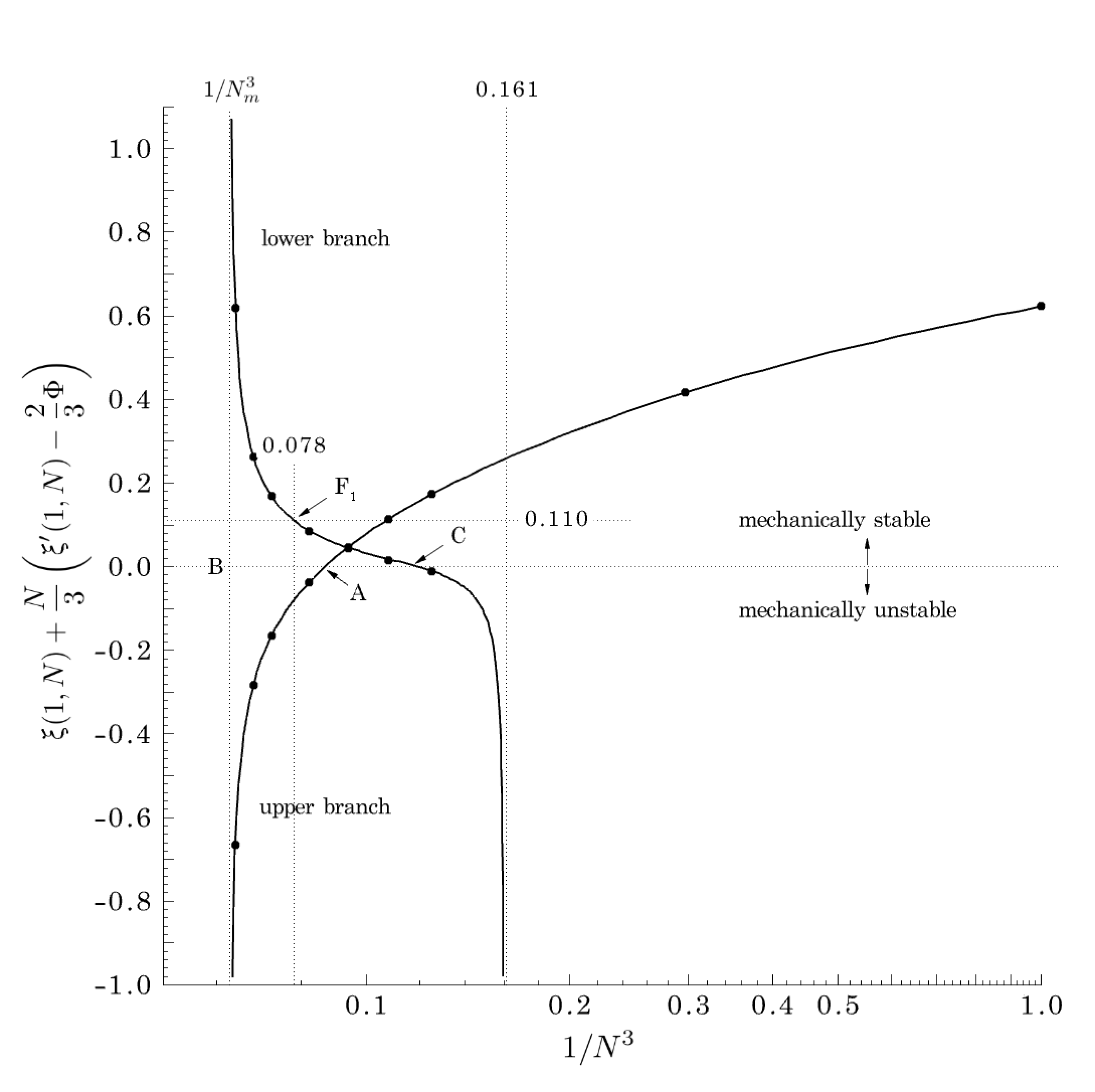}
  \caption{Nondimensional profile of negative isotherm's slope [\REqd{gcn.V}{der.pfrak}]; lower branch and upper branch between asymptotes (1.84,$N_{m}$) corresponding to (0.161,$1/N_{m}^{3}$)\hfill\ }
  \label{p.pwall.slope}
\end{figure}
Unfortunately, the zeros of \REq{der.pfrak} cannot be found analytically.
In our calculations, $\Phi\simeq 0.441$ [\REq{g.ener.shell.our.case}] and the numerically found values for the positions of A,C are indicated in \Rfi{p.pwall}; the values corresponding to Bonnor's case ($\Phi=0$) are indicated in \Rfi{p.pwall.b.nd}.
The profiles in \Rfid{p.pwall}{p.pwall.slope} give unequivocal evidence of the existence of a) configurations that can simultaneously be thermally stable/unstable and mechanically unstable/stable and b) upper-branch configurations that are mechanically unstable.
Indeed, the configurations to the right of A \mbox{($N_{\mbox{\tiny A}}\simeq 2.25$ or $1/N^{3}_{\mbox{\tiny A}}\simeq 0.088$)} and on the arc AB belong to the upper branch and are all thermally stable (\Rfid{p.phient1}{p.fr.ener.z.us}).
Those to the right of A are also mechanically stable because \mbox{$-\!\pdst{}{\frak{p}}{V}{T,m_{g}}>0$} for them;
instead, those on the arc AB have \mbox{$\,-\!\pdst{}{\frak{p}}{V}{T,m_{g}}<0$} and are mechanically unstable.
The stability condition swaps for the configurations on the arc BF$_{1}$: they are thermally unstable but mechanically stable because \mbox{$-\!\pdst{}{\frak{p}}{V}{T,m_{g}}>0$} again.
The described pattern repeats endlessly as the profile keeps spiraling.

We are aware that Bonnor's \cite{wb1956mnras} and our axiomatic-thermodynamics flavored conclusions differ. 
We have done our homework to follow his reasoning, presented at page 356, but have not been able to reconcile the differences.
First of all, we have verified analytically that his \mbox{Eq. (2.16)} is contained in our \REq{der.pfrak}; with a bit of attention to notation conversion, the verification is rather easily accomplished by substituting \REq{dxi.a} in \REq{der.pfrak} and then setting $\Phi=0$. 
After that, we could consider with reassurance the graph of Bonnor's \mbox{Eq. (2.16)}, unfortunately not shown in \cite{wb1956mnras}, adequately represented by the profile of \Rfi{p.pwall.slope}; the locations of A and C are slightly displaced if \mbox{$\Phi=0$} but the aspect of the profile in terms of slope and curvature remains similar.
Subsequently, we have proceeded to further derivation of \REq{der.pfrak} to determine the second derivative   
\begin{multline}\label{der2.pfrak}
   \pds{2}{\frak{p}}{V}{T,m_{g}} =  -\frac{1}{V} \pds{}{\frak{p}}{V}{T,m_{g}} \\[.25\baselineskip]
            -\frac{\bpres}{V^{2}N^{3}} \pd{}{}{(1/N^{3})}\left[ - \frac{V}{\bpres} \pds{}{\frak{p}}{V}{T,m_{g}}\right] 
\end{multline}
because it is a key element in Bonnor's argumentation.
The partial derivative on the bottom line of \REq{der2.pfrak} is the slope of the profile in \Rfi{p.pwall.slope}; it does not need to be expanded for the purpose of the present discussion.
\REqb{der2.pfrak} informs that, in A and C, such a slope and the second derivative $\pdst{2}{\frak{p}}{V}{T,m_{g}}$ are directly proportional because the first derivative $\pdst{}{\frak{p}}{V}{T,m_{g}}$ vanishes.
With these preliminaries in hand, we concentrated on Bonnor's text.
At page 356, just below \mbox{Eq. (3.2)}, he considered the situation at A and recognized the negativity of the second derivative
\begin{quote}
    Let $V=v$ be the smallest volume for which (3.1) (and therefore $\pdt{}{p}{V}$) becomes zero; then one easily finds  \[ \pds{2}{p}{V}{N,T(V=v)} < 0 .  \]
\end{quote}
although without specifying how to find it analytically. 
Geometrically, one observes the curvature at A in the profiles of \Rfi{p.pwall.b.d} or \Rfi{p.pwall}; analytically, \REq{der2.pfrak} settles the second-derivative negative sign because the slope in the profile of \Rfi{p.pwall.slope} at A is positive.
Then Bonnor considered the consequence of a small fluctuation and drew the correct conclusion
\begin{quote}
    Thus for a small fluctuation $dV$ in the volume of the sphere of mean volume $v$, we have
    \[ dp = \pds{2}{p}{V}{V=v} dV^{2} + O(dV^{3}) < 0.  \]
    Therefore if a fluctuation occurs which result in a slight decrease in the volume of the sphere, this will produce a decrease in the pressure inside its boundary which will lead to a further reduction in its volume.
\end{quote}
Now, the reasons that have lead Bonnor to draw from the quoted statements both the specific conclusion
\begin{quote}
    Thus an isothermal gas sphere of volume greater than $v$ will be unstable because small fluctuations will make the volume $v$ collapse towards the centre.
\end{quote}
and the general conclusion at page 357 
\begin{quote}
   \ldots we can now see that any point on the spiral part of the curve below A refers to a state of unstable equilibrium.
\end{quote}
meticulously supported by the remark
\begin{quote}
   It is necessary to point this out because along BC, for example, $\pdet{p}{V}$ is positive, so it might seem that this part of the curve refers to stable states.
\end{quote}
remain obscure to us.
As a matter of fact, the repetition of Bonnor's reasoning for the configuration C brings to the inescapable conclusion of mechanical stability.
The volume at C is greater than the volume at A (Bonnor's $v$). 
The slope of the profile in \Rfi{p.pwall.slope} is negative at C and \REq{der2.pfrak} yields a positive second derivative; therefore, a small volume-decreasing fluctuation around C will produce an increase in the peripheral pressure so that the gas sphere will \textit{not} "collapse towards the center" but will return to C. 

\subsubsection{State equation \mbox{$\pdst{}{E}{m_{g}}{S,V}$}}
For the sake of completeness, we recall that there is a third partial derivative \mbox{$\pdst{}{E}{m_{g}}{S,V}$} hardly considered or even mentioned in the literature.
We adapt one more time the recipe described in the paragraph just before \REq{se.T} and obtain
\begin{align}\label{se.g}
    \frak{g}  & =  \pds{}{E}{m_{g}}{S,V} \notag \\[.25\baselineskip]
               & = - R T \left\{ C - \frac{5}{2} + \frac{3}{2}\ln T + \ln\frac{V}{m_{g}} \rule{0pt}{3.5ex}\right. \notag \\[.25\baselineskip]
               & \hspace*{5.3em} + \frac{s_{gi}(T)}{R} \!-\! \frac{u_{gi}(T)}{RT} \! + \! \frac{m_{s}}{m_{g}} \left(\frac{s_{s}(T)}{R} \! - \! \frac{u_{s}(T)}{RT} \right)  \notag \\[.25\baselineskip]
               & \hspace*{5.3em} + N - \ln \xiy(1,N) + \Phi N \left.\rule{0pt}{3.5ex}\right\} 
\end{align}
The state equation defined by \REq{se.g} is important because its derivatives \mbox{$\pdst{}{\frak{g}}{V}{T,m_{g}}$},  \mbox{$\pdst{}{\frak{g}}{m_{g}}{T,V}$} together with \mbox{$\pdst{}{\frak{p}}{V}{T,m_{g}}$} should be involved in a third criterion of thermodynamic stability connected to the physical-system response to mass variations.
However, pending our task (last paragraph of \Rse{entr}) of carrying out a thermodynamic-stability analysis within an axiomatic-thermodynamics framework, here we simply limit ourselves to mention the existence of the third stability criterion and the necessity to explore it but prefer to refrain from engaging prematurely and intuitively into details, necessarily extrapolated from our experience with thermodynamic systems without gravitational fields, that could turn out to be incomplete and/or inaccurate. 

\subsubsection{Fundamental-relation inhomogeneity}
The state-equation definitions [\REq{se.T.1}, \REq{se.p} and \REq{se.g}] legitimate the Gibbs equation both in the energetic representation
\begin{equation}\label{gibbs.E}
   dE = TdS - \frak{p}dV + \frak{g} dm_{g}
\end{equation}
and in the entropic representation
\begin{equation}\label{gibbs.S}
   dS = \frac{1}{T}dE + \frac{\frak{p}}{T}dV - \frac{\frak{g}}{T} dm_{g}
\end{equation}
The Euler equation, however, does not hold because entropy and total energy do not possess first-order homogeneity; indeed, it requires a simple and straightforward algebra to obtain 
\begin{align}\label{euler.E}
   \frac{E  - \left(TS - \frak{p}V + \frak{g}\,m_{g} \right)}{m_{g}RT}  &  = - \frac{S - \dfrac{E}{T} - \dfrac{\frak{p}}{T}V + \dfrac{\frak{g}}{T}\,m_{g}}{m_{g}R}\notag \\[.5\baselineskip]
     &  = 2 \left( 1- \xiy(1,N) \right) + \frac{\Phi}{3} N 
\end{align}
The term on the bottom line of \REq{euler.E} measures the \textit{in}homogeneity of entropy and energy; its nondimensional profile is shown in \Rfi{p.hcheck}.
\begin{figure}[h]
  \includegraphics[keepaspectratio=true, trim= 1ex 8ex 4ex 18ex , clip , width=\columnwidth]{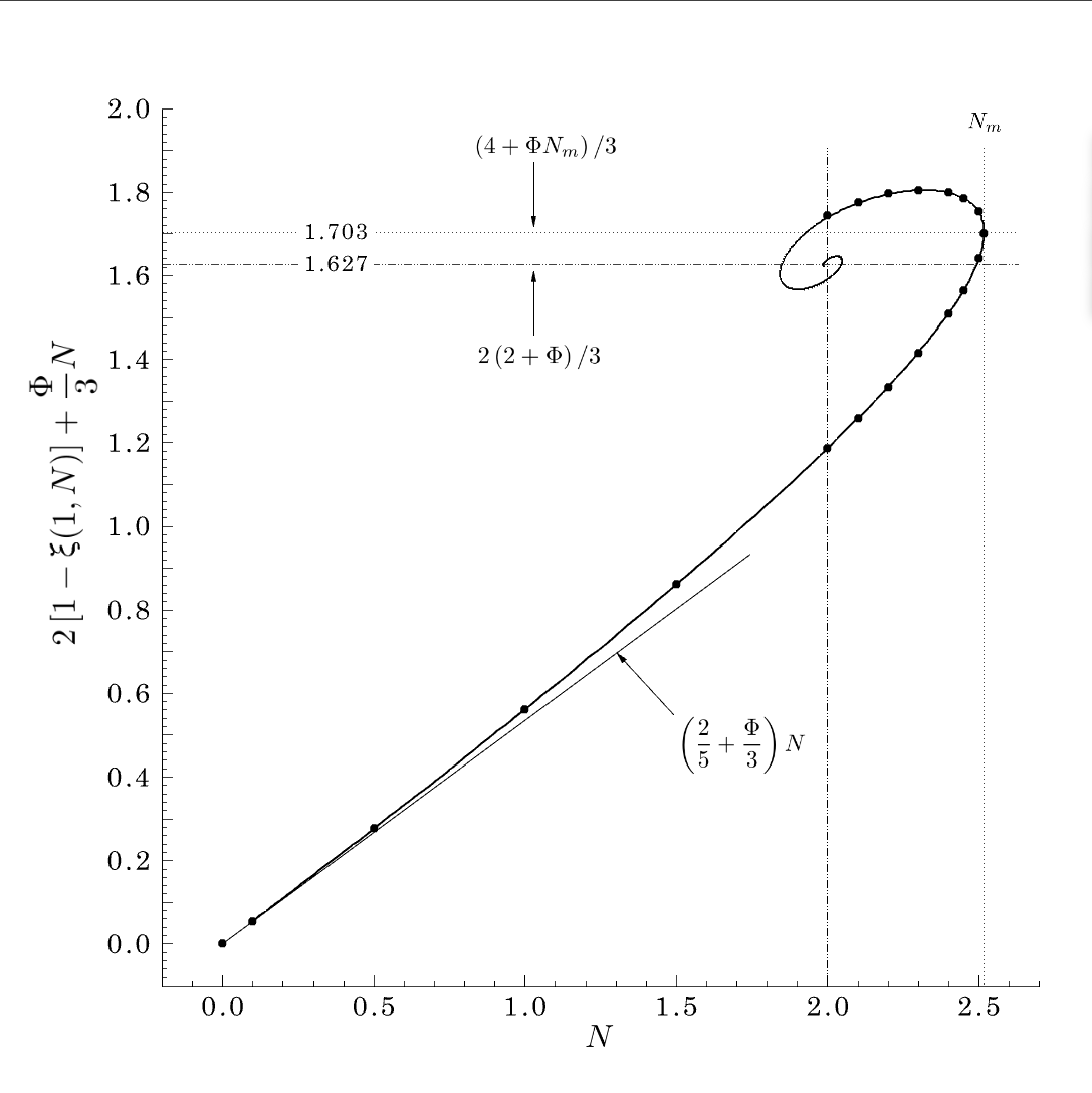}
  \caption{Nondimensional profile of energy and entropy inhomogeneity\hfill\ }
  \label{p.hcheck}
\end{figure}

\section{Conclusions and future work}\label{prel.con}
The main objective of this paper, described in the paragraph beginning after \Rfi{sphere}, has been achieved: the gravitofluid-static fields have been determined and they are at our disposal as initial fields to start gravitofluid-dynamic calculations. 
We have followed a solution pathway different [\REq{bc.rho.r=a} and \REq{bc.rho.r=a.nd.cn.m}] from the widely adopted one on which the Lane-Emden solution is based but most of our results are aligned with those of the latter approach. 
We have learned that the gravitofluid-static problem [\REqq{nd.prob.cn.m}] hinges on one single characteristic number [\REq{gcn1}] that we call the gravitational number because it measures the importance of gravity effects.
With respect to uniform gas conditions and gravitational-field linear profile prevailing when the gravitational number is vanishingly small, density/pressure and gravitational-field profiles (\Rfid{p.xi}{p.gf.g}) vary smoothly and monotonically with increasing gravitational number until, somewhat surprisingly, the latter reaches an upper bound \mbox{$(N_{m}\simeq2.51755148)$} beyond which no gravitofluid-static configuration exists.
For a prescribed gravitational number belonging to the interval \mbox{$[1.84 , N_{m}]$}, we have also found out (\Rfi{p.xi1}) the existence of multiple solutions.
These findings are in accordance with what already known in the literature and generated in our mind questions, formulated in the last paragraph of \Rse{ge+bc.nd.res}, whose answers' quest lead us to investigate the thermodynamics of the physical system, specifically entropy (\Rse{entr}) and total energy (\Rse{ener}) with the aim in mind to derive the all-governing fundamental relation according to the guidance of axiomatics thermodynamics \cite{hc1963,lt1966,ln1971,hc1985} (\Rse{thd.con}). 
In general, thermodynamic properties comply with a separation structure composed by a standard term when gravitational effects are absent plus a gravitational correction [\REqq{entr.g} for example] that removes first-order homogeneity because of the long-range nature of gravity forces.
This was our first contact with non-extensive thermodynamics and we really experienced in full the excitement noted in a comment of Prof. Landsberg \cite[end of first paragraph on page 49]{pl1987jnet}
\begin{quote}
   Here we have a thermodynamic system which is different and therefore of exciting novelty.
\end{quote}
With the fundamental relation in hand, we obtained both the first derivatives [\REq{se.T.1}, \REq{se.p}, \REq{se.g}], i.e. the state equations, and the second derivatives [\REq{p.fr.ener.c}, \REq{der.pfrak}] that, by conceptual extrapolation from our past familiarity with thermodynamics without gravitational fields, we believe should have a direct bearing on the thermodynamic stability of the physical system.
The gravitational correction to the gas entropy (\Rfi{p.Ient}) permits unambiguously to distinguish upper and lower branches of gravitofluid-static configurations.
It turns out that all upper-branch configurations are thermally stable (\Rfi{p.phient1}, \Rfi{p.fr.ener}). 
Regarding the lower-branch configurations (\Rfi{p.phient}, \Rfi{p.fr.ener.z}), some of them are thermally unstable and, therefore, not physically realizable; the first series begins with the configuration relative to the upper-bound value \mbox{$N_{m}$} of the gravitational number;  the latter's origin, therefore, appears to be of thermodynamic nature.
We could take the last statement as an, admittedly vague, answer to question (a) (last paragraph of \Rse{ge+bc.nd.res}) although we also have to surrender humbly to the same defeat acknowledged by Darwin at page 19 of \cite{gd1889ptrs}
\begin{quote}
   ... I am unable to find any analytical relationships by which the minimum value of $\frac{1}{3}\betay^{2}$ can be deduced.
\end{quote}
130 years ago; in Darwin's notation, $\frac{1}{3}\betay^{2}$ corresponds to our $1/N$.
Other lower-branch configurations are as legitimately thermally stable as those of the upper branch and we could not find any physical argument within the gravitofluid-statics context preventing their realizability.
We entrust our planned gravitofluid-dynamics study with the hope to shed light on the multiple-solution conundrum.
In the thermodynamic plane $E, S$ (\Rfi{p.fr.ener.z}), series of thermally stable and unstable configurations alternate and are separated by cusps; the cusp F$_{1}$ is important because it flags limiting values of entropy and total energy to the left of or below which the physical system must necessarily be in a gravitofluid-dynamic condition.
Moving on to mechanical stability (\Rfi{p.pwall}, \Rfi{p.pwall.slope}), the striking feature is the existence of upper-branch configurations (arc AB in \Rfi{p.pwall}) that are mechanically unstable, their thermal stability notwithstanding.
For the lower-branch configurations, we find on the isotherm's profile the same alternating pattern of mechanically stable and unstable series as we have seen to exist on the fundamental-relation profile regarding thermal stability.
We have mentioned the existence of a third criterion of thermodynamic stability [paragraph after \REq{se.g}] connected to mass variations that should be considered on equal footing with the other two criteria we have discussed.
We did not feel comfortable, though, to plunge into third-criterion details and proceed with intuitive elaborations; we preferred to hold on with this matter in view of our planned task to carry out a thorough thermodynamic-stability study along the axiomatic-thermodynamics guidelines.
The latter constitutes one step of our future work. 
The other steps will be gravitofluid-dynamics computational studies of our test case. 
The first one will concentrate on a spherical-symmetric motion of the gas; the targets are clearly the answers to questions (b) and (c) and, additionally, the verification of the consequences predicted by the thermodynamic-stability criteria.
The second one will focus on an axisymmetric motion of the gas compatible with both Netwon's and gravitomagnetic theories of gravity in the hope to bring to light, at least qualitatively, detectable differences in the respective flow fields.




\bibliographystyle{apsrev}        
\bibliography{mybibreflibrary,./ffp.bib}   

\end{document}